\documentclass{article}
\usepackage{amsfonts}
\usepackage{amsmath}
\usepackage{makeidx}
\usepackage{eurosym}
\usepackage{amssymb}
\usepackage{graphicx}
\usepackage{caption}
\usepackage{wrapfig}
\usepackage{subfig}
\usepackage{rotating}
\usepackage{natbib}
\usepackage{appendix}

\newtheorem{theorem}{Theorem}
\newtheorem{acknowledgement}[theorem]{Acknowledgement}

\input{tcilatex}
\begin{document}

\title{Kelvin Waves, Klein-Kramers and Kolmogorov Equations, Path-Dependent
Financial Instruments: Survey and New Results}
\author{A. Lipton \\
ADIA, Abu\ Dhabi, UAE\thanks{%
Global Head, Research and Development,\ Strategy and Planning Department} \\
ADIA\ Lab, Abu\ Dhabi, UAE\thanks{%
Advisory Board Member} \\
The Hebrew University of Jerusalem, Jerusalem, Israel\thanks{%
Visiting Professor and Dean's Fellow, Business School} \\
Khalifa University, Abu\ Dhabi, UAE\thanks{%
Professor of Practice, Department of Mathematics} \\
MIT, Cambridge, MA, USA\thanks{%
Connection Science Fellow, Media Lab}}
\maketitle

\begin{abstract}
We discover several surprising relationships between large classes of
seemingly unrelated foundational problems of financial engineering and
fundamental problems of hydrodynamics and molecular physics. Solutions in
all these domains can be reduced to solving affine differential equations
commonly used in various mathematical and scientific disciplines to model
dynamic systems. We have identified connections in these seemingly disparate
areas as we link together small wave-like perturbations of linear flows in
ideal and viscous fluids described in hydrodynamics by Kevin waves to
motions of free and harmonically bound particles described in molecular
physics by Klein-Kramers and Kolmogorov equations to Gaussian and
non-Gaussian affine processes, e.g., Ornstein-Uhlenbeck and Feller, arising
in financial engineering. To further emphasize the parallels between these
diverse fields, we build a coherent mathematical framework using Kevin waves
to construct transition probability density functions for problems in
hydrodynamics, molecular physics, and financial engineering. As one of the
outcomes of our analysis, we discover that the original solution of the
Kolmogorov equation contains an error, which we subsequently correct. We
apply our interdisciplinary approach to advance the understanding of various
financial engineering topics, such as pricing of Asian options, volatility
and variance swaps, options on stocks with path-dependent volatility, bonds,
and bond options. We also discuss further applications to other exciting
problems of financial engineering.


\end{abstract}

\tableofcontents

\section{Introduction\label{Sec1}}

\subsection{Background\label{Sec11}}

The Fourier transform is the most powerful tool of an applied mathematician,
who can use it to solve linear partial differential equations (PDEs) with
spatially constant coefficients, analyze time series and perform many other
vital tasks. The principal building block of the n-dimensional Fourier
method is the wave function:

\begin{equation}
\begin{array}{c}
H\left( t,\mathbf{x,k}\right) =a\left( t\right) \exp \left( i\mathbf{k}\cdot 
\mathbf{x}\right) ,%
\end{array}
\label{Eq1.1}
\end{equation}%
where $\mathbf{x}$\ and $\mathbf{k}$ are $n$-dimensional vectors, $\cdot $
denote the scalar product, $a\left( t\right) $ is the amplitude, and $%
\mathbf{k}\cdot \mathbf{x}$ is the phase. Depending on the circumstances,
the amplitude can be a scalar or a vector. Substituting $H$ into a PDE with
spatially constant coefficients, we reduce the problem of interest to a
system of ordinary differential equations (ODEs) or a single ODE when $%
a\left( t\right) $ is scalar.

In this paper, we study PDEs with coefficients linearly dependent on $x$,
which we call affine PDEs. Hence, we must use a more general approach and
consider the wave function with a time-dependent wave vector:

\begin{equation}
\begin{array}{c}
H\left( t,\mathbf{x,k}\right) =a\left( t\right) \exp \left( i\mathbf{k}%
\left( t\right) \cdot \mathbf{x}\right) ,%
\end{array}
\label{Eq1.2}
\end{equation}

To the best of our knowledge, \cite{Kelvin87} and \cite{Orr07} were the
first to use such building blocks to analyze the stability of the steady
motions of a perfect fluid. Subsequently and independently, affine PDEs and
the associated wave functions were used by many researchers in various
areas, including the theory of stochastic processes, physics, biology, and
mathematical finance, to mention a few. In this respect, the
Ornstein-Uhlenbeck (OU) and Feller processes are archetypal; see \cite%
{Uhlenbeck30, Chandrasekhar43, Feller51, Feller52}. For financial
applications of affine processes see \cite{Duffie96, Duffie00, Dai00,
Duffie03, Filipovic09}, among others.

We use Kelvin waves of the form (\ref{Eq1.2}) in order to study transition
probability density functions (t.p.d.fs) for affine stochastic processes,
which can be either degenerate, i.e., having more independent components
than the sources of uncertainty, or non-degenerate, when every component has
its source of uncertainty. Our main objective is to link various problems of
financial engineering with their counterparts occurring in fluid dynamics
and molecular physics and showcase the interdisciplinary nature of
quantitative finance and financial modeling. Finding such connections allows
us to understand better how to model, price, and risk-manage various
financial instruments and provide additional intuition regarding their
salient features. This work continues our previous efforts in this
direction; see \cite{Lipton08, Lipton18}, Chapter 12.

There are several approaches one can use to solve affine equations
efficiently. For instance, Lie symmetries are a powerful tool for studying
certain classes of affine equations. Numerous authors describe general
techniques based on Lie symmetries; see, e.g., \cite{Ovsiannikov82,
Ibragimov85, Olver86, Bluman1989}, while their specific applications are
covered by \cite{Berest93, Aksenov95, Craddock04, Craddock12, Kovalenko14},
among many others. Reductions of a given equation to a simpler, solvable
form is another powerful method that can be successfully used in many
instances; see, e.g., \cite{Chandrasekhar43, Carr02, Lipton14, Lipton18},
Chapter 9. Finally, Kelvin waves (or affine ansatz) provide another
approach, which is the focus of the present paper; see also \cite{Duffie96,
Dai00, Duffie03, Lipton08, Filipovic09, Lipton18}, Chapter 12.

\subsection{Main Results and Paper Structure\label{Sec12}}

This paper builds a coherent mathematical framework using Kelvin waves as a
powerful and versatile tool for studying t.p.d.fs in the context of generic
affine processes. We discover previously uncovered connections among large
classes of apparently unrelated problems from hydrodynamics, molecular
physics, and financial engineering. All these problems require solving
affine differential equations, i.e., equations with coefficients, which
linearly depend on spatial variables. We discuss some classical results and
derive new ones related to 1) small wave-like perturbations of linear flows
of ideal and viscous fluids described by Euler and Navier-Stokes equations,
respectively; 2) motions of free and harmonically bound particles described
by the Klein-Kramers equations and the hypoelliptic Kolmogorov equation; and
3) dynamics of financial assets and market movements driven by Gaussian and
non-Gaussian affine processes, such as the Ornstein-Uhlenbeck and Feller. We
augment these processes by introducing subordinate processes for auxiliary
variables, such as integrals over the original mean-reverting variable. We
develop a uniform mathematical formalism to construct t.p.d.fs for the
abovementioned problems.

Our analysis identifies an error in the original solution of the Kolmogorov
equation, which we subsequently rectify. This work, in addition to
recovering and improving several well-known results, also derives new ones.
For instance, we find a considerably more convenient expression for t.p.d.fs
in the case of processes with stochastic volatility. Moreover, our analysis
reveals an unexpected similarity between the propagation of vorticity in
two-dimensional flows of viscous incompressible fluid and the motion of a
harmonically bound particle, which we use to find a new explicit expression
for the vorticity of a two-dimensional flow in terms of the Gaussian density.

Finally, we apply the new methodology to various financial engineering
topics, such as pricing Asian options, volatility and variance swaps,
options on stocks with path-dependent volatility, and bonds and bond
options. In contrast to the classical treatment, we view fixed-income
products as path-dependent, which allows us to gain additional intuition
regarding their pricing and risk management. We also highlight the
flexibility of our interdisciplinary framework to incorporate additional
complexities and to have natural applications to other financial engineering
problems, such as jump-diffusion processes (verbatim) and, more generally,
processes driven by affine pseudo differential equations.

The paper is organized as follows. Section \ref{Sec2} introduces the Euler
equations describing the evolution of the perfect fluid. We show that these
equations have exact solutions with the velocity linearly and pressure
quadratically depending on spatial coordinates. The linearized Euler
equations govern small perturbations of these solutions known as the
celebrated Kelvin waves. We demonstrate how to use Kelvin waves to study the
stability of the exact linear solutions. Subsequently, we utilize Kevin
waves as our primary tool in our methodology and apply it to other areas of
study. We also briefly cover the Navier-Stokes equation for the viscous
fluid and show how to modify the ODEs for Kelvin waves to account for
viscosity. In Section \ref{Sec55}, we return to this topic, establishing
deep connections between the two-dimensional vorticity equations and the
Klein-Kramers equation underpinning molecular physics, which allows us to
derive a new formula expressing the vorticity as a Gaussian density, and the
stream function as a solution of the corresponding Poisson equation.

Section \ref{Sec3} deals with some representative stochastic problems
occurring in physics. Namely, Section \ref{Sec31} introduces the Langevin
equation describing the motion of the underdamped Brownian particle. The
Klein-Kramers equation describing the motion of such particle is derived in
Section \ref{Sec32}. Its solutions found by Chandrasekhar for a free and a
harmonic particle are given in Section \ref{Sec33}. The Klein-Kramers
equation is naturally degenerate since there is a source of white noise for
the particle velocity but not for its position. As we shall see later, many
path-dependent mathematical finance problems fall into this category,
provided the corresponding financial variables, such as the geometric price
averages naturally occurring as underlyings for Asian options, are
interpreted as path integrals. Section \ref{Sec4} studies the celebrated
degenerate stochastic process introduced by \cite{Kolmogorov34}, which is
closely linked with the Klein-Kramers equation studied in \ref{Sec3}, the
associated Fokker-Planck equation, and its solution proposed by Kolmogorov.
Section \ref{Sec41} summarizes original Kolmogorov's results. In Section \ref%
{Sec42}, to our surprise, we show that the Fokker-Planck equation used by
Kolmogorov in his original paper is inconsistent with his assumptions about
the underlying process. Moreover, his proposed solution lacks dimensional
consistency and must be rectified to satisfy the Fokker-Planck equation and
the initial condition. Thankfully, not all is lost, and Kolmogorov's
solution can be corrected, as we demonstrate in Sections \ref{Sec43}, \ref%
{Sec44}. We start with direct calculation in Section \ref{Sec43} and then
arrive at the same corrected Kolmogorov's solution by using Kelvin waves in
Section \ref{Sec44}.

Sections \ref{Sec5}, \ref{Sec6} constitute the heart of the paper. Section %
\ref{Sec5} performs a detailed analysis of t.p.d.fs for Gaussian affine
processes, whose drifts are linearly dependent on the spatial coordinates,
while covariance matrices are spatially (but not necessarily temporary)
invariant. Section \ref{Sec6} investigates t.p.d.fs for non-Gaussian
processes with drift and covariance matrix linearly dependent on spatial
coordinates. Our methodology demonstrates that Kelvin waves are ideally
suited for studying such affine processes as the OU and Feller processes
that often occur in finance, asset management, and trading.

Section \ref{Sec5} applies our methodology to Gaussian processes. Section %
\ref{Sec51} derives the general formula for the Gaussian dynamics, which
covers both degenerate (like in Kolmogorov's case) and non-degenerate
situations. In Section \ref{Sec52}, we derive the t.p.d.f. for the
Kolmogorov process with time-dependent coefficients. Section \ref{Sec53},
investigates the OU process with time-dependent coefficients and the
augmented OU process, which describes the joint dynamics of the process and
its integral. While the results are classical, their derivation based on
Kelvin wave expansions complements them naturally and adds value to the
analysis by providing a different perspective to understand and derive these
well-established results. Section \ref{Sec54} investigates the case of free
and harmonically bound particles and compares and contrasts the Kelvin wave
approach with the classical one developed by \cite{Chandrasekhar43}.
Finally, Section \ref{Sec55} goes back to our departure point described in
Section \ref{Sec2}. It shows that the temporal-spatial evolution of the
two-dimensional flow's vorticity for a viscous fluid is very similar to the
dynamics of a harmonically bound particle. This is a notable and novel
result that bridges seemingly disparate underlying physical mechanisms.

Non-Gaussian processes are covered in Section \ref{Sec6}. Section \ref{Sec61}
derives the general formula for the non-Gaussian dynamics, which covers both
degenerate and non-degenerate possibilities. Section \ref{Sec62} treats the
Feller process with constant coefficients and two augmented Feller processes
- a degenerate linking the Feller process with its integral and a
non-degenerate describing an arithmetic Brownian motion with stochastic
volatility driven by Feller process. As a notable byproduct of our analysis,
we develop an alternative approach for analyzing finite-time explosions for
exponential moments for t.p.d.fs of augmented Feller processes. Finally,
Section \ref{Sec63} applies our methodology to degenerate and non-degenerate
arithmetic Brownian motions with stochastic volatility driven by an OU
process.

Section \ref{Sec7} extends our framework to the realm of financial
applications. Section \ref{Sec71} briefly outlines the basics of financial
engineering based on the geometric Brownian motion paradigm. Since this
paradigm is too narrow to be practical, we introduce modifications to better
capture the complexities and nuances of financial markets. It turns out that
our new methodology is versatile enough to analyze numerous practically
important cases. In Section \ref{Sec72}, we describe how Asian options with
geometric averaging can be priced by using the corrected version of the
Kolmogorov solution. Section \ref{Sec73} applies the results of Sections \ref%
{Sec53} and \ref{Sec62} to the pricing of volatility and variance swaps and
swaptions, respectively. In Section \ref{Sec74}, we introduce a
path-dependent volatility model and show how to use an augmented OU process
discussed in Section \ref{Sec63} in the context of moving particles to
construct a novel analytically solvable model. Section \ref{Sec75} expands
our methodology to cover the Vasicek-Hull-White and Cox-Ingersoll-Ross (CIR)
frameworks for bonds and bond options. In contrast to the conventional
treatment based on solving the backward Kolmogorov equation for the killed
process, we prefer to treat bonds and bond options as path-dependent
instruments. This approach enables us to utilize formulas derived in
Sections \ref{Sec5}, \ref{Sec6} to price fixed-income products. These
formulas give us additional insights into pricing and risk management of
bonds and bond options and connect the corresponding pricing formulas with
the ones derived in physics for describing the motion of underdamped
Brownian particles. Section \ref{Sec76} briefly discusses the potential
applications of Kelvin waves to other financial engineering problems.

Section \ref{Sec8} contains our conclusions. Appendices \ref{AppA}, \ref%
{AppB}, discuss killed affine processes.

A word about notation is in order. By its very nature, papers like the
present one cannot use the unified notation. The reason is simple - the
ground is too broad and stretches from fluid dynamics to molecular physics,
probability theory, and financial engineering. Each of these fields uses its
own set of notations, which are not coherent across different areas. The
best we could do is to use consistent notation in each Section and, as much
as possible, across sections, but the reader must be alert to keep her train
of thought.

\section{Kelvin Waves\label{Sec2}}

\subsection{Euler Equations}

In what follows, we use Kelvin waves as our primary tool. We dedicate this
section to their brief description.

Recall that the Euler system of partial differential equations (PDEs)
describing the motion of an inviscid, incompressible fluid has the form:%
\begin{equation}
\begin{array}{c}
\frac{\partial \mathbf{V}}{\partial t}+(\mathbf{V\cdot }\nabla )\mathbf{V+}%
\nabla \left( \frac{P}{\rho }\right) =0, \\ 
\\ 
\nabla \cdot \mathbf{V=}0\mathbf{,}%
\end{array}
\label{Eq2.1}
\end{equation}%
where $t$ is time, $\mathbf{x}$ is the position, \textbf{$V$}$\left( t,%
\mathbf{x}\right) $ is the velocity vector, $P\left( t,\mathbf{x}\right) $
is the pressure, $\rho $ is the constant density, $\nabla $ is the gradient;
see, e.g., \cite{Chandrasekhar61}. In Cartesian coordinates, Eqs (\ref{Eq2.1}%
) can be written as follows:%
\begin{equation}
\begin{array}{c}
\frac{\partial V_{i}}{\partial t}+\dsum\limits_{j=1}^{3}V_{j}\frac{\partial
V_{i}}{\partial x_{j}}+\frac{\partial }{\partial x_{i}}\left( \frac{P}{\rho }%
\right) =0, \\ 
\\ 
\dsum\limits_{i=1}^{3}\frac{\partial V_{i}}{\partial x_{i}}=0.%
\end{array}
\label{Eq2.2}
\end{equation}

It can be shown that Eqs (\ref{Eq2.1}) have a family of solutions $\left( 
\mathbf{V}\left( t,\mathbf{x}\right) ,P\left( t,\mathbf{x}\right) \right) $,
linearly depending on spatial coordinates:%
\begin{equation}
\begin{array}{c}
\mathbf{V}\left( t,\mathbf{x}\right) =\mathfrak{L}\left( t\right) \mathbf{%
x,\ \ \ }\frac{P\left( t,\mathbf{x}\right) }{\rho }=\frac{1}{2}\mathfrak{M}%
\left( t\right) \mathbf{x\cdot x,}%
\end{array}
\label{Eq2.3}
\end{equation}%
where the matrices $\mathfrak{L}\left( t\right) $, $\mathfrak{M}\left(
t\right) $, such that 
\begin{equation}
\begin{array}{c}
Tr\left( \mathfrak{L}\left( t\right) \right) =0,\ \ \ \mathfrak{M}\left(
t\right) =\mathfrak{M}^{\ast }\left( t\right) ,%
\end{array}
\label{Eq2.4}
\end{equation}%
satisfy the following system of ordinary differential equations (ODEs):%
\begin{equation}
\begin{array}{c}
\frac{d\mathfrak{L}\left( t\right) }{dt}+\mathfrak{L}^{2}\left( t\right) +%
\mathfrak{M}\left( t\right) =0.%
\end{array}
\label{Eq2.5}
\end{equation}%
We are interested in the behavior of perturbations of solutions given by\
Eq. (\ref{Eq2.3}), which we denote by $\left( \mathbf{v}\left( t,\mathbf{x}%
\right) ,p\left( t,\mathbf{x}\right) \right) $. By neglecting the quadratic
term $($\textbf{$v$}$\mathbf{\cdot }\nabla )$\textbf{$v$}, we can write the
system of PDEs for $\left( \mathbf{v,}p\right) $ as follows:%
\begin{equation}
\begin{array}{c}
\frac{\partial \mathbf{v}}{\partial t}+(\mathfrak{L}\left( t\right) \mathbf{%
x\cdot }\nabla )\mathbf{v+}\mathfrak{L}\left( t\right) \mathbf{v+}\nabla
\left( \frac{p}{\rho }\right) =0, \\ 
\\ 
\nabla \cdot \mathbf{v=}0\mathbf{.}%
\end{array}
\label{Eq2.6}
\end{equation}%
It has been known for a long time that linear PDEs (\ref{Eq2.6}) have
wave-like solutions of the form:%
\begin{equation}
\begin{array}{c}
\left( \mathbf{v}\left( t,\mathbf{x}\right) ,\frac{p\left( t,\mathbf{x}%
\right) }{\rho }\right) \\ 
\\ 
=\left( \mathbf{a}\left( t\right) \exp \left( i\mathbf{\Gamma }\left(
t\right) \cdot \left( \mathbf{x-\xi }\left( t\right) \right) \right)
,b\left( t\right) \exp \left( i\mathbf{\Gamma }\left( t\right) \cdot \left( 
\mathbf{x-\xi }\left( t\right) \right) \right) \right) ,%
\end{array}
\label{Eq2.7}
\end{equation}%
where $\left( \mathbf{a}\left( t\right) ,b\left( t\right) \right) $ are
time-dependent amplitudes, and $\mathbf{\Gamma }\left( t\right) $ is the
time-dependent wave vector; see \cite{Kelvin87, Orr07, Craik86,
Friedlander03}. We call these solutions the Kelvin waves. The point we
emphasize below is that the so-called affine ansatz is a special instance of
Kelvin wave. This observation allows us to draw similarities among seemingly
unrelated topics, which, in turn, facilitates their holistic and
comprehensive study.

The pair $\mathbf{a}\left( t\right) ,\mathbf{\Gamma }\left( t\right) $
satisfies the following system of ODEs:%
\begin{equation}
\begin{array}{c}
\frac{d\mathbf{\xi }\left( t\right) }{dt}-\mathfrak{L}\left( t\right) 
\mathbf{\xi }\left( t\right) =0,\ \ \ \mathbf{\xi }\left( 0\right) =\mathbf{%
\xi }_{0}, \\ 
\\ 
\frac{d\mathbf{\Gamma }\left( t\right) }{dt}+\mathfrak{L}^{T}\left( t\right) 
\mathbf{\Gamma }\left( t\right) =0,\ \ \ \mathbf{\Gamma }\left( 0\right) =%
\mathbf{\Gamma }_{0}, \\ 
\\ 
\frac{d\mathbf{a}\left( t\right) }{dt}+\mathfrak{L}\left( t\right) \mathbf{a}%
\left( t\right) -2\frac{\mathfrak{L}\left( t\right) \mathbf{a}\left(
t\right) \cdot \mathbf{k}\left( t\right) }{\mathbf{k}\left( t\right) \cdot 
\mathbf{k}\left( t\right) }\mathbf{k}\left( t\right) =0,\ \mathbf{a}\left(
0\right) =\mathbf{a}_{0}, \\ 
\\ 
\ \mathbf{\Gamma }_{0}\cdot \mathbf{a}_{0}=0.%
\end{array}
\label{Eq2.8}
\end{equation}%
Here and below, the superscript $T$ stands for transpose, and $\cdot $
denotes the scalar product. The corresponding $p\left( t\right) $ can be
found via the incompressibility condition. It is easy to show that for $%
t\geq 0$, 
\begin{equation}
\begin{array}{c}
\mathbf{\Gamma }\left( t\right) \cdot \mathbf{\xi }\left( t\right) =\mathbf{%
\Gamma }_{0}\cdot \mathbf{\xi }_{0},\ \ \ \mathbf{\Gamma }\left( t\right)
\cdot \mathbf{a}\left( t\right) =0.%
\end{array}
\label{Eq2.9}
\end{equation}%
Typically, Eqs (\ref{Eq2.8}) are used to study the stability of the linear
flow. Such a flow is unstable whenever $\left\Vert \mathbf{a}\left( t\right)
\right\Vert \rightarrow \theta $ for some choices of $\mathbf{\Gamma }_{0},%
\mathbf{a}_{0}$; see \cite{Bayly86, Bayly96, Lifschitz95}. Moreover, it can
be shown that the same instabilities occur in general three-dimensional
flows; see \cite{Friedlander91, Lifschitz91b, Friedlander03}.

Interestingly, \cite{Chandrasekhar61} pointed out that the superposition of
the linear flow (\ref{Eq2.3}) and the Kelvin wave (\ref{Eq2.7}), i.e.,%
\begin{equation}
\begin{array}{c}
\mathbf{\tilde{V}}\left( t,\mathbf{x}\right) =\mathfrak{L}\left( t\right) 
\mathbf{x+v}\left( t,\mathbf{x}\right) , \\ 
\\ 
\frac{\tilde{P}\left( t,\mathbf{x}\right) }{\rho }=\frac{1}{2}\mathfrak{M}%
\left( t\right) \mathbf{x\cdot x+}\frac{p\left( t,\mathbf{x}\right) }{\rho },%
\end{array}
\label{Eq2.10}
\end{equation}%
satisfies the nonlinear Euler equations (\ref{Eq2.1}) since the nonlinear
term \textbf{$v$}$\cdot \nabla $\textbf{$v$} vanishes identically due to
incompressibility.\footnote{%
Thus, even the greatest minds occasionally can be myopic -- it took eighty
years for fluid dynamists to connect the dots and observe that $\left( 
\mathbf{\tilde{v}},\tilde{p}/\rho \right) $ solve the nonlinear Euler
equations.}

\subsection{Navier-Stokes Equations}

The motion of the incompressible viscous fluid is described by the
celebrated Navier-Stokes equations of the form:%
\begin{equation}
\begin{array}{c}
\frac{\partial \mathbf{V}}{\partial t}+(\mathbf{V\cdot }\nabla )\mathbf{V-}%
\nu \Delta \mathbf{V+}\nabla \left( \frac{P}{\rho }\right) =0, \\ 
\\ 
\nabla \cdot \mathbf{V=}0\mathbf{,}%
\end{array}
\label{Eq2.11}
\end{equation}%
where $\nu $ is the kinematic viscosity; see, e.g., \cite{Chandrasekhar61}.
It is clear that linear flows, given by Eq. (\ref{Eq2.3}), are unaffected by
viscosity, hence they satisfy Eqs (\ref{Eq2.12}). Viscosity does affect
their small perturbations, which are governed by the following equations:%
\begin{equation}
\begin{array}{c}
\frac{\partial \mathbf{v}}{\partial t}+(\mathfrak{L}\left( t\right) \mathbf{%
x\cdot }\nabla )\mathbf{v+}\mathfrak{L}\left( t\right) \mathbf{v\mathbf{-}}%
\nu \Delta \mathbf{v+}\nabla \left( \frac{p}{\rho }\right) =0, \\ 
\\ 
\nabla \cdot \mathbf{v=}0\mathbf{.}%
\end{array}
\label{Eq2.12}
\end{equation}%
The viscous version of Eqs (\ref{Eq2.8}) has the form; see \cite%
{Lifschitz91a}:%
\begin{equation}
\begin{array}{c}
\frac{d\mathbf{\xi }\left( t\right) }{dt}-\mathfrak{L}\left( t\right) 
\mathbf{\xi }\left( t\right) =0,\ \ \ \mathbf{\xi }\left( 0\right) =\mathbf{%
\xi }_{0}, \\ 
\\ 
\frac{d\mathbf{\Gamma }\left( t\right) }{dt}+\mathfrak{L}^{T}\left( t\right) 
\mathbf{\Gamma }\left( t\right) =0,\ \ \ \mathbf{\Gamma }\left( 0\right) =%
\mathbf{\Gamma }_{0}, \\ 
\\ 
\frac{d\mathbf{a}\left( t\right) }{dt}+\mathfrak{L}\left( t\right) \mathbf{a}%
\left( t\right) -2\frac{\mathfrak{L}\left( t\right) \mathbf{a}\left(
t\right) \cdot \mathbf{k}\left( t\right) }{\mathbf{k}\left( t\right) \cdot 
\mathbf{k}\left( t\right) }\mathbf{k}\left( t\right) +\nu \left\vert \mathbf{%
k}\left( t\right) \right\vert ^{2}\mathbf{a}\left( t\right) =0,\ \mathbf{a}%
\left( 0\right) =\mathbf{a}_{0}, \\ 
\\ 
\ \mathbf{\Gamma }_{0}\cdot \mathbf{a}_{0}=0.%
\end{array}
\label{Eq2.13}
\end{equation}%
We shall show in Section \ref{Sec55} that in the two-dimensional case, the
Navier-Stokes equations for small perturbations of linear flows are more or
less identical to the Fokker-Planck equations for harmonically bound
articles, which is surprising.

\section{Klein-Kramers Equation\label{Sec3}}

\subsection{Langevin Equation\label{Sec31}}

We start with the Langevin equation for particles moving in a potential
field and impacted by random forces; see \cite{Langevin08}. This section
uses the same notation as in \cite{Chandrasekhar43}.

Consider an underdamped Brownian particle. In contrast to the standard
Brownian motion, which is overdamped, we assume that the frictions are
finite, so that we must treat the momentum as an independent degree of
freedom, and describe a particle's state as a pair $\left( \mathbf{r},%
\mathbf{u}\right) $, where $\mathbf{r}$ and $\mathbf{u}$ are its position
and velocity, respectively. We consider a $d$-dimensional space, with $d=1$
and $d=3$ of particular interest. Hence, the corresponding Langevin
equations are%
\begin{equation}
\begin{array}{c}
\frac{d\mathbf{r}}{dt}=\mathbf{u}, \\ 
\\ 
\frac{d\mathbf{u}}{dt}=-\beta \mathbf{u-}\frac{\nabla V\left( \mathbf{r}%
\right) }{m}+\sqrt{\frac{2\beta k_{B}T}{m}}\mathbf{\eta }\left( t\right) ,
\\ 
\\ 
\left\langle \mathbf{\eta }^{T}\left( t\right) \mathbf{\eta }\left(
t^{\prime }\right) \right\rangle =\mathbf{I}\delta \left( t-t^{\prime
}\right) .%
\end{array}
\label{Eq3.1}
\end{equation}%
Here $m$ is the particle mass, $\beta $ is the friction coefficient, $k_{B}$
is the Boltzmann constant, $T$ is the temperature, $V\left( \mathbf{r}%
\right) $ is the external potential, $\mathbf{\eta }\left( t\right) $ is a $%
d $-dimensional Gaussian white noise, and $\mathbf{I}$ is the $d$%
-dimensional identity matrix. Below we denote the ratio $\beta k_{B}T/m$ as $%
q$.

Of course, we can rewrite Eqs (\ref{Eq3.1}) as a system of stochastic
differential equations (SDEs):%
\begin{equation}
\begin{array}{c}
d\mathbf{r}=\mathbf{u}dt, \\ 
\\ 
d\mathbf{u}=-\beta \mathbf{u}dt\mathbf{-}\frac{\nabla V\left( \mathbf{r}%
\right) }{m}dt+\sqrt{2q}d\mathbf{W}\left( t\right) ,%
\end{array}
\label{Eq3.2}
\end{equation}%
where $\mathbf{W}\left( t\right) $ is a standard $d$-dimensional Wiener
process.

\subsection{Klein-Kramers Equation\label{Sec32}}

Fokker, Planck, and their numerous followers derived and studied the forward
parabolic equation for the transition probability density $W\left( \tau ,%
\mathbf{\rho ,\upsilon },t,\mathbf{r,u}\right) $ associated with a
stochastic process. For the stochastic process governed by SDEs (\ref{Eq3.2}%
), the corresponding equation, called the Klein-Kramers equation, has the
form:%
\begin{equation}
\begin{array}{c}
W_{t}-qW_{\mathbf{uu}}+\mathbf{u}W_{\mathbf{r}}-\left( \left( \beta \mathbf{u%
}+\frac{\nabla V\left( \mathbf{r}\right) }{m}\right) W\right) _{\mathbf{u}%
}=0, \\ 
\\ 
W\left( \tau ,\mathbf{\rho ,\upsilon },\tau ,\mathbf{r,u}\right) =\delta
\left( \mathbf{r}-\mathbf{\rho }\right) \delta \left( \mathbf{u}-\mathbf{%
\upsilon }\right) .%
\end{array}
\label{Eq3.3}
\end{equation}%
The backward parabolic Kolmogorov equation has the form:%
\begin{equation}
\begin{array}{c}
W_{\tau }+qW_{\mathbf{\upsilon \upsilon }}+\mathbf{\upsilon }W_{\mathbf{\rho 
}}-\left( \beta \mathbf{\upsilon }+\frac{\nabla V\left( \mathbf{\rho }%
\right) }{m}\right) W_{\mathbf{\upsilon }}=0, \\ 
\\ 
W\left( t,\mathbf{\rho ,\upsilon },t,\mathbf{r,u}\right) =\delta \left( 
\mathbf{r}-\mathbf{\rho }\right) \delta \left( \mathbf{u}-\mathbf{\upsilon }%
\right) .%
\end{array}
\label{Eq3.4}
\end{equation}%
Details are given in \cite{Fokker14, Planck17, Klein21, Chapman28,
Kolmogorov31, Kolmogorov33, Kolmogorov34, Kramers40, Chandrasekhar43,
Risken89, Hanggi90}, as well as a multitude of subsequent sources. For
fascinating historical details see \cite{Ebeling08}.

\subsection{Chandrasekhar's Solutions\label{Sec33}}

In a well-known survey article, \cite{Chandrasekhar43} described elegant
solutions of Eq. (\ref{Eq3.3}) for a free particle and a harmonically bound
particle, which he derived by using clever changes of coordinates.

First, we shall consider a free particle in $1D$. For such a particle, \cite%
{Chandrasekhar43} writes the corresponding Klein-Kramers equation as follows:%
\begin{equation}
\begin{array}{c}
W_{t}-qW_{uu}+uW_{r}-\beta uW_{u}-W=0, \\ 
\\ 
W\left( \tau ,\rho \mathbf{,}\upsilon ,\tau ,r\mathbf{,}u\right) =\delta
\left( r-\rho \right) \delta \left( u-\upsilon \right) ,%
\end{array}
\label{Eq3.5}
\end{equation}%
and shows that%
\begin{equation}
\begin{array}{c}
W=\frac{1}{2\pi \left( FG-H^{2}\right) ^{1/2}}\exp \left( -\frac{\left(
FS^{2}-2HRS+GR^{2}\right) }{2\left( FG-H^{2}\right) }\right) ,%
\end{array}
\label{Eq3.6}
\end{equation}%
where%
\begin{equation}
\begin{array}{c}
S=u-e^{-\beta \left( t-\tau \right) }\upsilon ,\ \ \ R=r-\rho -\frac{%
1-e^{-\beta \left( t-\tau \right) }}{\beta }\upsilon , \\ 
\\ 
F=\frac{q}{\beta ^{3}}\left( -3+4e^{-\beta \left( t-\tau \right)
}-e^{-2\beta \left( t-\tau \right) }+2\beta \left( t-\tau \right) \right) ,
\\ 
\\ 
G=\frac{q}{\beta }\left( 1-e^{-2\beta \left( t-\tau \right) }\right) ,\ \ \
H=\frac{q}{\beta ^{2}}\left( 1-e^{-\beta \left( t-\tau \right) }\right) ^{2}.%
\end{array}
\label{Eq3.7}
\end{equation}%
Since we assumed that stochastic drivers are uncorrelated, the t.p.d.f. $%
W^{\left( 3\right) }$ can be presented as a product of three $1D$ t.p.d.f. $%
W^{\left( 1\right) }$:%
\begin{equation}
\begin{array}{c}
W^{\left( 3\right) }=W_{1}^{\left( 1\right) }W_{2}^{\left( 1\right)
}W_{3}^{\left( 1\right) }=\frac{1}{8\pi ^{3}\left( FG-H^{2}\right) ^{3/2}}%
\exp \left( -\frac{\left( F\left\vert \mathbf{S}\right\vert ^{2}-2H\mathbf{R}%
\cdot \mathbf{S}+G\left\vert \mathbf{R}\right\vert ^{2}\right) }{2\left(
FG-H^{2}\right) }\right) ,%
\end{array}
\label{Eq3.8}
\end{equation}%
where $\mathbf{u=}\left( u_{1},u_{2},u_{3}\right) $, etc.

To analyze a harmonically bound particle, we can proceed as follows; see 
\cite{Chandrasekhar43}. As usual, the harmonic potential $V\left( x\right) $
has the form:%
\begin{equation}
\begin{array}{c}
V\left( x\right) =\frac{m\omega ^{2}x^{2}}{2}.%
\end{array}
\label{Eq3.9}
\end{equation}%
Let $\mu _{\pm }$ be the roots of%
\begin{equation}
\begin{array}{c}
\mu ^{2}+\beta \mu +\omega ^{2}=0, \\ 
\\ 
\mu _{\pm }=\frac{-\beta \pm \sqrt{\beta ^{2}-4\omega ^{2}}}{2}.%
\end{array}
\label{Eq3.10}
\end{equation}%
Define new variables 
\begin{equation}
\begin{array}{c}
\xi =e^{-\mu _{-}\left( t-\tau \right) }\left( \mu _{+}r-u\right) ,\ \ \
\eta =e^{-\mu _{+}\left( t-\tau \right) }\left( \mu _{-}r-u\right) , \\ 
\\ 
\hat{\xi}=\left( \mu _{+}\rho -\upsilon \right) ,\ \ \ \hat{\eta}=\left( \mu
_{-}\rho -\upsilon \right) ,%
\end{array}
\label{Eq3.11}
\end{equation}%
and coefficients%
\begin{equation}
\begin{array}{c}
a=\frac{q}{\mu _{+}}\left( 1-e^{-2\mu _{+}\left( t-\tau \right) }\right) ,\
\ \ b=\frac{q}{\mu _{-}}\left( 1-e^{-2\mu _{-}\left( t-\tau \right) }\right)
, \\ 
\\ 
h=-\frac{2q}{\left( \mu _{+}+\mu _{-}\right) }\left( 1-e^{-\left( \mu
_{+}+\mu _{-}\right) \left( t-\tau \right) }\right) .\ \ 
\end{array}
\label{Eq3.12}
\end{equation}%
Then%
\begin{equation}
\begin{array}{c}
W=\frac{e^{\beta \left( t-\tau \right) }}{2\pi \sqrt{ab-h^{2}}}\exp \left( -%
\frac{a\left( \xi -\hat{\xi}\right) ^{2}+2h\left( \xi -\hat{\xi}\right)
\left( \eta -\hat{\eta}\right) +b\left( \eta -\hat{\eta}\right) ^{2}}{%
2\left( ab-h^{2}\right) }\right) .%
\end{array}
\label{Eq3.13}
\end{equation}%
While \cite{Chandrasekhar43} stopped at Eq. (\ref{Eq3.13}), for practical
applications, it is more useful to represent the exponent as an explicit
quadratic form of $r$ and $u$, which we do below.

\section{Kolmogorov Equation\label{Sec4}}

\subsection{Summary of Kolmogorov's paper\label{Sec41}}

In a remarkable (and remarkably concise) note, Kolmogorov considers a system
of particles in $n$-dimensional space with coordinates $q_{1},...,q_{n}$,
and velocities $\dot{q}_{1},...,\dot{q}_{n}$, assumes the probability
density function%
\begin{equation*}
\begin{array}{c}
g\left( t,q_{1},...,q_{n},\dot{q}_{1},...,\dot{q}_{n},t^{\prime
},q_{1}^{\prime },...,q_{n}^{\prime },\dot{q}_{1}^{\prime },...,\dot{q}%
_{n}^{\prime }\right)%
\end{array}%
\end{equation*}%
exist for some time $t^{\prime }>t$, and reveals (without any explanation)
an analytical expression for $g$ in the one-dimensional case; see \cite%
{Kolmogorov34}. This note is the third in a series of papers; the previous
two being \cite{Kolmogorov31, Kolmogorov33}.

Kolmogorov makes the following natural assumptions:%
\begin{equation}
\begin{array}{c}
\mathbf{E}\left\vert \Delta q_{i}-\dot{q}_{i}\Delta t\right\vert =o\left(
\Delta t\right) ,%
\end{array}
\label{Eq4.1}
\end{equation}%
\begin{equation}
\begin{array}{c}
\mathbf{E}\left( \Delta q_{i}\right) ^{2}=o\left( \Delta t\right) ,%
\end{array}
\label{Eq4.2}
\end{equation}%
where $\Delta t=t^{\prime }-t$. Eqs (\ref{Eq4.1}), (\ref{Eq4.2}) imply%
\begin{equation}
\begin{array}{c}
\mathbf{E}\left( \Delta q_{i}\right) =\dot{q}_{i}\Delta t+o\left( \Delta
t\right) ,%
\end{array}
\label{Eq4.3}
\end{equation}%
\begin{equation}
\begin{array}{c}
\mathbf{E}\left( \Delta q_{i}\Delta q_{j}\right) \leq \sqrt{\mathbf{E}\left(
\Delta q_{i}\right) ^{2}\mathbf{E}\left( \Delta q_{j}\right) ^{2}}=o\left(
\Delta t\right) .%
\end{array}
\label{Eq4.4}
\end{equation}%
Furthermore, under very general assumptions, the following relationships hold%
\begin{equation}
\begin{array}{c}
\mathbf{E}\left( \Delta \dot{q}_{i}\right) =f_{i}\left( t,q,\dot{q}\right)
\Delta t+o\left( \Delta t\right) ,%
\end{array}
\label{Eq4.5}
\end{equation}%
\begin{equation}
\begin{array}{c}
\mathbf{E}\left( \Delta \dot{q}_{i}\right) ^{2}=k_{ii}\left( t,q,\dot{q}%
\right) \Delta t+o\left( \Delta t\right) ,%
\end{array}
\label{Eq4.6}
\end{equation}%
\begin{equation}
\begin{array}{c}
\mathbf{E}\left( \Delta \dot{q}_{i}\Delta \dot{q}_{j}\right) =k_{ij}\left(
t,q,\dot{q}\right) \Delta t+o\left( \Delta t\right) ,%
\end{array}
\label{Eq4.7}
\end{equation}%
where $f,k$ are continuous functions. Eqs (\ref{Eq4.2}), (\ref{Eq4.6}) imply%
\begin{equation}
\begin{array}{c}
\mathbf{E}\left( \Delta \dot{q}_{i}\Delta \dot{q}_{j}\right) \leq \sqrt{%
\mathbf{E}\left( \Delta \dot{q}_{i}\right) ^{2}\mathbf{E}\left( \Delta \dot{q%
}_{j}\right) ^{2}}=o\left( \Delta t\right) .%
\end{array}
\label{Eq4.8}
\end{equation}%
Under some natural physical assumptions, it follows that $g$ satisfies the
following differential equation of the Fokker-Planck type:%
\begin{equation}
\begin{array}{c}
\frac{\partial g}{\partial t^{\prime }}=-\sum \dot{q}_{i}^{\prime }\frac{%
\partial }{\partial q_{i}^{\prime }}g-\sum \frac{\partial }{\partial \dot{q}%
_{i}^{\prime }}\left\{ f_{i}\left( t,q,\dot{q}\right) g\right\} +\sum \sum 
\frac{\partial ^{2}}{\partial \dot{q}_{i}^{\prime 2}}\left\{ \left\{ k\left(
t,q,\dot{q}\right) g\right\} \right\} .%
\end{array}
\label{Eq4.9}
\end{equation}%
In the one-dimensional case, we have%
\begin{equation}
\begin{array}{c}
\frac{\partial g}{\partial t^{\prime }}=-\dot{q}^{\prime }\frac{\partial }{%
\partial q^{\prime }}g-\frac{\partial }{\partial \dot{q}^{\prime }}\left\{
f\left( t,q,\dot{q}\right) g\right\} +\frac{\partial ^{2}}{\partial \dot{q}%
^{\prime 2}}\left\{ \left\{ k\left( t,q,\dot{q}\right) g\right\} \right\} .%
\end{array}
\label{Eq4.10}
\end{equation}%
These equations are known as ultra-parabolic Fokker-Plank- Kolmogorov
equations due to their degeneracy.

When $f$ and $k$ are constants, the fundamental solution of Eq. (\ref{Eq4.10}%
) has the form:%
\begin{equation}
\begin{array}{c}
g=\frac{2\sqrt{3}}{\pi k^{2}\left( t^{\prime }-t\right) ^{2}}\exp \left\{ -%
\frac{\left[ \dot{q}^{\prime }-\dot{q}-f\left( t^{\prime }-t\right) \right]
^{2}}{4k\left( t^{\prime }-t\right) }-\frac{3\left[ q^{\prime }-q-\frac{\dot{%
q}^{\prime }+\dot{q}}{2}\left( t^{\prime }-t\right) \right] ^{2}}{%
k^{3}\left( t^{\prime }-t\right) ^{3}}\right\} .%
\end{array}
\label{Eq4.11}
\end{equation}%
One can see that $\Delta \dot{q}$ is of the order $\left( \Delta t\right)
^{1/2}$. At the same time%
\begin{equation}
\begin{array}{c}
\Delta q=\dot{q}\Delta t+O\left( \Delta t\right) ^{3/2}.%
\end{array}
\label{Eq4.12}
\end{equation}%
One can prove that a similar relation holds for the general Eq. (\ref{Eq4.9}%
).

The previous Section shows that physicists derived Eqs (\ref{Eq4.9}), (\ref%
{Eq4.10}) at least a decade earlier than Kolmogorov, which puts the
Kolmogorov equation into a broader context.

Kolmogorov equations fascinated mathematicians for a long time and generated
a great deal of research; see, e.g., \cite{Weber51, Hormander67, Kuptsov72,
Lanconelli02, Pascucci05, Ivasishen10, Duong18}, among others.

\subsection{Challenge and Response\label{Sec42}}

Despite its undoubting brilliance, the original Kolmogorov's paper has
several issues.

First, Eqs (\ref{Eq4.9}), (\ref{Eq4.10}) \emph{are not} the Fokker-Planck
equations associated with the process described by Eqs (\ref{Eq4.5}) - (\ref%
{Eq4.7}), since they miss the prefactor $1/2$ in front of the diffusion
terms. The corrected multivariate equation has the form:%
\begin{equation}
\begin{array}{c}
\frac{\partial g}{\partial t^{\prime }}=-\sum \dot{q}_{i}^{\prime }\frac{%
\partial }{\partial q_{i}^{\prime }}g-\sum \frac{\partial }{\partial \dot{q}%
_{i}^{\prime }}\left\{ f_{i}\left( t,q,\dot{q}\right) g\right\} +\frac{1}{2}%
\sum \sum \frac{\partial ^{2}}{\partial \dot{q}_{i}^{\prime }\partial \dot{q}%
_{j}^{\prime }}\left\{ \left\{ k\left( t,q,\dot{q}\right) g\right\} \right\}
,%
\end{array}
\label{Eq4.13}
\end{equation}%
while the corresponding one-dimensional equation has the form:%
\begin{equation}
\begin{array}{c}
\frac{\partial g}{\partial t^{\prime }}=-\dot{q}^{\prime }\frac{\partial }{%
\partial q^{\prime }}g-\frac{\partial }{\partial \dot{q}^{\prime }}\left\{
f\left( t,q,\dot{q}\right) g\right\} +\frac{1}{2}\frac{\partial ^{2}}{%
\partial \dot{q}^{\prime 2}}\left\{ \left\{ k\left( t,q,\dot{q}\right)
g\right\} \right\} .%
\end{array}
\label{Eq4.14}
\end{equation}%
Alternatively, Eqs (\ref{Eq4.6}), (\ref{Eq4.7}) can be altered as follows:%
\begin{equation}
\begin{array}{c}
\mathbf{E}\left( \Delta \dot{q}_{i}\right) ^{2}=2k_{ii}\left( t,q,\dot{q}%
\right) \Delta t+o\left( \Delta t\right) ,%
\end{array}
\label{Eq4.6a}
\end{equation}%
\begin{equation}
\begin{array}{c}
\mathbf{E}\left( \Delta \dot{q}_{i}\Delta \dot{q}_{j}\right) =2k_{ij}\left(
t,q,\dot{q}\right) \Delta t+o\left( \Delta t\right) .%
\end{array}
\label{Eq4.7a}
\end{equation}%
We prefer to update the Fokker-Planck equation.

Second, $g$ given by Eq. (\ref{Eq4.11}) does not solve Eq. (\ref{Eq4.10}).
It also does not satisfy the (implicit) initial condition%
\begin{equation}
\begin{array}{c}
g\left( t,q,\dot{q},t,q^{\prime },\dot{q}^{\prime }\right) =\delta \left(
q^{\prime }-q\right) \delta \left( \dot{q}^{\prime }-\dot{q}\right) ,%
\end{array}
\label{Eq4.15}
\end{equation}%
where $\delta \left( .\right) $ is the Dirac $\delta $-function. The fact
that expression (\ref{Eq4.11}) does not solve Eq. (\ref{Eq4.10}) can be
verified by substitution. However, it is easier to verify this statement via
dimensional analysis. The dimensions of the corresponding variables and
coefficients are as follows:%
\begin{equation}
\begin{array}{c}
\left[ t\right] =\left[ t^{\prime }\right] =T,\ \ \ \left[ q\right] =\left[
q^{\prime }\right] =L,\ \ \ \left[ \dot{q}\right] =\left[ \dot{q}^{\prime }%
\right] =T^{-1}L,\ \ \ \left[ g\right] =T^{1}L^{-2}, \\ 
\\ 
\left[ f\right] =T^{-2}L,\ \ \ \left[ k\right] =T^{-3}L^{2}.%
\end{array}
\label{Eq4.16}
\end{equation}%
It is easy to show that $g$ is scale-invariant, so that%
\begin{equation}
\begin{array}{c}
g\left( \lambda ^{2}t,\lambda ^{3}q,\lambda \dot{q},\lambda ^{2}t^{\prime
},\lambda ^{3}q^{\prime },\lambda \dot{q}^{\prime };\lambda ^{-1}f,k\right)
=\lambda ^{-4}g\left( t,q,\dot{q},t^{\prime },q^{\prime },\dot{q}^{\prime
};f,k\right) .%
\end{array}
\label{Eq4.17}
\end{equation}%
The original Kolmogorov formula contains two typos, making it dimensionally
incorrect since the term 
\begin{equation*}
\frac{3\left[ q^{\prime }-q-\frac{\dot{q}^{\prime }+\dot{q}}{2}\left(
t^{\prime }-t\right) \right] ^{2}}{k^{3}\left( t^{\prime }-t\right) ^{3}}
\end{equation*}%
in the exponent is not nondimensional, as it should be, and has dimension $%
T^{6}L^{-1}$, while the prefactor 
\begin{equation*}
\begin{array}{c}
\frac{2\sqrt{3}}{\pi k^{2}\left( t^{\prime }-t\right) ^{2}}%
\end{array}%
\end{equation*}%
has dimension $T^{4}L^{-1}$, instead of the right dimension $TL^{-2}$.

Third, due to yet another typo, the solution given by Eq. (\ref{Eq4.11})
does not converge to the initial condition in the limit $t^{\prime
}\rightarrow t$. Indeed, asymptotically, we have%
\begin{equation}
\begin{array}{c}
g\sim P^{H}\left( \frac{k^{3}\left( t^{\prime }-t\right) ^{3}}{6},q^{\prime
}-q\right) P^{H}\left( 2k\left( t^{\prime }-t\right) ,\dot{q}^{\prime }-\dot{%
q}\right) \rightarrow 4\delta \left( q^{\prime }-q\right) \delta \left( \dot{%
q}^{\prime }-\dot{q}\right) .%
\end{array}
\label{Eq4.18}
\end{equation}%
Here $P^{H}$ is the standard heat kernel:%
\begin{equation}
\begin{array}{c}
P^{H}\left( \tau ,\eta \right) =\frac{\exp \left( -\frac{\eta ^{2}}{2\tau }%
\right) }{\sqrt{2\pi \tau }}.%
\end{array}
\label{Eq4.19}
\end{equation}

However, not all is lost. Dimensional analysis shows that the correct
solution $g\left( t,q,\dot{q},t^{\prime },q^{\prime },\dot{q}^{\prime
};f,k\right) $ of Eq. (\ref{Eq4.10}) has the form:%
\begin{equation}
\begin{array}{c}
g=\frac{\sqrt{3}}{2\pi k\left( t^{\prime }-t\right) ^{2}}\exp \left\{ -\frac{%
\left[ \dot{q}^{\prime }-\dot{q}-f\left( t^{\prime }-t\right) \right] ^{2}}{%
4k\left( t^{\prime }-t\right) }-\frac{3\left[ q^{\prime }-q-\frac{\dot{q}%
^{\prime }+\dot{q}}{2}\left( t^{\prime }-t\right) \right] ^{2}}{k\left(
t^{\prime }-t\right) ^{3}}\right\} ,%
\end{array}
\label{Eq4.20}
\end{equation}%
which is not far from Kolmogorov's formula. The correct solution of Eq. (\ref%
{Eq4.14}) has the form:%
\begin{equation}
\begin{array}{c}
g=\frac{\sqrt{3}}{\pi k\left( t^{\prime }-t\right) ^{2}}\exp \left\{ -\frac{%
\left[ \dot{q}^{\prime }-\dot{q}-f\left( t^{\prime }-t\right) \right] ^{2}}{%
2k\left( t^{\prime }-t\right) }-\frac{6\left[ q^{\prime }-q-\frac{\dot{q}%
^{\prime }+\dot{q}}{2}\left( t^{\prime }-t\right) \right] ^{2}}{k\left(
t^{\prime }-t\right) ^{3}}\right\} .%
\end{array}
\label{Eq4.21}
\end{equation}

\subsection{Direct Verification\label{Sec43}}

In order to avoid confusion, from now on, we shall change the notation to
make the formulas easier to read. Specifically, we assume that $x$
represents the position of a particle at time $t$ and $\xi $ its position at
time $\tau $, while $y$ represents its velocity at time $t$, and $\theta $
its velocity at time $\tau $, so that 
\begin{equation}
\begin{array}{c}
\left( t,q,\dot{q}\right) \rightarrow \left( \tau ,\xi ,\theta \right) ,\ \
\ \left( t^{\prime },q^{\prime },\dot{q}^{\prime }\right) \rightarrow \left(
t,x,y\right) .%
\end{array}
\label{Eq4.22}
\end{equation}

One of our objectives is deriving the (corrected) Kolmogorov formula from
first principles using Kelvin Waves. Subsequently, we show how to use it in
the financial mathematics context.

The governing SDE can be written as%
\begin{equation}
\begin{array}{c}
dx_{t}=y_{t}dt,\ \ \ x_{\tau }=\xi , \\ 
\\ 
dy_{t}=bdt+\sqrt{a}dW_{t},\ \ \ y_{\tau }=\theta .%
\end{array}
\label{Eq4.23}
\end{equation}%
The corresponding Fokker-Planck- Kolmogorov problem has the form: 
\begin{equation}
\begin{array}{c}
P_{t}\left( \tau ,\xi ,\theta ,t,x,y\right) -\frac{1}{2}aP_{yy}\left( \tau
,\xi ,\theta ,t,x,y\right) \\ 
\\ 
+yP_{x}\left( \tau ,\xi ,\theta ,t,x,y\right) +bP_{y}\left( \tau ,\xi
,\theta ,t,x,y\right) =0, \\ 
\\ 
P\left( \tau ,x,y,\tau ,\xi ,\theta \right) =\delta \left( x-\xi \right)
\delta \left( y-\theta \right) .%
\end{array}
\label{Eq4.24}
\end{equation}%
We claim that the solution of Eqs (\ref{Eq4.24}) is as follows:%
\begin{equation}
P\left( \tau ,\xi ,\theta ,t,x,y\right) =\frac{\sqrt{3}}{\pi at^{2}}\exp
\left( -\Phi \left( \tau ,\xi ,\theta ,t,x,y\right) \right) ,  \label{Eq4.25}
\end{equation}%
where%
\begin{equation}
\begin{array}{c}
\Phi \left( \tau ,\xi ,\theta ,t,x,y\right) =\frac{\left( y-\theta -b\left(
t-\tau \right) \right) ^{2}}{2at}+\frac{6\left( x-\xi -\frac{\left( y+\theta
\right) \left( t-\tau \right) }{2}\right) ^{2}}{at^{3}}=\frac{A^{2}}{2}%
+6B^{2},%
\end{array}
\label{Eq4.26}
\end{equation}%
and%
\begin{equation}
\begin{array}{c}
A=\frac{\left( y-\theta -b\left( t-\tau \right) \right) }{\sqrt{a\left(
t-\tau \right) }},\ \ \left[ A\right] =1,\  \\ 
\\ 
B=\frac{\left( x-\xi -\frac{\left( y+\theta \right) \left( t-\tau \right) }{2%
}\right) }{\sqrt{a\left( t-\tau \right) ^{3}}},\ \ \ \left[ B\right] =1.%
\end{array}
\label{Eq4.27}
\end{equation}%
Let us check that $P$ satisfies the Fokker-Plank equation and the initial
conditions. We have:%
\begin{equation}
\begin{array}{c}
\Phi _{t}=-\left( \frac{A^{2}}{2\left( t-\tau \right) }+\frac{bA}{\sqrt{%
a\left( t-\tau \right) }}+\frac{18B^{2}}{\left( t-\tau \right) }+\frac{%
6\left( y+\theta \right) B}{\sqrt{a\left( t-\tau \right) ^{3}}}\right) , \\ 
\\ 
\ \Phi _{x}=\frac{12B}{\sqrt{a\left( t-\tau \right) ^{3}}},\ \ \ \Phi _{y}=%
\frac{A-6B}{\sqrt{a\left( t-\tau \right) }},\ \ \ \Phi _{yy}=\frac{4}{%
a\left( t-\tau \right) },%
\end{array}
\label{Eq4.28}
\end{equation}%
\begin{equation}
\begin{array}{c}
\frac{P_{t}}{P}=-\frac{2}{\left( t-\tau \right) }-\Phi _{t},\ \ \ \frac{P_{x}%
}{P}=-\Phi _{x},\ \ \ \frac{P_{y}}{P}=-\Phi _{y},\ \ \ \frac{P_{yy}}{P}%
=-\Phi _{yy}+\Phi _{x}^{2},%
\end{array}
\label{Eq4.29}
\end{equation}%
so that 
\begin{equation}
\begin{array}{c}
P_{t}^{K}-\frac{1}{2}aP_{yy}^{K}+yP_{x}^{K}+bP_{y}^{K} \\ 
\\ 
=P^{K}\left( -\frac{2}{\left( t-\tau \right) }-\Phi _{t}+\frac{1}{2}a\Phi
_{yy}-\frac{1}{2}a\Phi _{y}^{2}-y\Phi _{x}-b\Phi _{y}\right) \\ 
\\ 
=P^{K}\left( -\frac{2}{\left( t-\tau \right) }+\frac{A^{2}}{2\left( t-\tau
\right) }+\frac{bA}{\sqrt{a\left( t-\tau \right) }}+\frac{18B^{2}}{\left(
t-\tau \right) }+\frac{6\left( y+\theta \right) B}{\sqrt{a\left( t-\tau
\right) ^{3}}}\right. \\ 
\\ 
\left. +\frac{2}{\left( t-\tau \right) }-\frac{\left( A-6B\right) ^{2}}{%
2\left( t-\tau \right) }-\frac{12yB}{\sqrt{a\left( t-\tau \right) ^{3}}}-%
\frac{b\left( A-6B\right) }{\sqrt{a\left( t-\tau \right) }}\right) \\ 
\\ 
=0.%
\end{array}
\label{Eq4.30}
\end{equation}%
When $\left( t-\tau \right) \rightarrow 0$ we have the following asymptotic
expression:%
\begin{equation}
\begin{array}{c}
P^{K}\left( \tau ,\xi ,\theta ,t,x,y\right) \sim P^{H}\left( \frac{a\left(
t-\tau \right) ^{3}}{12},x-\xi \right) P^{H}\left( a\left( t-\tau \right)
,y-\theta \right) \\ 
\\ 
\rightarrow \delta \left( x-\xi \right) \delta \left( y-\theta \right) .%
\end{array}
\label{Eq4.31}
\end{equation}

\subsection{Solution via Kelvin Waves\label{Sec44}}

Now, we wish to derive Kolmogorov's formula by using Kelvin waves (or an
affine ansatz). To this end, we start with the problem of the form:%
\begin{equation}
\begin{array}{c}
H_{t}\left( \tau ,\xi ,\theta ,t,x,y,k,l\right) -\frac{1}{2}aH_{yy}\left(
\tau ,\xi ,\theta ,t,x,y,k,l\right) \\ 
\\ 
+yH_{x}\left( \tau ,\xi ,\theta ,t,x,y,k,l\right) +bH_{y}\left( \tau ,\xi
,\theta ,t,x,y,k,l\right) =0, \\ 
\\ 
H\left( \tau ,x,y,\tau ,\xi ,\theta ,k,l\right) =\exp \left( ik\left( x-\xi
\right) +il\left( y-\theta \right) \right) .%
\end{array}
\label{Eq4.32}
\end{equation}%
Here%
\begin{equation}
\begin{array}{c}
\ \left[ k\right] =L^{-1},\ \ \ \left[ l\right] =TL^{-1},\ \ \left[ H\right]
=1.%
\end{array}
\label{Eq4.33}
\end{equation}%
By using the well-known results concerning the inverse Fourier transform of
the $\delta $-function, we get:%
\begin{equation}
\begin{array}{c}
P\left( \tau ,\xi ,\theta ,t,x,y\right) =\frac{1}{\left( 2\pi \right) ^{2}}%
\int_{-\infty }^{\infty }\int_{-\infty }^{\infty }H\left( \tau ,\xi ,\theta
,t,x,y,k,l\right) dkdl.%
\end{array}
\label{Eq4.34}
\end{equation}%
To calculate $H$, we use the affine ansatz and represent it in the form:%
\begin{equation}
\begin{array}{c}
H\left( \tau ,\xi ,\theta ,t,x,y,k,l\right) =\exp \left( \Psi \left( \tau
,\xi ,\theta ,t,x,y,k,l\right) \right) ,%
\end{array}
\label{Eq4.35}
\end{equation}%
where%
\begin{equation}
\begin{array}{c}
\Psi \left( \tau ,\xi ,\theta ,t,x,y,k,l\right) =\alpha \left( t\right)
+ik\left( x-\xi \right) +i\beta \left( t\right) y-il\theta .%
\end{array}
\label{Eq4.36}
\end{equation}%
and%
\begin{equation}
\begin{array}{c}
\frac{H_{t}}{H}=\Psi _{t}=\left( \alpha ^{\prime }\left( t\right) +i\beta
^{\prime }\left( t\right) y\right) ,\ \ \ \frac{H_{x}}{H}=\Psi _{x}=ik,\ \ \ 
\\ 
\\ 
\frac{H_{y}}{H}=\Psi _{y}=i\beta \left( t\right) ,\ \ \ \frac{H_{yy}}{H}%
=\Psi _{y}^{2}=-\beta ^{2}\left( t\right) .%
\end{array}
\label{Eq4.37}
\end{equation}%
Using this ansatz in Eqs (\ref{Eq4.32}) yields:%
\begin{equation}
\begin{array}{c}
\alpha ^{\prime }\left( t\right) +\frac{1}{2}a\beta ^{2}\left( t\right)
+i\beta ^{\prime }\left( t\right) y+iky+ib\beta \left( t\right) =0, \\ 
\\ 
\alpha \left( \tau \right) =0,\ \ \ \beta \left( \tau \right) =l,%
\end{array}
\label{Eq4.38}
\end{equation}%
so that%
\begin{equation}
\begin{array}{c}
\alpha ^{\prime }\left( t\right) +\frac{1}{2}a\beta ^{2}\left( t\right)
+ib\beta \left( t\right) =0,\ \ \ \alpha \left( \tau \right) =0, \\ 
\\ 
\beta ^{\prime }\left( t\right) +k=0,\ \ \ \beta \left( \tau \right) =l.%
\end{array}
\label{Eq4.39}
\end{equation}%
Straightforward calculation shows:%
\begin{equation}
\begin{array}{c}
\beta \left( t\right) =-k\left( t-\tau \right) +l, \\ 
\\ 
\alpha \left( t\right) =-\frac{1}{2}a\left( \frac{k^{2}\left( t-\tau \right)
^{3}}{3}-kl\left( t-\tau \right) ^{2}+l^{2}\left( t-\tau \right) \right)
-ib\left( -\frac{k\left( t-\tau \right) ^{2}}{2}+l\left( t-\tau \right)
\right) .%
\end{array}
\label{Eq4.40}
\end{equation}%
Equations (\ref{Eq4.34}, \ref{Eq4.35}, \ref{Eq4.36}, \ref{Eq4.40}) yield:%
\begin{equation}
\begin{array}{c}
P\left( \tau ,\xi ,\theta ,t,x,y\right) =\frac{1}{\left( 2\pi \right) ^{2}}%
\int_{-\infty }^{\infty }\int_{-\infty }^{\infty } \\ 
\\ 
\times \exp \left( -\frac{1}{2}a\left( \frac{k^{2}\left( t-\tau \right) ^{3}%
}{3}-kl\left( t-\tau \right) ^{2}+l^{2}\left( t-\tau \right) \right) \right.
\\ 
\\ 
\left. +ik\left( x-\xi -y\left( t-\tau \right) +\frac{b\left( t-\tau \right)
^{2}}{2}\right) +il\left( y-\theta -b\left( t-\tau \right) \right) \right)
dkdl.%
\end{array}
\label{Eq4.41}
\end{equation}%
It is clear that $P\left( \tau ,\xi ,\theta ,t,x,y\right) $ can be viewed as
the characteristic function of the Gaussian density in the $\left(
k,l\right) $ space, evaluated at the point $\left( x-\xi -y\left( t-\tau
\right) \right. $ $\left. +\frac{b\left( t-\tau \right) ^{2}}{2},y-\theta
-b\left( t-\tau \right) \right) $:%
\begin{equation}
\begin{array}{c}
P\left( \tau ,\xi ,\theta ,t,x,y\right) =\frac{\left( \det \left( \mathfrak{C%
}\right) \right) ^{1/2}}{2\pi }\int_{-\infty }^{\infty }\int_{-\infty
}^{\infty } \\ 
\\ 
\times G\left( t,k,l\right) \exp \left( ik\left( x-\xi -y\left( t-\tau
\right) +\frac{b\left( t-\tau \right) ^{2}}{2}\right) +il\left( y-\theta
-b\left( t-\tau \right) \right) \right) dkdl,%
\end{array}
\label{Eq4.42}
\end{equation}%
where%
\begin{equation}
\begin{array}{c}
G\left( t,k,l\right) =\frac{1}{2\pi \left( \det \left( \mathfrak{C}\right)
\right) ^{1/2}}\exp \left( -\frac{1}{2}\left( 
\begin{array}{c}
k \\ 
l%
\end{array}%
\right) \cdot \mathfrak{C}^{-1}\left( 
\begin{array}{c}
k \\ 
l%
\end{array}%
\right) \right) ,%
\end{array}
\label{Eq4.43}
\end{equation}%
and%
\begin{equation}
\begin{array}{c}
\mathfrak{C}\mathbb{=}\left( 
\begin{array}{cc}
\frac{12}{a\left( t-\tau \right) ^{3}} & \frac{6}{a\left( t-\tau \right) ^{2}%
} \\ 
\frac{6}{a\left( t-\tau \right) ^{2}} & \frac{4}{a\left( t-\tau \right) }%
\end{array}%
\right) , \\ 
\\ 
\det \left( \mathfrak{C}\right) =\frac{12}{a^{2}\left( t-\tau \right) ^{4}}.%
\end{array}
\label{Eq4.44}
\end{equation}%
As before, $\cdot $ denotes the scalar product. Accordingly,%
\begin{equation}
\begin{array}{c}
P\left( \tau ,\xi ,\theta ,t,x,y\right) =\frac{\sqrt{3}}{\pi a\left( t-\tau
\right) ^{2}}\exp \left( -\Omega \left( \tau ,\xi ,\theta ,t,x,y\right)
\right) ,%
\end{array}
\label{Eq4.45}
\end{equation}%
where%
\begin{equation}
\begin{array}{c}
\Omega \left( \tau ,\xi ,\theta ,t,x,y\right) = \\ 
\\ 
=\frac{1}{2}\left( 
\begin{array}{c}
x-\xi -y\left( t-\tau \right) +\frac{b\left( t-\tau \right) ^{2}}{2} \\ 
y-\theta -b\left( t-\tau \right)%
\end{array}%
\right) \cdot \mathfrak{C}\left( 
\begin{array}{c}
x-\xi -y\left( t-\tau \right) +\frac{b\left( t-\tau \right) ^{2}}{2} \\ 
y-\theta -b\left( t-\tau \right)%
\end{array}%
\right) \\ 
\\ 
=\frac{6\left( x-\xi -y\left( t-\tau \right) +\frac{b\left( t-\tau \right)
^{2}}{2}\right) ^{2}}{a\left( t-\tau \right) ^{3}}+\frac{6\left( x-\xi
-y\left( t-\tau \right) +\frac{b\left( t-\tau \right) ^{2}}{2}\right) \left(
y-\theta -b\left( t-\tau \right) \right) }{a\left( t-\tau \right) ^{2}}+%
\frac{2\left( y-\theta -b\left( t-\tau \right) \right) ^{2}}{a\left( t-\tau
\right) } \\ 
\\ 
=6\left( B-\frac{A}{2}\right) ^{2}+6\left( B-\frac{A}{2}\right) A+2A^{2} \\ 
\\ 
=6B^{2}-6AB+\frac{3A^{2}}{2}+6AB-3A^{2}+2A^{2} \\ 
\\ 
=\frac{A^{2}}{2}+6B^{2} \\ 
\\ 
=\Phi \left( t,x,y\,,\xi ,\theta \right) ,%
\end{array}
\label{Eq4.46}
\end{equation}%
as expected. This calculation completes our derivation of the corrected
Kolmogorov formula.

$P$ is a bivariate Gaussian distribution. Completing the square, we get:%
\begin{equation}
\begin{array}{c}
\Phi =\frac{\left( y-\theta -b\left( t-\tau \right) \right) ^{2}}{2a\left(
t-\tau \right) }+\frac{6\left( x-\xi -\frac{\left( y+\theta \right) \left(
t-\tau \right) }{2}\right) ^{2}}{a\left( t-\tau \right) ^{3}} \\ 
\\ 
=\frac{6}{a\left( t-\tau \right) ^{3}}\left( x-p\right) ^{2}-\frac{6}{%
a\left( t-\tau \right) ^{2}}\left( x-p\right) \left( y-q\right) +\frac{2}{%
a\left( t-\tau \right) }\left( y-q\right) ^{2},%
\end{array}
\label{Eq4.47}
\end{equation}%
and represent $P$ the form:%
\begin{equation}
\begin{array}{c}
P\left( \tau ,\xi ,\theta ,t,x,y\right) =\frac{\exp \left( -\frac{1}{2\left(
1-\rho ^{2}\right) }\left( \frac{\left( x-p\right) ^{2}}{\sigma _{x}^{2}}-%
\frac{2\rho \left( x-p\right) \left( y-q\right) }{\sigma _{x}\sigma _{y}}+%
\frac{\left( y-q\right) ^{2}}{\sigma _{y}^{2}}\right) \right) }{2\pi \sigma
_{x}\sigma _{y}\sqrt{1-\rho ^{2}}},%
\end{array}
\label{Eq4.48}
\end{equation}%
where%
\begin{equation}
\begin{array}{c}
\sigma _{x}=\sqrt{\frac{a\left( t-\tau \right) ^{3}}{3}},\ \ \ \sigma _{y}=%
\sqrt{a\left( t-\tau \right) },\ \ \ \rho =\frac{\sqrt{3}}{2}, \\ 
\\ 
p=\xi +\theta \left( t-\tau \right) +\frac{b\left( t-\tau \right) ^{2}}{2},\
\ q=\theta +b\left( t-\tau \right) .%
\end{array}
\label{Eq4.49}
\end{equation}

\section{Gaussian Processes\label{Sec5}}

\subsection{The General Case\label{Sec51}}

We now consider a more general Kolmogorov-type SDE solvable via the Kelvin
(or affine) ansatz:%
\begin{equation}
\begin{array}{c}
d\mathbf{x}_{t}=\left( \mathbf{b}^{\left( x\right) }+\mathfrak{B}^{\left(
xx\right) }\mathbf{x}_{t}+\mathfrak{B}^{\left( xy\right) }\mathbf{y}%
_{t}\right) dt,\ \ \ \mathbf{x}_{\tau }=\mathbf{\xi ,} \\ 
\\ 
d\mathbf{y}_{t}=\left( \mathbf{b}^{\left( y\right) }+\mathfrak{B}^{\left(
yx\right) }\mathbf{x}_{t}+\mathfrak{B}^{\left( yy\right) }\mathbf{y}%
_{t}\right) dt+\mathfrak{\mathbf{\Sigma }}^{\left( yy\right) }d\mathbf{W}%
_{t}^{\left( y\right) },\ \ \ \mathbf{y}_{\tau }=\mathbf{\theta ,}%
\end{array}
\label{Eq5.1}
\end{equation}%
where $\mathbf{x}_{_{t}}=\left( x_{m}\right) $ and $\mathbf{b}^{\left(
x\right) }=\left( b_{m}^{\left( x\right) }\right) $ are $\left( M\times
1\right) $ column vectors, $\mathbf{y}_{t}=\left( y_{n}\right) ^{T}$ and $%
\mathbf{b}^{\left( y\right) }=\left( b_{n}^{\left( y\right) }\right) $ are $%
\left( N\times 1\right) $ column vectors, $\mathfrak{B}^{\left( xx\right)
}=\left( b_{mm^{\prime }}^{\left( xx\right) }\right) $ is $\left( M\times
M\right) $ matrix, $\mathfrak{B}^{\left( xy\right) }=\left( b_{mn}^{\left(
xy\right) }\right) $ is $\left( M\times N\right) $ matrix, $\mathfrak{B}%
^{\left( yx\right) }=\left( b_{nm}^{\left( xx\right) }\right) $ is $\left(
N\times M\right) $ matrix, $\mathfrak{B}^{\left( yy\right) }=\left(
b_{nn^{\prime }}^{\left( xx\right) }\right) $ and $\mathbf{\Sigma }^{\left(
yy\right) }=\left( \sigma _{nn^{\prime }}\right) $ are $\left( N\times
N\right) $ matrices. Below, we assume that the corresponding coefficients
are time-dependent. As usual, $\mathbf{W}_{t}^{\left( y\right) }$ is a
standard $N$-dimensional Brownian motion.

More compactly, we can write the system of SDEs as follows:%
\begin{equation}
\begin{array}{cc}
d\mathbf{z}_{t}=\left( \mathbf{b}^{\left( z\right) }+\mathfrak{B}^{\left(
zz\right) }\mathbf{z}_{t}\right) dt+\left( 
\begin{array}{c}
0 \\ 
\mathbf{\Sigma }^{\left( yy\right) }d\mathbf{W}_{t}^{\left( y\right) }%
\end{array}%
\right) , & \mathbf{z}_{\tau }=\left( 
\begin{array}{c}
\mathbf{\xi } \\ 
\mathbf{\theta }%
\end{array}%
\right) ,%
\end{array}
\label{Eq5.2}
\end{equation}%
where%
\begin{equation}
\begin{array}{c}
\mathbf{z}=\left( 
\begin{array}{c}
\mathbf{x} \\ 
\mathbf{y}%
\end{array}%
\right) ,\ \ \ \mathbf{b}^{\left( z\right) }=\left( 
\begin{array}{c}
\mathbf{b}^{\left( x\right) } \\ 
\mathbf{b}^{\left( y\right) }%
\end{array}%
\right) ,\ \ \ \mathfrak{B}^{\left( zz\right) }=\left( 
\begin{array}{cc}
\mathfrak{B}^{\left( xx\right) } & \mathfrak{B}^{\left( xy\right) } \\ 
\mathfrak{B}^{\left( yx\right) } & \mathfrak{B}^{\left( yy\right) }%
\end{array}%
\right) ,%
\end{array}
\label{Eq5.3}
\end{equation}%
so that $\mathbf{z}$ and $\mathbf{b}^{\left( z\right) }$ are $\left( I\times
1\right) $ column vector, and $\mathfrak{B}^{\left( zz\right) }$ is a $%
\left( I\times I\right) $ matrix, where $I=M+N$. We also introduce a scalar $%
b^{\left( z\right) }=\mathrm{Tr}\left( \mathfrak{B}^{\left( zz\right)
}\right) =\mathrm{Tr}\left( \mathfrak{B}^{\left( xx\right) }\right) +\mathrm{%
Tr}\left( \mathfrak{B}^{\left( yx\right) }\right) $.

The corresponding Fokker-Plank problem has the form:%
\begin{equation}
\begin{array}{c}
P_{t}\left( \tau ,\mathbf{\zeta ,}t,\mathbf{z}\right) -\frac{1}{2}\dsum
\dsum \mathfrak{A}P_{\mathbf{yy}}\left( \tau ,\mathbf{\zeta ,}t,\mathbf{z}%
\right) \\ 
\\ 
+\left( \mathbf{b}^{\left( z\right) }+\mathfrak{B}^{\left( z\right) }\mathbf{%
z}\right) \cdot P_{\mathbf{z}}\left( \tau ,\mathbf{\zeta ,}t,\mathbf{z}%
\right) +b^{\left( z\right) }P\left( \tau ,\mathbf{\zeta ,}t,\mathbf{z}%
\right) =0, \\ 
\\ 
P\left( \tau ,\mathbf{z},\tau ,\mathbf{\zeta }\right) =\delta \left( \mathbf{%
x}-\mathbf{\xi }\right) \delta \left( \mathbf{y}-\mathbf{\theta }\right) ,%
\end{array}
\label{Eq5.4}
\end{equation}%
where $\mathfrak{A}$ is the covariance matrix, 
\begin{equation}
\begin{array}{c}
\mathfrak{A}=\left( a_{nn^{\prime }}\right) =\sum\limits_{k=1}^{N}\sigma
_{nk}\sigma _{n^{\prime }k}=\mathbf{\Sigma }^{\left( yy\right) }\mathbf{%
\Sigma }^{\left( yy\right) T}.%
\end{array}
\label{Eq5.5}
\end{equation}%
Explicitly,%
\begin{equation}
\begin{array}{c}
P_{t}-\frac{1}{2}\sum\limits_{n=1}^{N}\sum\limits_{n^{\prime
}=1}^{N}a_{nn^{\prime }}P_{y_{n}y_{n^{\prime }}}+\sum\limits_{i=1}^{I}\left(
b_{i}^{\left( z\right) }+\sum\limits_{i^{\prime }=1}^{I}b_{ii^{\prime
}}^{\left( zz\right) }x_{i^{\prime }}\right) P_{z_{i}}+b^{\left( z\right)
}P=0,%
\end{array}
\label{Eq5.6}
\end{equation}

We use the same Kelvin ansatz as before and represent $P$ in the form:%
\begin{equation}
\begin{array}{c}
P\left( \tau ,\mathbf{\zeta ,}t,\mathbf{z}\right) =\frac{1}{\left( 2\pi
\right) ^{I}}\int_{-\infty }^{\infty }...\int_{-\infty }^{\infty }H\left(
\tau ,\mathbf{\zeta ,}t,\mathbf{z,m}\right) d\mathbf{m}, \\ 
\\ 
H\left( \tau ,\mathbf{\zeta ,}t,\mathbf{z,m}\right) =\exp \left( \Psi \left(
\tau ,\mathbf{\zeta ,}t,\mathbf{z,m}\right) \right) , \\ 
\\ 
\Psi \left( \tau ,\mathbf{\zeta ,}t,\mathbf{z,m}\right) =\alpha \left(
t\right) +i\Upsilon \left( t\right) \cdot \mathbf{z}-i\mathbf{m}\cdot 
\mathbf{\zeta },%
\end{array}
\label{Eq5.7}
\end{equation}%
where $\mathbf{m}\mathcal{=}\left( m_{i}\right) $ is an $\left( I\times
1\right) $ column vector, $\mathbf{m}=\left( \mathbf{k,l}\right) ^{T}$, $%
\mathbf{k}\mathcal{=}\left( k_{m}\right) $ is an $\left( M\times 1\right) $
column vector, $\mathbf{l}=\left( l_{n}\right) ~$is an $\left( N\times
1\right) $ column vector, $\mathbf{\Upsilon =}\left( \upsilon _{i}\right) $
is an $\left( I\times 1\right) $ column vector, $\ \mathbf{\Upsilon }=\left( 
\mathbf{\Gamma },\mathbf{\Delta }\right) ^{T}$, $\mathbf{\Gamma }=\left(
\gamma _{m}^{\left( x\right) }\right) ~$is an $\left( M\times 1\right) $
column vector$,$ $\mathbf{\Delta }=\left( \delta _{n}^{\left( y\right)
}\right) ~$is an $\left( N\times 1\right) $ column vector, and%
\begin{equation}
\begin{array}{c}
\alpha \left( \tau \right) =0,\ \ \ \mathbf{\Upsilon }\left( \tau \right)
=\left( \mathbf{\Gamma }\left( \tau \right) ,\mathbf{\Delta }\left( \tau
\right) \right) ^{T}=\mathbf{m}=\left( \mathbf{k,l}\right) ^{T}.%
\end{array}
\label{Eq5.8}
\end{equation}%
As before:%
\begin{equation}
\begin{array}{c}
\frac{H_{t}}{H}=\Psi _{t}=\left( \alpha ^{\prime }\left( t\right) +i\mathbf{%
\Upsilon }^{\prime }\left( t\right) \cdot \mathbf{z}\right) ,\ \ \ \frac{H_{%
\mathbf{x}}}{H}=\Psi _{\mathbf{x}}=i\mathbf{\Gamma }\left( t\right) , \\ 
\\ 
\frac{H_{\mathbf{y}}}{H}=\Psi _{\mathbf{y}}=i\mathbf{\Delta }\left( t\right)
,\ \ \ \frac{H_{\mathbf{yy}}}{H}=\Psi _{\mathbf{y}}^{2}=-\mathbf{\Delta }%
\left( t\right) \mathbf{\Delta }^{T}\left( t\right) .%
\end{array}
\label{Eq5.9}
\end{equation}%
The equations for $\alpha ,\mathbf{\Upsilon }$ have the form:%
\begin{equation}
\begin{array}{c}
\alpha ^{\prime }\left( t\right) +i\mathbf{\Upsilon }^{\prime }\left(
t\right) \cdot \mathbf{z}+\frac{1}{2}\mathbf{\Delta }\left( t\right) \cdot 
\mathfrak{A}\mathbf{\Delta }\left( t\right) +i\mathbf{\Upsilon }\left(
t\right) \cdot \left( \mathbf{b}^{\left( z\right) }+\mathfrak{B}^{\left(
zz\right) }\mathbf{z}\right) +b^{\left( z\right) }=0.%
\end{array}
\label{Eq5.10}
\end{equation}%
Accordingly,%
\begin{equation}
\begin{array}{c}
\mathbf{\Upsilon }^{\prime }\left( t\right) +\mathfrak{B}^{\left( zz\right)
T}\mathbf{\Upsilon }\left( t\right) =0,\ \ \ \mathbf{\Upsilon }\left( \tau
\right) =\mathbf{m}=\left( \mathbf{k,l}\right) ^{T},%
\end{array}
\label{Eq5.11}
\end{equation}%
and%
\begin{equation}
\begin{array}{c}
\alpha ^{\prime }\left( t\right) +\frac{1}{2}\mathbf{\Delta }\left( t\right)
\cdot \mathfrak{A}\mathbf{\Delta }\left( t\right) +i\mathbf{\Upsilon }\left(
t\right) \cdot \mathbf{b}^{\left( z\right) }+b^{\left( z\right) }=0,\ \ \
\alpha \left( \tau \right) =0.%
\end{array}
\label{Eq5.12}
\end{equation}%
Let $\mathfrak{L}\left( \tau ,t\right) $ is the fundamental solution of the
homogeneous ODE (\ref{Eq5.12}), i.e., the matrix such that%
\begin{equation}
\begin{array}{c}
\mathbf{\mathfrak{L}}^{\prime }\left( \tau ,t\right) +\mathfrak{B}^{\left(
zz\right) T}\mathbf{\mathfrak{L}}\left( \tau ,t\right) =0,\ \ \ \mathbf{%
\mathfrak{L}}\left( \tau ,\tau \right) =\mathfrak{I},%
\end{array}
\label{Eq5.13}
\end{equation}%
where $\mathfrak{I}$ is the identity matrix. The well-known Liouville's
formula, which we shall use shortly, yields%
\begin{equation}
\begin{array}{c}
\det \left( \mathfrak{L}\left( \tau ,t\right) \right) =\exp \left(
-\int_{\tau }^{t}b^{\left( z\right) }\left( s\right) ds\right) .%
\end{array}
\label{Eq5.14}
\end{equation}%
The solution of Eq. (\ref{Eq5.12}) has the form:%
\begin{equation}
\begin{array}{c}
\mathbf{\Upsilon }\left( t\right) =\mathbf{\mathfrak{L}}\left( \tau
,t\right) \mathbf{m}.%
\end{array}
\label{Eq5.15}
\end{equation}%
It is convenient to write $\mathfrak{L}\left( \tau ,t\right) $ in the block
form:%
\begin{equation}
\begin{array}{c}
\mathfrak{L}\left( \tau ,t\right) =\left( 
\begin{array}{cc}
\mathfrak{L}^{\left( xx\right) }\left( \tau ,t\right) & \mathfrak{L}^{\left(
xy\right) }\left( \tau ,t\right) \\ 
\mathfrak{L}^{\left( yx\right) }\left( \tau ,t\right) & \mathfrak{L}^{\left(
yy\right) }\left( \tau ,t\right)%
\end{array}%
\right) .%
\end{array}
\label{Eq5.16}
\end{equation}%
It follows from Eq. (\ref{Eq5.11}) that%
\begin{equation}
\begin{array}{c}
\alpha \left( t\right) =-\frac{1}{2}\mathbf{m}\cdot \mathfrak{C}^{-1}\left(
\tau ,t\right) \mathbf{m}-i\mathbf{m\cdot d}^{\left( z\right) }\left( \tau
,t\right) -\varpi \left( \tau ,t\right) ,%
\end{array}
\label{Eq5.17}
\end{equation}%
where $\mathfrak{C}^{-1}$ is an $I\times I$ positive-definite matrix split
into four blocks of the form:%
\begin{equation}
\begin{array}{c}
\mathfrak{C}^{-1}\left( \tau ,t\right) \\ 
\\ 
=\left( 
\begin{array}{cc}
\int_{\tau }^{t}\mathfrak{L}^{\left( yx\right) T}\left( \tau ,s\right) 
\mathfrak{A}\left( s\right) \mathfrak{L}^{\left( yx\right) }\left( \tau
,s\right) ds & \int_{\tau }^{t}\mathfrak{L}^{\left( yx\right) T}\left( \tau
,s\right) \mathfrak{A}\left( s\right) \mathfrak{L}^{\left( yy\right) }\left(
\tau ,s\right) ds \\ 
\int_{\tau }^{t}\mathfrak{L}^{\left( yy\right) T}\left( \tau ,s\right) 
\mathfrak{A}\left( s\right) \mathfrak{L}^{\left( yx\right) }\left( \tau
,s\right) ds & \int_{\tau }^{t}\mathfrak{L}^{\left( yy\right) T}\left( \tau
,s\right) \mathfrak{A}\left( s\right) \mathfrak{L}^{\left( yy\right) }\left(
\tau ,s\right) ds%
\end{array}%
\right) ,%
\end{array}
\label{Eq5.18}
\end{equation}%
while $\mathbf{d}^{\left( z\right) }=\left( \mathbf{d}^{\left( x\right) },%
\mathbf{d}^{\left( y\right) }\right) ^{T}$, $\mathbf{d}^{\left( x\right) }$
and $\mathbf{d}^{\left( y\right) }$ are $\left( M\times 1\right) $ and $%
\left( N\times 1\right) $ column vectors, and $\varpi $ is a scalar:%
\begin{equation}
\begin{array}{c}
\mathbf{d}^{\left( z\right) }\left( \tau ,t\right) =\int_{\tau }^{t}%
\mathfrak{L}^{T}\left( \tau ,s\right) \mathbf{b}^{\left( z\right) }\left(
s\right) ds,%
\end{array}
\label{Eq5.19}
\end{equation}%
\begin{equation}
\begin{array}{c}
\varpi \left( \tau ,t\right) =\int_{\tau }^{t}b^{\left( z\right) }\left(
s\right) ds.%
\end{array}
\label{Eq5.19a}
\end{equation}%
Accordingly,%
\begin{equation}
\begin{array}{c}
\Psi \left( \tau ,\mathbf{\zeta ,}t,\mathbf{z},\mathbf{m}\right) \\ 
\\ 
=-\frac{1}{2}\mathbf{m}\cdot \mathfrak{C}^{-1}\left( \tau ,t\right) \mathbf{m%
}+i\mathbf{\mathfrak{L}}\left( \tau ,t\right) \mathbf{m}\cdot \mathbf{z}-i%
\mathbf{m}\cdot \left( \mathbf{d}^{\left( z\right) }\left( \tau ,t\right) +%
\mathbf{\zeta }\right) -\varpi \left( \tau ,t\right) \\ 
\\ 
=-\frac{1}{2}\mathbf{m}\cdot \mathfrak{C}^{-1}\left( \tau ,t\right) \mathbf{m%
}+i\mathbf{m}\cdot \left( \mathfrak{L}^{T}\left( \tau ,t\right) \mathbf{z-d}%
^{\left( z\right) }\left( \tau ,t\right) \mathbf{-\zeta }\right) -\varpi
\left( \tau ,t\right) .%
\end{array}
\label{Eq5.20}
\end{equation}%
Thus, 
\begin{equation}
\begin{array}{c}
P\left( \tau ,\mathbf{\zeta ,}t,\mathbf{z}\right) =\frac{\det \left( 
\mathfrak{C}\right) ^{1/2}\exp \left( -\varpi \left( \tau ,t\right) \right) 
}{\left( 2\pi \right) ^{I/2}}\int_{-\infty }^{\infty }...\int_{-\infty
}^{\infty }G\left( t,\mathbf{m}\right) \\ 
\\ 
\times \exp \left( i\mathbf{m}\cdot \left( \mathfrak{L}^{T}\left( \tau
,t\right) \mathbf{z-d}^{\left( z\right) }\left( \tau ,t\right) \mathbf{%
-\zeta }\right) \right) d\mathbf{m},%
\end{array}
\label{Eq5.21}
\end{equation}%
where $G\left( t,\mathbf{m}\right) $ is the density of a multivariate
Gaussian distribution in the $\mathbf{m}$-space. It is clear that $P\left(
\tau ,\mathbf{\zeta ,}t,\mathbf{z}\right) $ is proportional to the
characteristic function of $G$ evaluated at the point $\left( \mathfrak{L}%
^{T}\left( \tau ,t\right) \mathbf{z-d}^{\left( z\right) }\left( \tau
,t\right) \mathbf{-\zeta }\right) $, so that%
\begin{equation}
\begin{array}{c}
P\left( \tau ,\mathbf{\zeta ,}t,\mathbf{z}\right) =\frac{\det \left( 
\mathfrak{C}\right) ^{1/2}\exp \left( -\varpi \left( \tau ,t\right) \right) 
}{\left( 2\pi \right) ^{I/2}} \\ 
\\ 
\times \exp \left( -\frac{1}{2}\left( \mathfrak{L}^{T}\left( \tau ,t\right) 
\mathbf{z-d}^{\left( z\right) }\left( \tau ,t\right) \mathbf{-\zeta }\right)
\cdot \mathfrak{C}\left( \mathfrak{L}^{T}\left( \tau ,t\right) \mathbf{z-d}%
^{\left( z\right) }\left( \tau ,t\right) \mathbf{-\zeta }\right) \right) .%
\end{array}
\label{Eq5.22}
\end{equation}%
Simple calculation using Eq. (\ref{Eq5.14}) allows one to rewrite (\ref%
{Eq5.22}) it the standard Gaussian form:%
\begin{equation}
\begin{array}{c}
P\left( \tau ,\mathbf{\zeta ,}t,\mathbf{z}\right) =\mathrm{N}\left( \mathbf{r%
}\left( \tau ,t\right) ,\mathfrak{H}\left( \tau ,t\right) \right) ,%
\end{array}
\label{Eq5.23}
\end{equation}%
where%
\begin{equation}
\begin{array}{c}
\mathfrak{H}\left( \tau ,t\right) =\left( \mathfrak{L}^{T}\left( \tau
,t\right) \right) ^{-1}\mathfrak{C}^{-1}\left( \tau ,t\right) \mathfrak{L}%
^{-1}\left( \tau ,t\right) , \\ 
\\ 
\mathbf{r}\left( \tau ,t\right) \mathbf{=}\left( \mathfrak{L}^{T}\left( \tau
,t\right) \right) ^{-1}\left( \mathbf{d}^{\left( z\right) }\left( \tau
,t\right) +\mathbf{\zeta }\right) .%
\end{array}
\label{Eq5.24}
\end{equation}

\subsection{Example: Kolmogorov Process\label{Sec52}}

We now wish to extend the Kolmogorov formula to the case when $b$ and $a$
are functions of time. In order to make connections with other cases a bit
easier to draw, we use the specific notation: $b\left( t\right) $ and $%
a\left( t\right) $. The corresponding SDE has the form:%
\begin{equation}
\begin{array}{c}
dx_{t}=y_{t}dt,\ \ \ x_{\tau }=\xi , \\ 
\\ 
dy_{t}=b\left( t\right) dt+\sqrt{a\left( t\right) }dW_{t},\ \ \ y_{\tau
}=\theta .%
\end{array}
\label{Eq5.26}
\end{equation}%
Accordingly, Eq. (\ref{Eq5.13}) can be written as follows:%
\begin{equation}
\begin{array}{c}
\mathfrak{L}^{\prime }\left( \tau ,t\right) +\left( 
\begin{array}{cc}
0 & 0 \\ 
1 & 0%
\end{array}%
\right) \mathfrak{L}\left( \tau ,t\right) =0,\ \ \mathfrak{L}\left( \tau
,\tau \right) =\left( 
\begin{array}{cc}
1 & 0 \\ 
0 & 1%
\end{array}%
\right) ,%
\end{array}
\label{Eq5.27}
\end{equation}%
so that%
\begin{equation}
\begin{array}{c}
\mathfrak{L}\left( \tau ,t\right) =\left( 
\begin{array}{cc}
1 & 0 \\ 
-\left( t-\tau \right) & 1%
\end{array}%
\right) , \\ 
\\ 
\mathfrak{L}^{-1}\left( \tau ,t\right) =\left( 
\begin{array}{cc}
1 & 0 \\ 
\left( t-\tau \right) & 1%
\end{array}%
\right) .%
\end{array}
\label{Eq5.28}
\end{equation}%
Once we know $\mathfrak{L}\left( \tau ,t\right) $, we can compute $\mathfrak{%
C}^{-1}\left( \tau ,t\right) $, $\mathbf{d}^{\left( z\right) }\left( \tau
,t\right) =\left( d^{\left( x\right) }\left( \tau ,t\right) ,d^{\left(
y\right) }\left( \tau ,t\right) \right) ^{T}$, $\varpi \left( \tau ,t\right) 
$: 
\begin{equation}
\begin{array}{c}
\mathfrak{C}^{-1}\left( \tau ,t\right) =\left( 
\begin{array}{cc}
\psi _{2}\left( \tau ,t\right) & -\psi _{1}\left( \tau ,t\right) \\ 
-\psi _{1}\left( \tau ,t\right) & \psi _{0}\left( \tau ,t\right)%
\end{array}%
\right) , \\ 
\\ 
\mathbf{d}^{\left( z\right) }\left( \tau ,t\right) =\left( 
\begin{array}{c}
d^{\left( x\right) }\left( \tau ,t\right) \\ 
d^{\left( y\right) }\left( \tau ,t\right)%
\end{array}%
\right) =\left( 
\begin{array}{c}
-\phi _{1}\left( \tau ,t\right) \\ 
\phi _{0}\left( \tau ,t\right)%
\end{array}%
\right) , \\ 
\\ 
\varpi \left( \tau ,t\right) =0,%
\end{array}
\label{Eq5.29}
\end{equation}%
where%
\begin{equation}
\begin{array}{c}
\phi _{i}\left( \tau ,t\right) =\int_{\tau }^{t}\left( s-\tau \right)
^{i}b\left( s\right) ds,\ \ \ \ \ \psi _{i}\left( \tau ,t\right) =\int_{\tau
}^{t}\left( s-\tau \right) ^{i}a\left( s\right) ds.%
\end{array}
\label{Eq5.30}
\end{equation}%
Next, we calculate $\mathfrak{H}\left( \tau ,t\right) $, and $\mathbf{r}%
\left( \tau ,t\right) $:%
\begin{equation}
\begin{array}{c}
\mathfrak{H}\left( \tau ,t\right) =\left( \mathfrak{L}^{T}\left( \tau
,t\right) \right) ^{-1}\mathfrak{C}^{-1}\left( \tau ,t\right) \mathfrak{L}%
^{-1}\left( \tau ,t\right) \\ 
\\ 
=\left( 
\begin{array}{cc}
1 & \left( t-\tau \right) \\ 
0 & 1%
\end{array}%
\right) \left( 
\begin{array}{cc}
\psi _{2}\left( \tau ,t\right) & -\psi _{1}\left( \tau ,t\right) \\ 
-\psi _{1}\left( \tau ,t\right) & \psi _{0}\left( \tau ,t\right)%
\end{array}%
\right) \left( 
\begin{array}{cc}
1 & 0 \\ 
\left( t-\tau \right) & 1%
\end{array}%
\right) \\ 
\\ 
=\left( 
\begin{array}{cc}
\psi _{0}\left( \tau ,t\right) \left( t-\tau \right) ^{2}-2\psi _{1}\left(
\tau ,t\right) \left( t-\tau \right) +\psi _{2}\left( \tau ,t\right) & \psi
_{0}\left( \tau ,t\right) \left( t-\tau \right) -\psi _{1}\left( \tau
,t\right) \\ 
\psi _{0}\left( \tau ,t\right) \left( t-\tau \right) -\psi _{1}\left( \tau
,t\right) & \psi _{0}\left( \tau ,t\right)%
\end{array}%
\right) ,%
\end{array}
\label{Eq5.31}
\end{equation}%
\begin{equation}
\begin{array}{c}
\mathbf{r}\left( \tau ,t\right) \mathbf{=}\left( \mathfrak{L}^{T}\left( \tau
,t\right) \right) ^{-1}\left( \mathbf{d}^{\left( z\right) }\left( \tau
,t\right) +\mathbf{\zeta }\right) \\ 
\\ 
=\left( 
\begin{array}{c}
-\phi _{1}\left( \tau ,t\right) +\xi +\left( \phi _{0}\left( \tau ,t\right)
+\theta \right) \left( t-\tau \right) \\ 
\phi _{0}\left( \tau ,t\right) +\theta%
\end{array}%
\right) .%
\end{array}
\label{Eq5.32}
\end{equation}

Accordingly, $P\left( \tau ,\xi ,\theta ,t,x,y\right) $ is a bivariate
Gaussian distribution of the form (\ref{Eq5.23}), with 
\begin{equation}
\begin{array}{c}
\sigma _{x}^{2}\left( \tau ,t\right) =\psi _{0}\left( \tau ,t\right) \left(
t-\tau \right) ^{2}-2\psi _{1}\left( \tau ,t\right) \left( t-\tau \right)
+\psi _{2}\left( \tau ,t\right) ,\ \ \ \sigma _{y}^{2}=\psi _{0}\left( \tau
,t\right) , \\ 
\\ 
\rho \left( \tau ,t\right) =\frac{\left( \psi _{0}\left( \tau ,t\right)
\left( t-\tau \right) -\psi _{1}\left( \tau ,t\right) \right) }{\sqrt{\psi
_{0}\left( \tau ,t\right) \left( \psi _{0}\left( \tau ,t\right) \left(
t-\tau \right) ^{2}-2\psi _{1}\left( \tau ,t\right) \left( t-\tau \right)
+\psi _{2}\left( \tau ,t\right) \right) }}, \\ 
\\ 
p\left( \tau ,t\right) =-\phi _{1}\left( \tau ,t\right) +\xi +\left( \phi
_{0}\left( \tau ,t\right) +\theta \right) \left( t-\tau \right) ,\ \ \
q\left( \tau ,t\right) =\phi _{0}\left( \tau ,t\right) +\theta .%
\end{array}
\label{Eq5.33}
\end{equation}%
It is easy to see that Eq. (\ref{Eq5.33}) coincides with Eq. (\ref{Eq4.48})
when $a,b$ are constant.

\subsection{Example: OU Process\label{Sec53}}

\subsubsection{OU Process}

For benchmarking purposes, we start with deriving the with the well-known
t.p.d.f. for the OU process:%
\begin{equation}
\begin{array}{c}
dy=\left( \chi \left( t\right) -\kappa \left( t\right) y\right)
dt+\varepsilon \left( t\right) dZ,\ \ \ y\left( \tau \right) =\theta .%
\end{array}
\label{Eq5.100}
\end{equation}%
The corresponding Fokker-Planck problem has the form:

\begin{equation}
\begin{array}{c}
P_{t}\left( \tau ,\theta ,t,y\right) -\frac{1}{2}\varepsilon
^{2}P_{yy}\left( \tau ,\theta ,t,y\right) +\left( \chi -\kappa y\right)
P_{y}\left( \tau ,\theta ,t,y\right) -\kappa P\left( \tau ,\theta
,t,y\right) =0, \\ 
\\ 
P\left( \tau ,\theta ,t,y\right) =\delta \left( y-\theta \right) .%
\end{array}
\label{Eq5.101}
\end{equation}%
The associated function $H\left( \tau ,\theta ,t,y,l\right) $ has the form:%
\begin{equation}
\begin{array}{c}
H=\exp \left( \alpha \left( t\right) +i\Delta \left( t\right) y-il\theta
\right) ,%
\end{array}
\label{Eq5.102}
\end{equation}%
so that%
\begin{equation}
\begin{array}{c}
\alpha ^{\prime }+\frac{1}{2}\varepsilon ^{2}\Delta ^{2}+i\chi \Delta
-\kappa =0,\ \ \ \alpha \left( \tau \right) =0, \\ 
\\ 
\Delta ^{\prime }-\kappa \Delta =0,\ \ \ \Delta \left( \tau \right) =l.%
\end{array}
\label{Eq5.103}
\end{equation}%
It is clear that 
\begin{equation}
\begin{array}{c}
\Delta \left( \tau ,t\right) =e^{\eta \left( \tau ,t\right) }l, \\ 
\\ 
\alpha \left( \tau ,t\right) =-\frac{1}{2}\psi _{0}\left( \tau ,t\right)
l^{2}-\left( \int_{\tau }^{t}e^{\eta \left( \tau ,s\right) }\chi \left(
s\right) ds\right) il+\eta \left( \tau ,t\right) .%
\end{array}
\label{Eq5.104}
\end{equation}%
where%
\begin{equation}
\begin{array}{c}
\eta \left( \tau ,t\right) =\dint\limits_{\tau }^{t}\kappa \left( s\right)
ds,%
\end{array}
\label{Eq5.104a}
\end{equation}%
and%
\begin{equation}
\begin{array}{c}
\psi _{0}\left( \tau ,t\right) =\dint\limits_{\tau }^{t}e^{2\eta \left( \tau
,s\right) }\varepsilon ^{2}\left( s\right) ds.%
\end{array}
\label{Eq5.105}
\end{equation}%
Thus, we obtain the following well-known expression:%
\begin{equation}
\begin{array}{c}
P\left( \tau ,\theta ,t,y\right) =\frac{1}{2\pi }\int_{-\infty }^{\infty
}\exp \left( -\frac{\psi _{0}\left( \tau ,t\right) l^{2}}{2}\right. \\ 
\\ 
\left. +\left( e^{\eta \left( \tau ,t\right) }y-\int_{\tau }^{t}e^{\eta
\left( \tau ,s\right) }\chi \left( s\right) ds-\theta \right) il+\eta \left(
\tau ,t\right) \right) dl \\ 
\\ 
=\frac{\exp \left( \eta \left( \tau ,t\right) \right) }{\sqrt{2\pi \psi
_{0}\left( \tau ,t\right) }}\exp \left( -\frac{\left( e^{\eta \left( \tau
,t\right) }y-\int_{\tau }^{t}e^{\eta \left( \tau ,s\right) }\chi \left(
s\right) ds-\theta \right) ^{2}}{2\psi _{0}\left( \tau ,t\right) }\right) \\ 
\\ 
=\frac{1}{\sqrt{2\pi \hat{\psi}_{0}\left( \tau ,t\right) }}\exp \left( -%
\frac{\left( y-\int_{\tau }^{t}e^{-\eta \left( s,t\right) }\chi \left(
s\right) ds-e^{-\eta \left( \tau ,t\right) }\theta \right) ^{2}}{2\hat{\psi}%
_{0}\left( \tau ,t\right) }\right) ,%
\end{array}
\label{Eq5.106}
\end{equation}%
where%
\begin{equation}
\begin{array}{c}
\hat{\psi}_{0}\left( \tau ,t\right) =\frac{\psi _{0}\left( \tau ,t\right) }{%
\exp \left( 2\eta \left( \tau ,t\right) \right) }=\dint\limits_{\tau
}^{t}e^{-2\eta \left( s,t\right) }\varepsilon ^{2}\left( s\right) ds.%
\end{array}
\label{Eq5.107}
\end{equation}%
For further discussion, see the original paper by \cite{Uhlenbeck30}, as
well as \cite{Chandrasekhar43, Risken89},\ and references therein.

For time-independent parameters, Eq. (\ref{Eq5.106}) has the form:%
\begin{equation}
\begin{array}{c}
P\left( \tau ,\theta ,t,y\right) =\frac{1}{\sqrt{\frac{\pi \varepsilon
^{2}\left( 1-e^{-2\kappa \left( t-\tau \right) }\right) }{\kappa }}}\exp
\left( -\frac{\left( y-\frac{\chi }{\kappa }-e^{-\kappa \left( t-\tau
\right) }\left( \theta -\frac{\chi }{\kappa }\right) \right) ^{2}}{\frac{%
\varepsilon ^{2}\left( 1-e^{-2\kappa \left( t-\tau \right) }\right) }{\kappa 
}}\right) .%
\end{array}
\label{Eq5.107a}
\end{equation}

\subsubsection{Degenerate Augmented OU Process}

In this subsection, we consider an augmented one-dimensional OU process:

\begin{equation}
\begin{array}{c}
dx_{t}=y_{t}dt,\ \ \ x_{\tau }=\xi , \\ 
\\ 
dy_{t}=\left( \chi \left( t\right) -\kappa \left( t\right) y_{t}\right)
dt+\varepsilon \left( t\right) dW_{t},\ \ \ y_{\tau }=\theta .%
\end{array}
\label{Eq5.34}
\end{equation}%
To align our analysis with the existing body of work, we switch from the
general notation, used above, to a specific one customary used for the OU
process. Here and below, the word \textquotedblleft
augmentation\textquotedblright\ means that we expand the original process by
incorporating its integral or other path-dependent characteristics, such as
running maximum or minimum as part of the process. The augmentation is very
useful for handling large classes of path-dependent options; details can be
found in \cite{Lipton01}, Chapter 13.

In the case in question, Eq. (\ref{Eq5.13}) can be written as follows:%
\begin{equation}
\begin{array}{c}
\mathfrak{L}^{\prime }\left( \tau ,t\right) +\left( 
\begin{array}{cc}
0 & 0 \\ 
1 & -\kappa \left( t\right)%
\end{array}%
\right) \mathfrak{L}\left( \tau ,t\right) =0,\ \ \mathfrak{L}\left( \tau
,\tau \right) =\left( 
\begin{array}{cc}
1 & 0 \\ 
0 & 1%
\end{array}%
\right) ,%
\end{array}
\label{Eq5.35}
\end{equation}%
so that 
\begin{equation}
\begin{array}{c}
\mathfrak{L}\left( \tau ,t\right) =\left( 
\begin{array}{cc}
1 & 0 \\ 
-\int_{\tau }^{t}e^{\eta \left( s,t\right) }ds & e^{\eta \left( \tau
,t\right) }%
\end{array}%
\right) , \\ 
\\ 
\mathfrak{L}^{-1}\left( \tau ,t\right) =\left( 
\begin{array}{cc}
1 & 0 \\ 
\int_{\tau }^{t}e^{-\eta \left( \tau ,s\right) }ds & e^{-\eta \left( \tau
,t\right) }%
\end{array}%
\right) ,%
\end{array}
\label{Eq5.36}
\end{equation}%
where $\eta $ is given by Eq. (\ref{Eq5.104a}).

Now, we can compute $\mathfrak{C}^{-1}\left( \tau ,t\right) $, $\mathbf{d}%
^{\left( z\right) }\left( \tau ,t\right) =\left( d^{\left( x\right) }\left(
\tau ,t\right) ,d^{\left( y\right) }\left( \tau ,t\right) \right) ^{T}$, $%
\varpi \left( \tau ,t\right) $:%
\begin{equation}
\begin{array}{c}
\mathfrak{C}^{-1}\left( \tau ,t\right) =\left( 
\begin{array}{cc}
\psi _{2}\left( \tau ,t\right) & -\psi _{1}\left( \tau ,t\right) \\ 
-\psi _{1}\left( \tau ,t\right) & \psi _{0}\left( \tau ,t\right)%
\end{array}%
\right) ,%
\end{array}
\label{Eq5.38}
\end{equation}%
where%
\begin{equation}
\begin{array}{c}
\psi _{0}\left( \tau ,t\right) =\dint\limits_{\tau }^{t}e^{2\eta \left( \tau
,s\right) }\varepsilon ^{2}\left( s\right) ds, \\ 
\\ 
\psi _{1}\left( \tau ,t\right) =-\dint\limits_{\tau }^{t}\left(
\dint\limits_{\tau }^{s}e^{\eta \left( s^{\prime },t\right) }ds^{\prime
}\right) e^{\eta \left( \tau ,s\right) }\varepsilon ^{2}\left( s\right) ds,
\\ 
\\ 
\psi _{2}\left( \tau ,t\right) =\dint\limits_{\tau }^{t}\left(
\dint\limits_{\tau }^{s}e^{\eta \left( s^{\prime },t\right) }ds^{\prime
}\right) ^{2}\varepsilon ^{2}\left( s\right) ds.%
\end{array}
\label{Eq5.39}
\end{equation}%
\begin{equation}
\begin{array}{c}
\mathbf{d}^{\left( z\right) }\left( \tau ,t\right) =\left( 
\begin{array}{c}
d^{\left( x\right) }\left( \tau ,t\right) \\ 
d^{\left( y\right) }\left( \tau ,t\right)%
\end{array}%
\right) =\left( 
\begin{array}{c}
-\int_{\tau }^{t}\left( \int_{\tau }^{s}e^{\eta \left( s^{\prime },s\right)
}ds^{\prime }\right) \chi \left( s\right) ds \\ 
\int_{\tau }^{t}e^{\eta \left( \tau ,s\right) }\chi \left( s\right) ds,%
\end{array}%
\right) ,%
\end{array}
\label{Eq5.40}
\end{equation}%
\begin{equation}
\begin{array}{c}
\varpi \left( \tau ,t\right) =-\eta \left( \tau ,t\right) .%
\end{array}
\label{Eq5.41}
\end{equation}%
Next, we calculate $\mathfrak{H}\left( \tau ,t\right) $, and $\mathbf{r}%
\left( \tau ,t\right) $:%
\begin{equation}
\begin{array}{c}
\mathfrak{H}\left( \tau ,t\right) =\left( \mathfrak{L}^{T}\left( \tau
,t\right) \right) ^{-1}\mathfrak{C}^{-1}\left( \tau ,t\right) \mathfrak{L}%
^{-1}\left( \tau ,t\right) \\ 
\\ 
=\left( 
\begin{array}{cc}
1 & \int_{\tau }^{t}e^{-\eta \left( \tau ,s\right) }ds \\ 
0 & e^{-\eta \left( \tau ,t\right) }%
\end{array}%
\right) \left( 
\begin{array}{cc}
\psi _{2}\left( \tau ,t\right) & -\psi _{1}\left( \tau ,t\right) \\ 
-\psi _{1}\left( \tau ,t\right) & \psi _{0}\left( \tau ,t\right)%
\end{array}%
\right) \left( 
\begin{array}{cc}
1 & 0 \\ 
\int_{\tau }^{t}e^{-\eta \left( \tau ,s\right) }ds & e^{-\eta \left( \tau
,t\right) }%
\end{array}%
\right) \\ 
\\ 
=\left( 
\begin{array}{cc}
h_{0}\left( \tau ,t\right) & h_{1}\left( \tau ,t\right) \\ 
h_{1}\left( \tau ,t\right) & h_{2}\left( \tau ,t\right)%
\end{array}%
\right) ,%
\end{array}
\label{Eq5.42}
\end{equation}%
where%
\begin{equation}
\begin{array}{c}
h_{0}\left( \tau ,t\right) =\psi _{0}\left( \int_{\tau }^{t}e^{-\eta \left(
\tau ,s\right) }ds\right) ^{2}-2\psi _{1}\int_{\tau }^{t}e^{-\eta \left(
\tau ,s\right) }ds+\psi _{2}, \\ 
\\ 
h_{1}\left( \tau ,t\right) =\left( \psi _{0}\int_{\tau }^{t}e^{-\eta \left(
\tau ,s\right) }ds-\psi _{1}\right) e^{-\eta \left( \tau ,t\right) }, \\ 
\\ 
h_{2}\left( \tau ,t\right) =\psi _{0}e^{-2\eta \left( \tau ,t\right) }.%
\end{array}
\label{Eq5.43}
\end{equation}%
\begin{equation}
\begin{array}{c}
\mathbf{r}\left( \tau ,t\right) =\left( 
\begin{array}{c}
p\left( \tau ,t\right) \\ 
q\left( \tau ,t\right)%
\end{array}%
\right) =\left( \mathfrak{L}^{T}\left( \tau ,t\right) \right) ^{-1}\left( 
\mathbf{d}^{\left( z\right) }\left( \tau ,t\right) +\mathbf{\zeta }\right)
\\ 
\\ 
=\left( 
\begin{array}{c}
-\int_{\tau }^{t}\left( \int_{\tau }^{s}e^{\eta \left( s^{\prime },s\right)
}ds^{\prime }\right) \chi \left( s\right) ds+\xi \\ 
\\ 
+\int_{\tau }^{t}e^{-\eta \left( \tau ,s\right) }ds\left( \int_{\tau
}^{t}e^{\eta \left( \tau ,s\right) }\chi \left( s\right) ds+\theta \right) ,
\\ 
\\ 
e^{-\eta \left( \tau ,t\right) }\left( \int_{\tau }^{t}e^{\eta \left( \tau
,s\right) }\chi \left( s\right) ds+\theta \right) ,%
\end{array}%
\right) .%
\end{array}
\label{Eq5.44}
\end{equation}%
Thus, $P\left( \tau ,\xi ,\theta ,t,x,y\right) $ is a bivariate Gaussian
distribution of the form (\ref{Eq5.23}) with the covariance matrix $%
\mathfrak{H}$, given by Eq. (\ref{Eq5.43}) centered at the point $\mathbf{r=}%
\left( p,q\right) ^{T}$ given by Eq. (\ref{Eq5.44}). Explicitly, we have%
\begin{equation}
\begin{array}{c}
\sigma _{x}^{2}\left( \tau ,t\right) =h_{0}\left( \tau ,t\right) ,\ \ \
\sigma _{y}^{2}\left( \tau ,t\right) =h_{2}\left( \tau ,t\right) ,\ \ \ \rho
\left( \tau ,t\right) =\frac{h_{1}\left( \tau ,t\right) }{\sqrt{h_{0}\left(
\tau ,t\right) h_{2}\left( \tau ,t\right) }}.%
\end{array}
\label{Eq5.45}
\end{equation}

When $\chi ,\kappa ,\varepsilon $ are constant, the above formulas become
significantly simpler. Namely,%
\begin{equation}
\begin{array}{c}
\mathfrak{L}\left( \tau ,t\right) =\left( 
\begin{array}{cc}
1 & 0 \\ 
\frac{1-e^{\kappa \left( t-\tau \right) }}{\kappa } & e^{\kappa \left(
t-\tau \right) }%
\end{array}%
\right) , \\ 
\\ 
\mathfrak{L}^{-1}\left( \tau ,t\right) =\left( 
\begin{array}{cc}
1 & 0 \\ 
\frac{1-e^{-\kappa \left( t-\tau \right) }}{\kappa } & e^{-\kappa \left(
t-\tau \right) }%
\end{array}%
\right) .%
\end{array}
\label{Eq5.46}
\end{equation}%
\begin{equation}
\begin{array}{c}
\mathfrak{C}^{-1}\left( \tau ,t\right) =\left( 
\begin{array}{cc}
\psi _{2}\left( \tau ,t\right) & -\psi _{1}\left( \tau ,t\right) \\ 
-\psi _{1}\left( \tau ,t\right) & \psi _{0}\left( \tau ,t\right)%
\end{array}%
\right) , \\ 
\\ 
\psi _{0}\left( \tau ,t\right) =\frac{\varepsilon ^{2}\left( e^{2\kappa
\left( t-\tau \right) }-1\right) }{2\kappa }, \\ 
\\ 
\psi _{1}\left( \tau ,t\right) =\frac{\varepsilon ^{2}\left( e^{\kappa
\left( t-\tau \right) }-1\right) ^{2}}{2\kappa ^{2}}, \\ 
\\ 
\psi _{2}\left( \tau ,t\right) =\frac{\varepsilon ^{2}\left( e^{2\kappa
\left( t-\tau \right) }-4e^{\kappa \left( t-\tau \right) }+3+2\kappa \left(
t-\tau \right) \right) }{2\kappa ^{3}},%
\end{array}
\label{Eq5.47}
\end{equation}%
\begin{equation}
\begin{array}{c}
\mathbf{d}^{\left( z\right) }\left( \tau ,t\right) =\left( 
\begin{array}{c}
d^{\left( x\right) }\left( \tau ,t\right) \\ 
d^{\left( y\right) }\left( \tau ,t\right)%
\end{array}%
\right) =\left( 
\begin{array}{c}
\left( \left( t-\tau \right) -\frac{e^{\kappa \left( t-\tau \right) }-1}{%
\kappa }\right) \frac{\chi }{\kappa } \\ 
\left( e^{\kappa \left( t-\tau \right) }-1\right) \frac{\chi }{\kappa }%
\end{array}%
\right) ,%
\end{array}
\label{Eq5.48}
\end{equation}%
\begin{equation}
\begin{array}{c}
\varpi \left( \tau ,t\right) =-\kappa \left( t-\tau \right) ,%
\end{array}
\label{Eq5.49}
\end{equation}%
\begin{equation}
\begin{array}{c}
\mathfrak{H}\left( \tau ,t\right) =\left( \mathfrak{L}^{T}\left( \tau
,t\right) \right) ^{-1}\mathfrak{C}^{-1}\left( \tau ,t\right) \mathfrak{L}%
^{-1}\left( \tau ,t\right) =\left( 
\begin{array}{cc}
h_{0}\left( \tau ,t\right) & h_{1}\left( \tau ,t\right) \\ 
h_{1}\left( \tau ,t\right) & h_{2}\left( \tau ,t\right)%
\end{array}%
\right) , \\ 
\\ 
h_{0}\left( \tau ,t\right) =\frac{\varepsilon ^{2}}{2\kappa ^{3}}\left(
-3+4e^{-\kappa \left( t-\tau \right) }-e^{-2\kappa \left( t-\tau \right)
}+2\kappa \left( t-\tau \right) \right) , \\ 
\\ 
h_{1}\left( \tau ,t\right) =\frac{\varepsilon ^{2}}{2\kappa ^{2}}\left(
1-e^{-\kappa \left( t-\tau \right) }\right) ^{2}, \\ 
\\ 
h_{2}\left( \tau ,t\right) =\frac{\varepsilon ^{2}\left( 1-e^{-2\kappa
\left( t-\tau \right) }\right) }{2\kappa }.%
\end{array}
\label{Eq5.51}
\end{equation}%
\begin{equation}
\begin{array}{c}
\mathbf{r}\left( \tau ,t\right) =\left( 
\begin{array}{c}
\xi +\frac{\left( \kappa \left( t-\tau \right) -\left( 1-e^{-\kappa \left(
t-\tau \right) }\right) \right) \chi }{\kappa ^{2}}+\frac{\left(
1-e^{-\kappa \left( t-\tau \right) }\right) \theta }{\kappa } \\ 
\\ 
\frac{\left( 1-e^{-\kappa \left( t-\tau \right) }\right) \chi }{\kappa }%
+e^{-\kappa \left( t-\tau \right) }\theta%
\end{array}%
\right) .%
\end{array}
\label{Eq5.52}
\end{equation}%
Thus, $P\left( \tau ,\xi ,\theta ,t,x,y\right) $ is a bivariate Gaussian
distribution of the form (\ref{Eq5.23}) with the covariance matrix $%
\mathfrak{H}$, given by Eq. (\ref{Eq5.51}) and the expected value $\mathbf{r=%
}\left( p,q\right) ^{T}$ given by Eq. (\ref{Eq5.52}).

Below, we need to find the marginal distribution of $x$, which we denote by $%
Q^{\left( x\right) }\left( \tau ,\theta ,t,x\right) $. It is well known that
marginal distributions of a multivariate Gaussian distribution are also
Gaussian, so that:%
\begin{equation}
\begin{array}{c}
Q^{\left( x\right) }\left( \tau ,\theta ,t,x\right) =\frac{1}{\sqrt{2\pi
h_{0}\left( \tau ,t\right) }}\exp \left( \frac{\left( x-p\left( \tau
,t\right) \right) ^{2}}{2h_{0}\left( \tau ,t\right) }\right) ,%
\end{array}
\label{Eq5.53}
\end{equation}%
where $h_{0}$ is given by Eqs (\ref{Eq5.51}). At the same time, the density
of marginal distribution for $y$ has the form:%
\begin{equation}
\begin{array}{c}
Q^{\left( y\right) }\left( \tau ,\theta ,t,y\right) =\frac{1}{\sqrt{2\pi
h_{2}\left( \tau ,t\right) }}\exp \left( \frac{\left( y-q\left( \tau
,t\right) \right) ^{2}}{2h_{2}\left( \tau ,t\right) }\right) ,%
\end{array}
\label{Eq5.54}
\end{equation}%
where $h_{2}$ is given by Eqs (\ref{Eq5.51}), which is the familiar density
of the OU process derived in the previous section.

\subsection{Example: Evolution of Free and Harmonically Bound Particles\label%
{Sec54}}

We can use the above results to revisit the motion of free and harmonically
bound particles considered in Section \ref{Sec3}.

To describe a free particle, we assume that $\chi =0$. Eq. (\ref{Eq5.51})
does not change, while Eq. (\ref{Eq5.52}) becomes a bit simpler:%
\begin{equation}
\begin{array}{c}
\left( p\left( \tau ,t\right) ,q\left( \tau ,t\right) \right) ^{T}=\left(
\xi +\frac{\left( 1-e^{-\kappa \left( t-\tau \right) }\right) \theta }{%
\kappa },e^{-\kappa \left( t-\tau \right) }\theta \right) ^{T}.%
\end{array}
\label{Eq5.55}
\end{equation}%
It is clear that Eqs (\ref{Eq3.6}), (\ref{Eq3.7}) and (\ref{Eq5.51}), (\ref%
{Eq5.55}) are in agreement.

Analysis of a harmonically bound particle requires additional efforts. In
the case in question, Eq. (\ref{Eq5.13}) can be written as follows:%
\begin{equation}
\begin{array}{c}
\mathfrak{L}^{\prime }\left( \tau ,t\right) +\left( 
\begin{array}{cc}
0 & -\omega ^{2} \\ 
1 & -\kappa%
\end{array}%
\right) \mathfrak{L}\left( \tau ,t\right) =0,\ \ \mathfrak{L}\left( \tau
,\tau \right) =\left( 
\begin{array}{cc}
1 & 0 \\ 
0 & 1%
\end{array}%
\right) .%
\end{array}
\label{Eq5.56}
\end{equation}%
The corresponding second-order equation is%
\begin{equation}
\begin{array}{c}
l^{\prime \prime }-\kappa l^{\prime }+\omega ^{2}l=0,%
\end{array}
\label{Eq5.57}
\end{equation}%
while the characteristic equation and its solutions can be written as
follows:%
\begin{equation}
\begin{array}{c}
\lambda ^{2}-\kappa \lambda +\omega ^{2}=0, \\ 
\\ 
\lambda _{\pm }=\frac{\kappa \pm \sqrt{\kappa ^{2}-4\omega ^{2}}}{2}=-\mu
_{\mp },%
\end{array}
\label{Eq5.58}
\end{equation}%
where $\mu _{\pm }$ are given by Eq. (\ref{Eq3.10}). It is easy to check that%
\begin{equation}
\begin{array}{c}
\mathfrak{L}\left( \tau ,t\right) =\frac{1}{\sqrt{\kappa ^{2}-4\omega ^{2}}}%
\left( 
\begin{array}{cc}
-\left( \lambda _{-}\mathsf{E}_{+}-\lambda _{+}\mathsf{E}_{-}\right) & 
\omega ^{2}\left( \mathsf{E}_{+}-\mathsf{E}_{-}\right) \\ 
-\left( \mathsf{E}_{+}-\mathsf{E}_{-}\right) & \left( \lambda _{+}\mathsf{E}%
_{+}-\lambda _{-}\mathsf{E}_{-}\right)%
\end{array}%
\right) ,%
\end{array}
\label{Eq5.59}
\end{equation}%
\begin{equation}
\begin{array}{c}
\mathfrak{L}^{-1}\left( \tau ,t\right) =\frac{1}{\sqrt{\kappa ^{2}-4\omega
^{2}}}\left( 
\begin{array}{cc}
\left( \lambda _{+}\mathsf{E}_{-}^{-1}-\lambda _{-}\mathsf{E}_{+}^{-1}\right)
& -\omega ^{2}\left( \mathsf{E}_{-}^{-1}-\mathsf{E}_{+}^{-1}\right) \\ 
\left( \mathsf{E}_{-}^{-1}-\mathsf{E}_{+}^{-1}\right) & -\left( \lambda _{-}%
\mathsf{E}_{-}^{-1}-\lambda _{+}\mathsf{E}_{+}^{-1}\right)%
\end{array}%
\right) ,%
\end{array}
\label{Eq5.60}
\end{equation}%
\begin{equation}
\begin{array}{c}
\det \mathfrak{L=}\left( \det \mathfrak{L}^{-1}\right) ^{-1}\mathfrak{=}%
\mathsf{E}_{+}\mathsf{E}_{-}=\exp \left( \kappa \left( t-\tau \right)
\right) ,%
\end{array}
\label{Eq5.61}
\end{equation}%
where%
\begin{equation}
\begin{array}{c}
\mathsf{E}_{+}=e^{\lambda _{\pm }\left( t-\tau \right) }.%
\end{array}
\label{Eq5.62}
\end{equation}%
Accordingly,%
\begin{equation}
\begin{array}{c}
P\left( \tau ,\mathbf{\zeta ,}t,\mathbf{z}\right) =\mathrm{N}\left( \mathbf{r%
}\left( \tau ,t\right) ,\mathfrak{H}\left( \tau ,t\right) \right) ,%
\end{array}
\label{Eq5.63}
\end{equation}%
with%
\begin{equation}
\begin{array}{c}
\mathfrak{H}\left( \tau ,t\right) =\left( \mathfrak{L}^{T}\left( \tau
,t\right) \right) ^{-1}\mathfrak{C}^{-1}\left( \tau ,t\right) \mathfrak{L}%
^{-1}\left( \tau ,t\right) , \\ 
\\ 
\mathbf{r}\left( \tau ,t\right) \mathbf{=}\left( \mathfrak{L}^{T}\left( \tau
,t\right) \right) ^{-1}\mathbf{\zeta }.%
\end{array}
\label{Eq5.64}
\end{equation}%
Here%
\begin{equation}
\begin{array}{c}
\mathbf{r=}\left( p,q\right) ^{T}=\left( \frac{\left( \left( \lambda _{+}%
\mathsf{E}_{-}^{-1}-\lambda _{-}\mathsf{E}_{+}^{-1}\right) \xi +\left( 
\mathsf{E}_{-}^{-1}-\mathsf{E}_{+}^{-1}\right) \theta \right) }{\sqrt{\kappa
^{2}-4\omega ^{2}}},-\frac{\left( \omega ^{2}\left( \mathsf{E}_{-}^{-1}-%
\mathsf{E}_{+}^{-1}\right) \xi +\left( \lambda _{-}\mathsf{E}%
_{-}^{-1}-\lambda _{+}\mathsf{E}_{+}^{-1}\right) \theta \right) }{\sqrt{%
\kappa ^{2}-4\omega ^{2}}}\right) ^{T}.%
\end{array}
\label{Eq5.65}
\end{equation}%
In the limit $\omega ^{2}\rightarrow 0$,%
\begin{equation}
\begin{array}{c}
\mathbf{r=}\left( p,q\right) ^{T}=\left( \xi +\frac{\left( 1-e^{-\kappa
\left( t-\tau \right) }\right) \theta }{\kappa },e^{-\kappa \left( t-\tau
\right) }\theta \right) ^{T}.%
\end{array}
\label{Eq5.66}
\end{equation}%
Next, 
\begin{equation}
\begin{array}{c}
\mathfrak{C}^{-1}\left( \tau ,t\right) =\varepsilon ^{2}\left( 
\begin{array}{cc}
\int_{\tau }^{t}\mathfrak{L}^{\left( yx\right) T}\left( s\right) \mathfrak{L}%
^{\left( yx\right) }\left( s\right) ds & \int_{\tau }^{t}\mathfrak{L}%
^{\left( yx\right) T}\left( s\right) \mathfrak{L}^{\left( yy\right) }\left(
s\right) ds \\ 
\int_{\tau }^{t}\mathfrak{L}^{\left( yy\right) T}\left( s\right) \mathfrak{L}%
^{\left( yx\right) }\left( s\right) ds & \int_{\tau }^{t}\mathfrak{L}%
^{\left( yy\right) T}\left( s\right) \mathfrak{L}^{\left( yy\right) }\left(
s\right) ds%
\end{array}%
\right) \\ 
\\ 
=\left( 
\begin{array}{cc}
\psi _{2}\left( \tau ,t\right) & -\psi _{1}\left( \tau ,t\right) \\ 
-\psi _{1}\left( \tau ,t\right) & \psi _{0}\left( \tau ,t\right)%
\end{array}%
\right) \equiv \frac{\varepsilon ^{2}}{\left( \kappa ^{2}-4\omega
^{2}\right) }\left( 
\begin{array}{cc}
\bar{\psi}_{2}\left( \tau ,t\right) & -\bar{\psi}_{1}\left( \tau ,t\right)
\\ 
-\bar{\psi}_{1}\left( \tau ,t\right) & \bar{\psi}_{0}\left( \tau ,t\right)%
\end{array}%
\right) ,%
\end{array}
\label{Eq5.67}
\end{equation}%
where%
\begin{equation}
\begin{array}{c}
\bar{\psi}_{0}\left( \tau ,t\right) =\int_{\tau }^{t}\left( \lambda _{+}%
\mathsf{E}_{+}\left( s\right) -\lambda _{-}\mathsf{E}_{-}\left( s\right)
\right) ^{2}ds \\ 
\\ 
=\frac{1}{2\kappa }\left( \kappa \lambda _{+}\mathsf{E}_{+}^{2}\left(
t\right) -4\omega ^{2}\mathsf{E}_{+}\left( t\right) \mathsf{E}_{-}\left(
t\right) +\kappa \lambda _{-}\mathsf{E}_{-}^{2}\left( t\right) -\left(
\kappa ^{2}-4\omega ^{2}\right) \right) ,%
\end{array}
\label{Eq5.68}
\end{equation}%
\begin{equation}
\begin{array}{c}
\bar{\psi}_{1}\left( \tau ,t\right) =\int_{\tau }^{t}\left( \mathsf{E}%
_{+}\left( s\right) -\mathsf{E}_{-}\left( s\right) \right) \left( \lambda
_{+}\mathsf{E}_{+}\left( s\right) -\lambda _{-}\mathsf{E}_{-}\left( s\right)
\right) ds \\ 
\\ 
=\frac{1}{2}\left( \mathsf{E}_{+}^{2}\left( t\right) -2\mathsf{E}_{+}\left(
t\right) \mathsf{E}_{-}\left( t\right) +\mathsf{E}_{-}^{2}\left( t\right)
\right) ,%
\end{array}
\label{Eq5.69}
\end{equation}%
\begin{equation}
\begin{array}{c}
\bar{\psi}_{2}\left( \tau ,t\right) =\int_{\tau }^{t}\left( \mathsf{E}%
_{+}\left( s\right) -\mathsf{E}_{-}\left( s\right) \right) ^{2}ds \\ 
\\ 
=\frac{1}{2\kappa \omega ^{2}}\left( \kappa \lambda _{-}\mathsf{E}%
_{+}^{2}\left( t\right) -4\omega ^{2}\mathsf{E}_{+}\left( t\right) \mathsf{E}%
_{-}\left( t\right) +\kappa \lambda _{+}\mathsf{E}_{-}^{2}\left( t\right)
-\left( \kappa ^{2}-4\omega ^{2}\right) \right) .%
\end{array}
\label{Eq5.70}
\end{equation}%
Further,%
\begin{equation}
\begin{array}{c}
\mathfrak{H}\left( \tau ,t\right) =\left( \mathfrak{L}^{T}\left( \tau
,t\right) \right) ^{-1}\mathfrak{C}^{-1}\left( \tau ,t\right) \mathfrak{L}%
^{-1}\left( \tau ,t\right) \\ 
\\ 
=\left( 
\begin{array}{cc}
h_{0}\left( \tau ,t\right) & h_{1}\left( \tau ,t\right) \\ 
h_{1}\left( \tau ,t\right) & h_{2}\left( \tau ,t\right)%
\end{array}%
\right) \equiv \frac{\varepsilon ^{2}}{\left( \kappa ^{2}-4\omega
^{2}\right) ^{2}}\left( 
\begin{array}{cc}
\bar{h}_{0}\left( \tau ,t\right) & \bar{h}_{1}\left( \tau ,t\right) \\ 
\bar{h}_{1}\left( \tau ,t\right) & \bar{h}_{2}\left( \tau ,t\right)%
\end{array}%
\right) ,%
\end{array}
\label{Eq5.71}
\end{equation}%
\begin{equation}
\begin{array}{c}
\left( 
\begin{array}{cc}
\bar{h}_{0} & \bar{h}_{1} \\ 
\bar{h}_{1} & \bar{h}_{2}%
\end{array}%
\right) =\left( 
\begin{array}{cc}
\left( \lambda _{+}\mathsf{E}_{-}^{-1}-\lambda _{-}\mathsf{E}_{+}^{-1}\right)
& \left( \mathsf{E}_{-}^{-1}-\mathsf{E}_{+}^{-1}\right) \\ 
-\omega ^{2}\left( \mathsf{E}_{-}^{-1}-\mathsf{E}_{+}^{-1}\right) & -\left(
\lambda _{-}\mathsf{E}_{-}^{-1}-\lambda _{+}\mathsf{E}_{+}^{-1}\right)%
\end{array}%
\right) \\ 
\\ 
\times \left( 
\begin{array}{cc}
\bar{\psi}_{2} & -\bar{\psi}_{1} \\ 
-\bar{\psi}_{1} & \bar{\psi}_{0}%
\end{array}%
\right) \left( 
\begin{array}{cc}
\left( \lambda _{+}\mathsf{E}_{-}^{-1}-\lambda _{-}\mathsf{E}_{+}^{-1}\right)
& -\omega ^{2}\left( \mathsf{E}_{-}^{-1}-\mathsf{E}_{+}^{-1}\right) \\ 
\left( \mathsf{E}_{-}^{-1}-\mathsf{E}_{+}^{-1}\right) & -\left( \lambda _{-}%
\mathsf{E}_{-}^{-1}-\lambda _{+}\mathsf{E}_{+}^{-1}\right)%
\end{array}%
\right) .%
\end{array}
\label{Eq5.72}
\end{equation}%
Straightforward but very tedious calculation yields%
\begin{equation}
\begin{array}{c}
\bar{h}_{0}=\frac{\left( \kappa ^{2}-4\omega ^{2}\right) }{2\kappa \omega
^{2}}\left( -\kappa \lambda _{-}\mathsf{E}_{+}^{-2}+4\omega ^{2}\mathsf{E}%
_{-}^{-1}\mathsf{E}_{+}^{-1}-\kappa \lambda _{+}\mathsf{E}_{-}^{-2}+\kappa
^{2}-4\omega ^{2}\right) , \\ 
\\ 
\bar{h}_{1}=\frac{\left( \kappa ^{2}-4\omega ^{2}\right) }{2}\left( \mathsf{E%
}_{-}^{-2}-2\mathsf{E}_{+}^{-1}\mathsf{E}_{-}^{-1}+\mathsf{E}%
_{+}^{-2}\right) , \\ 
\\ 
\bar{h}_{2}=\frac{\left( \kappa ^{2}-4\omega ^{2}\right) }{2\kappa }\left(
-\kappa \lambda _{+}\mathsf{E}_{+}^{-2}+4\omega ^{2}\mathsf{E}_{-}^{-1}%
\mathsf{E}_{+}^{-1}-\kappa \lambda _{-}\mathsf{E}_{-}^{-2}+\kappa
^{2}-4\omega ^{2}\right) ,%
\end{array}
\label{Eq5.73}
\end{equation}%
and%
\begin{equation}
\begin{array}{c}
h_{0}=\frac{\varepsilon ^{2}}{2\kappa \omega ^{2}\left( \kappa ^{2}-4\omega
^{2}\right) }\left( -\kappa \lambda _{-}\mathsf{E}_{+}^{-2}+4\omega ^{2}%
\mathsf{E}_{-}^{-1}\mathsf{E}_{+}^{-1}-\kappa \lambda _{+}\mathsf{E}%
_{-}^{-2}+\kappa ^{2}-4\omega ^{2}\right) , \\ 
\\ 
h_{1}=\frac{\varepsilon ^{2}}{2\left( \kappa ^{2}-4\omega ^{2}\right) }%
\left( \mathsf{E}_{-}^{-2}-2\mathsf{E}_{+}^{-1}\mathsf{E}_{-}^{-1}+\mathsf{E}%
_{+}^{-2}\right) , \\ 
\\ 
h_{2}=\frac{\varepsilon ^{2}}{2\kappa \left( \kappa ^{2}-4\omega ^{2}\right) 
}\left( -\kappa \lambda _{+}\mathsf{E}_{+}^{-2}+4\omega ^{2}\mathsf{E}%
_{-}^{-1}\mathsf{E}_{+}^{-1}-\kappa \lambda _{-}\mathsf{E}_{-}^{-2}+\kappa
^{2}-4\omega ^{2}\right) .%
\end{array}
\label{Eq5.74}
\end{equation}%
In the limit $\omega ^{2}\rightarrow 0$,%
\begin{equation}
\begin{array}{c}
h_{0}=\frac{\varepsilon ^{2}}{2\kappa ^{3}}\left( -3+2\kappa \left( t-\tau
\right) +4e^{-\kappa \left( t-\tau \right) }-e^{-2\kappa \left( t-\tau
\right) }\right) , \\ 
\\ 
h_{1}=\frac{\varepsilon ^{2}}{2\kappa ^{2}}\left( \mathsf{1}-2e^{-\kappa
\left( t-\tau \right) }+e^{-2\kappa \left( t-\tau \right) }\right) , \\ 
\\ 
h_{2}=\frac{\varepsilon ^{2}}{2\kappa }\left( 1-e^{-2\kappa \left( t-\tau
\right) }\right) ,%
\end{array}
\label{Eq5.75}
\end{equation}%
so that Eqs. (\ref{Eq5.51}) and (\ref{Eq5.75}) are in agreement.

Moreover, while it is easy to show that Chandrasekhar's solution given by
Eq. (\ref{Eq3.13}) is in agreement with our solution given by Eq. (\ref%
{Eq5.63}), our solution is more convenient from a practical standpoint,
since it is \emph{explicitly} written as a Gaussian density in the $\left(
x,y\right) $ space.

\subsection{Example: Evolution of Vorticity of Two-Dimensional Flows\label%
{Sec55}}

We shall briefly return to our point of departure and consider strictly
two-dimensional flows; see \cite{Friedlander03}. Velocity fields of such
flows have the form:%
\begin{equation}
\begin{array}{c}
\mathbf{V}\left( t,x_{1},x_{2}\right) =\left( V_{1}\left(
t,x_{1},x_{2}\right) ,V_{2}\left( t,x_{1},x_{2}\right) \right) , \\ 
\\ 
\mathbf{v}\left( t,x_{1},x_{2}\right) =\left( v_{1}\left(
t,x_{1},x_{2}\right) ,v_{2}\left( t,x_{1},x_{2}\right) \right) .%
\end{array}
\label{Eq5.76}
\end{equation}%
By virtue of incompressibility, we can introduce the so-called stream
functions such that%
\begin{equation}
\begin{array}{c}
V_{1}=-\frac{\partial \Psi }{\partial x_{2}},\ \ \ V_{2}=\frac{\partial \Psi 
}{\partial x_{1}},\ \ \ v_{1}=-\frac{\partial \psi }{\partial x_{2}},\ \ \
v_{2}=\frac{\partial \psi }{\partial x_{1}},%
\end{array}
\label{Eq5.77}
\end{equation}%
and define the (scalar) vorticity as follows:%
\begin{equation}
\begin{array}{c}
\Omega =\Delta \Psi ,\ \ \ \omega =\Delta \psi .%
\end{array}
\label{Eq5.78}
\end{equation}%
Contour lines of $\Psi $ are called streamlines of the flow.

By using the above definitions, we write the two-dimensional the
Navier-Stokes equations as equations for the stream and vorticity: 
\begin{equation}
\begin{array}{c}
\frac{\partial \Omega }{\partial t}-\frac{\partial \Psi }{\partial x_{2}}%
\frac{\partial \Omega }{\partial x_{1}}+\frac{\partial \Psi }{\partial x_{1}}%
\frac{\partial \Omega }{\partial x_{2}}-\nu \Delta \Omega =0, \\ 
\\ 
\Delta \Psi -\Omega =0.%
\end{array}
\label{Eq5.79}
\end{equation}%
It is clear that time-independent stream functions $\Psi $, quadratically
dependent of $\left( x_{1},x_{2}\right) $ generate exact solutions of Eqs (%
\ref{Eq5.79}). We are interested in fields consisting of pure strain and
pure rotation. The corresponding $\Psi $ have the form:%
\begin{equation*}
\begin{array}{c}
\Psi \left( x_{1},x_{2}\right) =\frac{1}{4}\left( w\left(
x_{1}^{2}+x_{2}^{2}\right) -2sx_{1}x_{2}\right) ,%
\end{array}%
\end{equation*}%
where $\omega >s$, to ensure that streamlines are elliptic rather than
hyperbolic, so that%
\begin{equation}
\begin{array}{c}
V_{1}=-\frac{\partial \Psi }{\partial x_{2}}=\frac{1}{2}\left(
sx_{1}-wx_{2}\right) ,\ \ \ V_{2}=\frac{\partial \Psi }{\partial x_{1}}=%
\frac{1}{2}\left( wx_{1}-sx_{2}\right) .%
\end{array}
\label{Eq5.80}
\end{equation}

Small perturbations $\psi $ of the time-independent quadratic stream
function $\Psi $ satisfy the following equations:%
\begin{equation}
\begin{array}{c}
\frac{\partial \omega }{\partial t}-\frac{\partial \Psi }{\partial x_{2}}%
\frac{\partial \omega }{\partial x_{1}}+\frac{\partial \Psi }{\partial x_{1}}%
\frac{\partial \omega }{\partial x_{2}}-\nu \Delta \omega =0, \\ 
\\ 
\Delta \psi -\omega =0.%
\end{array}
\label{Eq5.81}
\end{equation}%
We wish to study the first equation (\ref{Eq5.81}) in isolation, which we
write explicitly as follows:%
\begin{equation}
\begin{array}{c}
\frac{\partial \omega }{\partial t}+\frac{1}{2}\left( sx_{1}-wx_{2}\right) 
\frac{\partial \omega }{\partial x_{1}}+\frac{1}{2}\left(
wx_{1}-sx_{2}\right) \frac{\partial \omega }{\partial x_{2}}-\nu \Delta
\omega =0,%
\end{array}
\label{Eq5.82}
\end{equation}%
and supply with the initial condition at time $\tau $:%
\begin{equation}
\begin{array}{c}
\omega \left( \tau ,x_{1},x_{2}\right) =\delta \left( x_{1}-\xi _{1}\right)
\delta \left( x_{2}-\xi _{2}\right) .%
\end{array}
\label{Eq5.83}
\end{equation}%
Once the solution of Eqs (\ref{Eq5.82}), (\ref{Eq5.83}) is found, we can
find $\psi $ by solving the corresponding Laplace equation.

Surprisingly, we can view this equation as the Fokker-Planck equation
associated with the following SDEs for $\mathbf{z}_{t}=\left(
x_{1t},x_{2t}\right) $:%
\begin{equation}
\begin{array}{c}
d\mathbf{z}_{t}=\mathfrak{B}\mathbf{z}_{t}dt+\mathbf{\Sigma }d\mathbf{W}%
_{t},\ \ \ \mathbf{z}_{\tau }=\left( \xi _{1},\xi _{2}\right) ^{T},%
\end{array}
\label{Eq5.84}
\end{equation}%
where%
\begin{equation}
\begin{array}{c}
\mathfrak{B=}\frac{1}{2}\left( 
\begin{array}{cc}
s & -w \\ 
w & -s%
\end{array}%
\right) ,\ \ \ \mathbf{\Sigma =}\sqrt{2\nu }\left( 
\begin{array}{cc}
1 & 0 \\ 
0 & 1%
\end{array}%
\right) .%
\end{array}
\label{Eq5.85}
\end{equation}%
Thus, we can use Section \ref{Sec51} results. Eq. (\ref{Eq5.13}) becomes%
\begin{equation}
\begin{array}{c}
\mathfrak{L}^{\prime }\left( \tau ,t\right) +\frac{1}{2}\left( 
\begin{array}{cc}
s & w \\ 
-w & -s%
\end{array}%
\right) \mathfrak{L}\left( \tau ,t\right) =0,\ \ \mathfrak{L}\left( \tau
,\tau \right) =\left( 
\begin{array}{cc}
1 & 0 \\ 
0 & 1%
\end{array}%
\right) .%
\end{array}
\label{Eq5.86}
\end{equation}%
Simple but tedious calculations omitted for the sake of brevity show that%
\begin{equation}
\begin{array}{c}
\mathfrak{L}\left( \tau ,t\right) =\left( 
\begin{array}{cc}
\mathsf{C}_{1}\left( \tau ,t\right) -\frac{s}{2\zeta }\mathsf{S}_{1}\left(
\tau ,t\right) & -\frac{w}{2\zeta }\mathsf{S}_{1}\left( \tau ,t\right) \\ 
\frac{w}{2\zeta }\mathsf{S}_{1}\left( \tau ,t\right) & \mathsf{C}_{1}\left(
\tau ,t\right) +\frac{s}{2\zeta }\mathsf{S}_{1}\left( \tau ,t\right)%
\end{array}%
\right) ,%
\end{array}
\label{Eq5.87}
\end{equation}%
and%
\begin{equation}
\begin{array}{c}
\mathfrak{L}^{-1}\left( \tau ,t\right) =\left( 
\begin{array}{cc}
\mathsf{C}_{1}\left( \tau ,t\right) +\frac{s}{2\zeta }\mathsf{S}_{1}\left(
\tau ,t\right) & \frac{w}{2\zeta }\mathsf{S}_{1}\left( \tau ,t\right) \\ 
-\frac{w}{2\zeta }\mathsf{S}_{1}\left( \tau ,t\right) & \mathsf{C}_{1}\left(
\tau ,t\right) -\frac{s}{2\zeta }\mathsf{S}_{1}\left( \tau ,t\right)%
\end{array}%
\right) ,%
\end{array}
\label{Eq5.87a}
\end{equation}%
where%
\begin{equation}
\begin{array}{c}
\zeta =\frac{\sqrt{w^{2}-s^{2}}}{2},\ \ \ \mathsf{C}_{1}\left( \tau
,t\right) =\cos \left( \zeta \left( t-\tau \right) \right) ,\ \ \ \mathsf{S}%
_{1}\left( \tau ,t\right) =\sin \left( \zeta \left( t-\tau \right) \right) ,
\\ 
\\ 
\det \left( \mathfrak{L}\right) =\det \left( \mathfrak{L}^{-1}\right) =1.%
\end{array}
\label{Eq5.88}
\end{equation}%
Next, Eq. (\ref{Eq5.18}) yields%
\begin{equation}
\begin{array}{c}
\mathfrak{C}^{-1}\left( \tau ,t\right) =\int_{\tau }^{t}\mathfrak{L}%
^{T}\left( \tau ,s\right) \mathfrak{A}\left( s\right) \mathfrak{L}\left(
\tau ,s\right) ds \\ 
\\ 
=2\nu \int_{\tau }^{t}\mathfrak{L}^{T}\left( \tau ,s\right) \mathfrak{L}%
\left( \tau ,s\right) ds=\left( 
\begin{array}{cc}
\psi _{2}\left( \tau ,t\right) & -\psi _{1}\left( \tau ,t\right) \\ 
-\psi _{1}\left( \tau ,t\right) & \psi _{0}\left( \tau ,t\right)%
\end{array}%
\right) ,%
\end{array}
\label{Eq5.89}
\end{equation}%
where%
\begin{equation}
\begin{array}{c}
\psi _{0}\left( \tau ,t\right) =2\nu \int_{\tau }^{t}\left( 1+\frac{s}{%
2\zeta }\mathsf{S}_{2}\left( \tau ,s\right) +\frac{s^{2}}{4\zeta ^{2}}\left(
1-\mathsf{C}_{2}\left( \tau ,s\right) \right) \right) ds \\ 
\\ 
=2\nu \left( \left( 1+\frac{s^{2}}{4\zeta ^{2}}\right) \left( t-\tau \right)
-\frac{s}{4\zeta ^{2}}\mathsf{C}_{2}\left( \tau ,t\right) -\frac{s^{2}}{%
8\zeta ^{3}}\mathsf{S}_{2}\left( \tau ,t\right) \right) ,%
\end{array}
\label{Eq5.90}
\end{equation}%
\begin{equation}
\begin{array}{c}
\psi _{1}\left( \tau ,t\right) =-\frac{\nu sw}{2\zeta ^{2}}\int_{\tau
}^{t}\left( 1-\mathsf{C}_{2}\left( \tau ,s\right) \right) ds \\ 
\\ 
=-\frac{\nu sw}{2\zeta ^{2}}\left( \left( t-\tau \right) -\frac{1}{2\zeta }%
\mathsf{S}_{2}\left( \tau ,t\right) \right) ,%
\end{array}
\label{Eq5.91}
\end{equation}%
\begin{equation}
\begin{array}{c}
\psi _{2}\left( \tau ,t\right) =2\nu \int_{\tau }^{t}\left( 1-\frac{s}{%
2\zeta }\mathsf{S}_{2}\left( \tau ,s\right) +\frac{s^{2}}{4\zeta ^{2}}\left(
1-\mathsf{C}_{2}\left( \tau ,s\right) \right) \right) ds \\ 
\\ 
=2\nu \left( \left( 1+\frac{s^{2}}{4\zeta ^{2}}\right) \left( t-\tau \right)
+\frac{s}{4\zeta ^{2}}\mathsf{C}_{2}\left( \tau ,t\right) -\frac{s^{2}}{%
8\zeta ^{3}}\mathsf{S}_{2}\left( \tau ,t\right) \right) ,%
\end{array}
\label{Eq5.92}
\end{equation}%
where%
\begin{equation}
\begin{array}{c}
\mathsf{C}_{2}\left( \tau ,t\right) =\cos \left( 2\zeta \left( t-\tau
\right) \right) ,\ \ \ \mathsf{S}_{2}\left( \tau ,t\right) =\sin \left(
2\zeta \left( t-\tau \right) \right) .%
\end{array}
\label{Eq5.93}
\end{equation}

Finally, Eqs (\ref{Eq5.23}), (\ref{Eq5.24}) yield:%
\begin{equation}
\begin{array}{c}
\omega \left( \tau ,\mathbf{\zeta ,}t,\mathbf{z}\right) =\mathrm{N}\left( 
\mathbf{r}\left( \tau ,t\right) ,\mathfrak{H}\left( \tau ,t\right) \right) ,%
\end{array}
\label{Eq5.93a}
\end{equation}%
\begin{equation}
\begin{array}{c}
\mathfrak{H}\left( \tau ,t\right) =\left( \mathfrak{L}^{T}\left( \tau
,t\right) \right) ^{-1}\mathfrak{C}^{-1}\left( \tau ,t\right) \mathfrak{L}%
^{-1}\left( \tau ,t\right) =\left( 
\begin{array}{cc}
h_{0}\left( \tau ,t\right) & h_{1}\left( \tau ,t\right) \\ 
h_{1}\left( \tau ,t\right) & h_{2}\left( \tau ,t\right)%
\end{array}%
\right) ,%
\end{array}
\label{Eq5.94}
\end{equation}%
where%
\begin{equation}
\begin{array}{c}
h_{0}\left( \tau ,t\right) =\left( \frac{w^{2}}{8\zeta ^{2}}+\frac{4\zeta
^{2}-s^{2}}{8\zeta ^{2}}\mathsf{C}_{2}\left( \tau ,t\right) +\frac{s}{2\zeta 
}\mathsf{S}_{2}\left( \tau ,t\right) \right) \psi _{2}\left( \tau ,t\right)
\\ 
\\ 
+\frac{w}{2\zeta }\left( \frac{s}{2\zeta }\left( 1-\mathsf{C}_{2}\left( \tau
,t\right) \right) +\mathsf{S}_{2}\left( \tau ,t\right) \right) \psi
_{1}\left( \tau ,t\right) +\frac{w^{2}}{8\zeta ^{2}}\left( 1-\mathsf{C}%
_{2}\left( \tau ,t\right) \right) \psi _{0}\left( \tau ,t\right) , \\ 
\\ 
h_{1}\left( \tau ,t\right) =\frac{w}{4\zeta }\left( \frac{s}{2\zeta }\left(
1-\mathsf{C}_{2}\left( \tau ,t\right) \right) +\mathsf{S}_{2}\left( \tau
,t\right) \right) \psi _{2}\left( \tau ,t\right) \\ 
\\ 
-\left( 1+\frac{w^{2}}{4\zeta ^{2}}\left( 1-\mathsf{C}_{2}\left( \tau
,t\right) \right) \right) \psi _{1}\left( \tau ,t\right) +\frac{w}{4\zeta }%
\left( \frac{s}{2\zeta }\left( 1-\mathsf{C}_{2}\left( \tau ,t\right) \right)
-\mathsf{S}_{2}\left( \tau ,t\right) \right) \psi _{0}\left( \tau ,t\right) ,
\\ 
\\ 
h_{2}\left( \tau ,t\right) =\frac{w^{2}}{8\zeta ^{2}}\left( 1-\mathsf{C}%
_{2}\left( \tau ,t\right) \right) \psi _{2}\left( \tau ,t\right) +\frac{w}{%
2\zeta }\left( \frac{s}{2\zeta }\left( 1-\mathsf{C}_{2}\left( \tau ,t\right)
\right) -\mathsf{S}_{2}\left( \tau ,t\right) \right) \psi _{1}\left( \tau
,t\right) \\ 
\\ 
+\left( \frac{w^{2}}{8\zeta ^{2}}+\frac{4\zeta ^{2}-s^{2}}{8\zeta ^{2}}%
\mathsf{C}_{2}\left( \tau ,t\right) -\frac{s}{2\zeta }\mathsf{S}_{2}\left(
\tau ,t\right) \right) \psi _{0}\left( \tau ,t\right) ,%
\end{array}
\label{Eq5.95}
\end{equation}%
and%
\begin{equation}
\begin{array}{c}
\mathbf{r}\left( \tau ,t\right) \mathbf{=}\left( \mathfrak{L}^{T}\left( \tau
,t\right) \right) ^{-1}\mathbf{\zeta =}\left( 
\begin{array}{c}
r_{1}\left( \tau ,t\right) \\ 
r_{2}\left( \tau ,t\right)%
\end{array}%
\right) \\ 
\\ 
=\left( 
\begin{array}{c}
\left( \mathsf{C}_{1}\left( \tau ,t\right) +\frac{s}{2\zeta }\mathsf{S}%
_{1}\left( \tau ,t\right) \right) \xi _{1}-\frac{w}{2\zeta }\mathsf{S}%
_{1}\left( \tau ,t\right) \xi _{2} \\ 
\frac{w}{2\zeta }\mathsf{S}_{1}\left( \tau ,t\right) \xi _{1}+\left( \mathsf{%
C}_{1}\left( \tau ,t\right) -\frac{s}{2\zeta }\mathsf{S}_{1}\left( \tau
,t\right) \right) \xi _{2}%
\end{array}%
\right) .%
\end{array}
\label{Eq5.96}
\end{equation}%
Eqs (\ref{Eq5.95}) are symmetric, namely $h_{0}\rightarrow h_{2}$ when $%
\left( a,b\right) \rightarrow \left( -a,-b\right) $, $\left( \psi _{0},\psi
_{2}\right) \rightarrow \left( \psi _{2},\psi _{0}\right) $. The second Eq. (%
\ref{Eq5.81}), which is a static Poisson equation, allows us to find $\psi $%
. Its analytical solution is not easy to derive and we shall not pursue this
line of research here. The special case of purely rotational flow, $s=0$,
can be done easily; see Eq. (\ref{Eq5.99e}) below.

It is interesting to note that 
\begin{equation}
\begin{array}{c}
\Psi \left( r_{1},r_{2}\right) =\Psi \left( \xi _{1},\xi _{2}\right) ,%
\end{array}
\label{Eq5.97}
\end{equation}%
so that the location of the Gaussian distribution $\omega $ moves along
streamlines of the flow defined by the stream function $\Psi $.

When the flow is purely rotational, so that $s=0$, the above formulas
considerably simplify.\ Specifically, we have:%
\begin{equation}
\begin{array}{c}
\psi _{0}\left( \tau ,t\right) =2\nu \left( t-\tau \right) ,\ \ \psi
_{1}\left( \tau ,t\right) =0,\ \ \ \ \ \psi _{2}\left( \tau ,t\right) =2\nu
\left( t-\tau \right) , \\ 
\\ 
h_{0}\left( \tau ,t\right) =2\nu \left( t-\tau \right) ,\ \ \psi _{1}\left(
\tau ,t\right) =0,\ \ \ \ \ h_{2}\left( \tau ,t\right) =2\nu \left( t-\tau
\right) , \\ 
\\ 
r_{1}\left( \tau ,t\right) =\mathsf{C}_{1}\left( \tau ,t\right) \xi _{1}-%
\mathsf{S}_{1}\left( \tau ,t\right) \xi _{2},\ \ \ r_{2}\left( \tau
,t\right) =\mathsf{S}_{1}\left( \tau ,t\right) \xi _{1}+\mathsf{C}_{1}\left(
\tau ,t\right) \xi _{2},%
\end{array}
\label{Eq5.98}
\end{equation}%
so that%
\begin{equation}
\begin{array}{c}
\omega \left( \tau ,\xi _{1},\xi _{2},t,x_{1},x_{2}\right) \\ 
\\ 
=\frac{1}{4\pi \nu \left( t-\tau \right) }\exp \left( -\frac{\left( x_{1}-%
\mathsf{C}_{1}\left( \tau ,t\right) \xi _{1}+\mathsf{S}_{1}\left( \tau
,t\right) \xi _{2}\right) ^{2}+\left( x_{2}-\mathsf{S}_{1}\left( \tau
,t\right) \xi _{1}-\mathsf{C}_{1}\left( \tau ,t\right) \xi _{2}\right) ^{2}}{%
4\nu \left( t-\tau \right) }\right) .%
\end{array}
\label{Eq5.99}
\end{equation}%
We can directly calculate the stream function $\psi $ by solving the
corresponding Poisson equation.\footnote{%
We are grateful to Andrey Itkin for pointing this out.} To start, we notice
that \emph{both} $\omega $ and $\psi $ are rotationally symmetric around the
point $\left( \xi _{1},\xi _{2}\right) $. Thus, we can introduce 
\begin{equation}
\begin{array}{c}
R^{2}=\frac{\left( x_{1}-\mathsf{C}_{1}\left( \tau ,t\right) \xi _{1}+%
\mathsf{S}_{1}\left( \tau ,t\right) \xi _{2}\right) ^{2}+\left( x_{2}-%
\mathsf{S}_{1}\left( \tau ,t\right) \xi _{1}-\mathsf{C}_{1}\left( \tau
,t\right) \xi _{2}\right) ^{2}}{2\nu \left( t-\tau \right) },%
\end{array}
\label{Eq5.99a}
\end{equation}%
representing $\omega $ and $\psi $ as follows:%
\begin{equation}
\begin{array}{c}
\omega =\omega \left( R\right) =\frac{1}{4\pi \nu \left( t-\tau \right) }%
\exp \left( -\frac{R^{2}}{2}\right) ,\ \ \ \psi =\psi \left( R\right) .%
\end{array}
\label{Eq5.99b}
\end{equation}%
Then $\psi \left( R\right) $ solves a radially symmetric Poisson equation of
the form:%
\begin{equation}
\begin{array}{c}
\frac{1}{R}\left( R\psi _{R}\left( R\right) \right) _{R}=\frac{1}{2\pi }\exp
\left( -\frac{R^{2}}{2}\right) .%
\end{array}
\label{Eq5.99c}
\end{equation}%
Thus,%
\begin{equation}
\begin{array}{c}
R\psi _{R}\left( R\right) =-\frac{1}{2\pi }\exp \left( -\frac{R^{2}}{2}%
\right) +C,%
\end{array}
\label{Eq5.99d}
\end{equation}%
where $C$ is an arbitrary constant. Next,%
\begin{equation}
\begin{array}{c}
\psi \left( R\right) =\frac{1}{4\pi }E_{1}\left( \frac{R^{2}}{2}\right) +%
\frac{1}{2\pi }\ln \left( R\right) ,%
\end{array}
\label{Eq5.99e}
\end{equation}%
where our choice of $C$ guarantees that $\psi $ has the right behavior when $%
R\rightarrow 0$ and $R\rightarrow \infty $. Here $E_{1}\left( \eta \right) $
is the exponential integral of the form:%
\begin{equation}
\begin{array}{c}
E_{1}\left( \eta \right) =\int\limits_{\eta }^{\infty }\frac{e^{-\eta
^{\prime }}}{\eta ^{\prime }}d\eta ^{\prime }.%
\end{array}
\label{Eq5.99f}
\end{equation}

\section{Non-Gaussian Processes\label{Sec6}}

\subsection{The General Case\label{Sec61}}

For financial applications, it is useful to consider a more general equation
governing SDE of the form:%
\begin{equation}
\begin{array}{c}
d\mathbf{x}_{t}=\left( \mathbf{b}^{\left( x\right) }\left( t\right) +%
\mathfrak{B}^{\left( xx\right) }\left( t\right) \mathbf{x}_{t}+\mathfrak{B}%
^{\left( xy\right) }\left( t\right) \mathbf{y}_{t}\right) dt, \\ 
\\ 
d\mathbf{y}_{t}=\left( \mathbf{b}^{\left( y\right) }\left( t\right) +%
\mathfrak{B}^{\left( yx\right) }\left( t\right) \mathbf{x}_{t}+\mathfrak{B}%
^{\left( yy\right) }\left( t\right) \mathbf{y}_{t}\right) dt \\ 
\\ 
+\mathfrak{\mathbf{\Sigma }}^{\left( yy\right) }\left( t\right) \left( 
\mathrm{diag}\left( \mathbf{d}^{\left( 0\right) }\left( t\right) +\mathfrak{D%
}\left( t\right) \mathbf{z}_{t}\right) \right) ^{1/2}d\mathbf{W}_{t}^{\left(
y\right) }, \\ 
\\ 
\mathbf{x}_{\tau }=\mathbf{\xi ,\ \ \ y}_{\tau }=\mathbf{\theta ,}%
\end{array}
\label{Eq6.1}
\end{equation}%
or, more compactly,%
\begin{equation}
\begin{array}{c}
d\mathbf{z}_{t}=\left( \mathbf{b}^{\left( z\right) }\left( t\right) +%
\mathfrak{B}^{\left( zz\right) }\left( t\right) \mathbf{z}_{t}\right) dt \\ 
\\ 
+\left( 
\begin{array}{c}
0 \\ 
\mathbf{\Sigma }^{\left( yy\right) }\left( t\right) \left( \mathrm{diag}%
\left( \mathbf{d}^{\left( 0\right) }\left( t\right) +\mathfrak{D}\left(
t\right) \mathbf{z}_{t}\right) \right) ^{1/2}d\mathbf{W}_{t}^{\left(
y\right) }%
\end{array}%
\right) , \\ 
\\ 
\mathbf{z}_{\tau }=\mathbf{\zeta =}\left( 
\begin{array}{c}
\mathbf{\xi } \\ 
\mathbf{\theta }%
\end{array}%
\right) ,%
\end{array}
\label{Eq6.2}
\end{equation}%
Here, in addition to the quantities introduced in the previous section, we
define an $\left( N\times 1\right) $ column vector $\mathbf{d}^{\left(
0\right) }$, and an $\left( N\times I\right) $ matrix $\mathfrak{D}$. It is
convenient to introduce vectors $\mathbf{d}^{\left( i\right) }$ equal to the 
$i$-th column of $\mathfrak{D}$.

Since the corresponding $\left( N\times N\right) $ covariance matrix $%
\mathfrak{A}$ has the form:%
\begin{equation}
\begin{array}{c}
\mathfrak{A=}\mathbf{\Sigma }\left( \mathrm{diag}\left( \mathbf{d}^{\left(
0\right) }\left( t\right) +\mathfrak{D}\left( t\right) \mathbf{z}_{t}\right)
\right) ^{1/2}\left( \mathbf{\Sigma }\left( \mathrm{diag}\left( \mathbf{d}%
^{\left( 0\right) }\left( t\right) +\mathfrak{D}\left( t\right) \mathbf{z}%
_{t}\right) \right) ^{1/2}\right) ^{T},%
\end{array}
\label{Eq6.3}
\end{equation}%
it linearly depends on $\mathbf{z}$:%
\begin{equation}
\begin{array}{c}
\mathfrak{A=}\mathbf{\Sigma }\left( \mathrm{diag}\left( \mathbf{d}^{\left(
0\right) }+\mathfrak{D}\mathbf{z}\right) \right) \mathbf{\Sigma }^{T}=%
\mathfrak{A}^{\left( 0\right) }+\dsum\limits_{i=1}^{M+N}\mathfrak{A}^{\left(
i\right) }z_{i},%
\end{array}
\label{Eq6.4}
\end{equation}%
where%
\begin{equation}
\begin{array}{c}
\mathfrak{A}^{\left( 0\right) }=\mathbf{\Sigma }\mathrm{diag}\left( \mathbf{d%
}^{\left( 0\right) }\right) \Sigma ^{T},\ \ \ \mathfrak{A}^{\left( i\right)
}=\mathbf{\Sigma }\mathrm{diag}\left( \mathbf{d}^{\left( i\right) }\right)
\Sigma ^{T}.%
\end{array}
\label{Eq6.5}
\end{equation}%
In contrast to the Gaussian case, Eqs (\ref{Eq6.2}) have to be defined in
the domain $\boldsymbol{D}$ such that 
\begin{equation}
\boldsymbol{D=}\left\{ \left. \mathbf{z}\right\vert \mathbf{d}^{\left(
0\right) }+\mathfrak{D}\mathbf{z\geq 0}\right\} ,  \label{Eq6.6}
\end{equation}%
rather than in the whole space.

In financial engineering, covariance matrices of the form (\ref{Eq6.3}) were
introduced by \cite{Dai00}, they are discussed by \cite{Duffie03,
Filipovic09} and many others.

The corresponding Fokker-Plank problem has the form: 
\begin{equation}
\begin{array}{c}
P_{t}\left( \tau ,\mathbf{\xi },\mathbf{\theta ,}t,\mathbf{x},\mathbf{y}%
\right) -\frac{1}{2}\dsum \dsum \mathfrak{A}P_{\mathbf{zz}}\left( \tau ,%
\mathbf{\xi },\mathbf{\theta ,}t,\mathbf{x},\mathbf{y}\right) \\ 
\\ 
+\left( \mathbf{b}+\mathfrak{B}\mathbf{z}_{t}\right) \cdot P_{\mathbf{z}%
}\left( \tau ,\mathbf{\xi },\mathbf{\theta ,}t,\mathbf{x},\mathbf{y}\right)
+b^{\left( z\right) }P\left( \tau ,\mathbf{\xi },\mathbf{\theta ,}t,\mathbf{x%
},\mathbf{y}\right) =0, \\ 
\\ 
P\left( \tau ,\mathbf{z},\tau ,\mathbf{\zeta }\right) =\delta \left( \mathbf{%
z}-\mathbf{\zeta }\right) ,%
\end{array}
\label{Eq6.7}
\end{equation}%
Integral (\ref{Eq5.7}) expressing $P$ in terms of $H$ of the form:%
\begin{equation}
\begin{array}{c}
H\left( \tau ,\mathbf{\zeta ,}t,\mathbf{z,m}\right) =\exp \left( \alpha
\left( t\right) +i\Upsilon \left( t\right) \cdot \mathbf{z}-i\mathbf{m}\cdot 
\mathbf{\zeta }\right) ,%
\end{array}
\label{Eq6.7a}
\end{equation}%
holds. The equations for $\alpha ,\mathbf{\Upsilon }$ have the form:%
\begin{equation}
\begin{array}{c}
\alpha ^{\prime }\left( t\right) +i\mathbf{\Upsilon }^{\prime }\left(
t\right) \cdot \mathbf{z}+\frac{1}{2}\mathbf{\Delta }\left( t\right) \cdot 
\mathfrak{A}\mathbf{\Delta }\left( t\right) +i\mathbf{\Upsilon }\left(
t\right) \cdot \left( \mathbf{b}^{\left( z\right) }+\mathfrak{B}^{\left(
zz\right) }\mathbf{z}\right) +b^{\left( z\right) }=0,%
\end{array}
\label{Eq6.8}
\end{equation}%
or, more explicitly,%
\begin{equation}
\begin{array}{c}
\alpha ^{\prime }\left( t\right) +i\mathbf{\Upsilon }^{\prime }\left(
t\right) \cdot \mathbf{z}+\frac{1}{2}\mathbf{\Delta }\left( t\right) \cdot 
\mathfrak{A}^{\left( 0\right) }\mathbf{\Delta }\left( t\right)
+\sum\limits_{i=1}^{I}\frac{1}{2}\mathbf{\Delta }\left( t\right) \cdot 
\mathfrak{A}^{\left( i\right) }\mathbf{\Delta }\left( t\right) z_{i} \\ 
\\ 
+i\mathbf{\Upsilon }\left( t\right) \cdot \left( \mathbf{b}^{\left( z\right)
}+\mathfrak{B}^{\left( zz\right) }\mathbf{z}\right) +b^{\left( z\right) }=0.%
\end{array}
\label{Eq6.9}
\end{equation}%
Thus, the system of ODEs for $\alpha ,\mathbf{\Upsilon }$ can be written as
follows:%
\begin{equation}
\begin{array}{c}
\alpha ^{\prime }\left( t\right) +\frac{1}{2}\mathbf{\Delta }\left( t\right)
\cdot \mathfrak{A}^{\left( 0\right) }\left( t\right) \mathbf{\Delta }\left(
t\right) +i\mathbf{\Upsilon }\left( t\right) \cdot \mathbf{b}^{\left(
z\right) }\left( t\right) +b^{\left( z\right) }\left( t\right) =0,\ \ \
\alpha \left( \tau \right) =0, \\ 
\\ 
i\mathbf{\Upsilon }_{i}^{\prime }\left( t\right) +\frac{1}{2}\mathbf{\Delta }%
\left( t\right) \cdot \mathfrak{A}^{\left( i\right) }\mathbf{\Delta }\left(
t\right) +\sum\limits_{i=1}^{I}i\mathfrak{B}_{ij}^{\left( zz\right) }\mathbf{%
\Upsilon }_{j}\left( t\right) =0,\ \ \ \mathbf{\Upsilon }_{i}\left( \tau
\right) =m_{i}.%
\end{array}
\label{Eq6.10}
\end{equation}%
In the case in question, the equation for $\mathbf{\Upsilon }$ is no longer
linear. Instead, $\mathbf{\Upsilon }$ satisfies the so-called matrix Riccati
equation. Such equations are important for several applications, such as
optimal control. Solving a matrix Riccati equation is quite hard, so it is
more an art than a science. We shall report some of the results in this
direction below. However, in the one-dimensional case, the corresponding
Riccati equation can be converted into the second-order ODE, which can be
solved explicitly when the coefficients $\mathfrak{A}$, $\mathbf{b}$, $b$
are time-independent.

\subsection{Example: Feller Process\label{Sec62}}

\subsubsection{Feller Process}

For benchmarking purposes, we start with deriving the with the well-known
t.p.d.f. for the Feller process; see \cite{Feller51, Feller52}:%
\begin{equation}
\begin{array}{c}
dy=\left( \chi -\kappa y\right) dt+\varepsilon \sqrt{y}dZ,\ \ \ y\left( \tau
\right) =\theta .%
\end{array}
\label{Eq6.11}
\end{equation}%
To guarantee that the process does not hit zero, we assume that%
\begin{equation}
\begin{array}{c}
\vartheta =\frac{2\chi }{\varepsilon ^{2}}-1>0.%
\end{array}
\label{Eq6.12}
\end{equation}%
The corresponding Fokker-Planck problem has the form:

\begin{equation}
\begin{array}{c}
P_{t}-\frac{1}{2}\varepsilon ^{2}yP_{yy}+\left( \chi -\varepsilon
^{2}-\kappa y\right) P_{y}-\kappa P=0, \\ 
\\ 
P\left( \tau ,\theta ,t,y\right) =\delta \left( y-\theta \right) .%
\end{array}
\label{Eq6.13}
\end{equation}%
The associated function $H\left( \tau ,\theta ,t,y,l\right) $ has the form:%
\begin{equation}
\begin{array}{c}
H=\exp \left( \alpha \left( t\right) +i\Delta \left( t\right) y-il\theta
\right) ,%
\end{array}
\label{Eq6.14}
\end{equation}%
so that%
\begin{equation}
\begin{array}{c}
\alpha ^{\prime }+i\left( \chi -\varepsilon ^{2}\right) \Delta -\kappa =0,\
\ \ \alpha \left( \tau \right) =0, \\ 
\\ 
\ \ i\Delta ^{\prime }+\frac{1}{2}\varepsilon ^{2}\Delta ^{2}-i\kappa \Delta
=0,\ \ \ \Delta \left( \tau \right) =l.%
\end{array}
\label{Eq6.15}
\end{equation}%
Thus, $\Delta $ solves a nonlinear Riccati equation, which can be linearized
via a familiar substitution%
\begin{equation}
\begin{array}{c}
\Delta =\frac{2i\Omega ^{\prime }}{\varepsilon ^{2}\Omega }.%
\end{array}
\label{Eq6.16}
\end{equation}%
As a result, we get:%
\begin{equation}
\begin{array}{c}
\Omega ^{\prime \prime }-\kappa \Omega ^{\prime }=0,\ \ \ \Omega \left( \tau
\right) =1,\ \ \ \Omega ^{\prime }\left( \tau \right) =-\frac{i\varepsilon
^{2}l}{2},\ \ 
\end{array}
\label{Eq6.17}
\end{equation}%
\begin{equation}
\begin{array}{c}
\alpha ^{\prime }-2\left( \frac{\chi }{\varepsilon ^{2}}-1\right) \left( \ln
\left( \Omega \right) \right) ^{\prime }-\kappa =0,\ \ \ \alpha \left( \tau
\right) =0.%
\end{array}
\label{Eq6.18}
\end{equation}%
Accordingly,%
\begin{equation}
\begin{array}{c}
\Omega =\frac{2\kappa -i\varepsilon ^{2}\left( e^{\kappa \left( t-\tau
\right) }-1\right) l}{2\kappa },%
\end{array}
\label{Eq6.19}
\end{equation}%
\begin{equation}
\begin{array}{c}
\Omega ^{\prime }=-\frac{i\varepsilon ^{2}e^{\kappa \left( t-\tau \right) }l%
}{2},%
\end{array}
\label{Eq6.19a}
\end{equation}%
\begin{equation}
\begin{array}{c}
\Delta =\frac{2i\Omega ^{\prime }}{\varepsilon ^{2}\Omega }=\frac{2\kappa
le^{-\kappa \left( t-\tau \right) }}{2\kappa e^{-\kappa \left( t-\tau
\right) }-i\varepsilon ^{2}\left( 1-e^{-\kappa \left( t-\tau \right)
}\right) l},%
\end{array}
\label{Eq6.20}
\end{equation}%
\begin{equation}
\begin{array}{c}
\alpha =\kappa \left( t-\tau \right) +2\left( \frac{\chi }{\varepsilon ^{2}}%
-1\right) \ln \left( \frac{2\kappa -i\varepsilon ^{2}l\left( e^{\kappa
\left( t-\tau \right) }-1\right) }{2\kappa }\right) ,%
\end{array}
\label{Eq6.21}
\end{equation}%
and%
\begin{equation}
\begin{array}{c}
H=\exp \left( \kappa \left( t-\tau \right) +2\left( \frac{\chi }{\varepsilon
^{2}}-1\right) \ln \left( \frac{2\kappa -i\varepsilon ^{2}l\left( e^{\kappa
\left( t-\tau \right) }-1\right) }{2\kappa }\right) \right. \\ 
\\ 
\left. +il\left( \frac{2\kappa }{2\kappa e^{-\kappa \left( t-\tau \right)
}-i\varepsilon ^{2}l\left( 1-e^{-\kappa \left( t-\tau \right) }\right) }%
y-\theta \right) \right) .%
\end{array}
\label{Eq6.22}
\end{equation}%
New variable, $l\rightarrow \eta $:%
\begin{equation}
\begin{array}{c}
\eta =\frac{\varepsilon ^{2}\left( 1-e^{-\kappa \left( \tau -\tau \right)
}\right) }{4\kappa e^{-\kappa \left( t-\tau \right) }}l,\ \ \ l=\frac{%
4\kappa e^{-\kappa \left( t-\tau \right) }}{\varepsilon ^{2}\left(
1-e^{-\kappa \left( \tau -\tau \right) }\right) }\eta .%
\end{array}
\label{Eq6.23}
\end{equation}%
and rescaled $H$, $Hdl\rightarrow \tilde{H}d\eta $:%
\begin{equation}
\begin{array}{c}
\tilde{H}=2M\exp \left( 2\left( \frac{\chi }{\varepsilon ^{2}}-1\right) \ln
\left( 1-2i\eta \right) +2Mi\eta \left( \frac{y}{1-2i\eta }-\bar{\theta}%
\right) \right) ,%
\end{array}
\label{Eq6.24}
\end{equation}%
where $M$ appears due to the change of variables, and $N$ is a prefactor:%
\begin{equation}
\begin{array}{c}
M=\frac{2\kappa }{\varepsilon ^{2}\left( 1-e^{-\kappa \left( \tau -\tau
\right) }\right) }, \\ 
\\ 
N=\frac{4\kappa e^{-\kappa \left( t-\tau \right) }}{\varepsilon ^{2}\left(
1-e^{-\kappa \left( \tau -\tau \right) }\right) }=2MZ, \\ 
\\ 
Z=e^{-\kappa \left( t-\tau \right) }, \\ 
\\ 
\bar{\theta}=e^{-\kappa \left( t-\tau \right) }\theta =Z\theta .%
\end{array}
\label{Eq6.25}
\end{equation}%
Finally,%
\begin{equation}
\begin{array}{c}
Q\left( \tau ,\theta ,t,y\right) =\frac{2M}{2\pi }\int_{-\infty }^{\infty
}\exp \left( 2\left( \frac{\chi }{\varepsilon ^{2}}-1\right) \ln \left(
1-2i\eta \right) +2Mi\eta \left( \frac{y}{1-2i\eta }-\bar{\theta}\right)
\right) d\eta ,%
\end{array}
\label{Eq6.26}
\end{equation}%
A well-known formula yields%
\begin{equation}
\begin{array}{c}
Q\left( \tau ,\theta ,t,y\right) =Me^{-M\left( y+\bar{\theta}\right) }\left( 
\frac{\bar{\theta}}{y}\right) ^{-\frac{\vartheta }{2}}I_{-\vartheta }\left(
2M\sqrt{y\bar{\theta}}\right) \\ 
\\ 
=Me^{-M\left( y+\bar{\theta}\right) }\left( \frac{y}{\bar{\theta}}\right) ^{%
\frac{\vartheta }{2}}I_{\vartheta }\left( 2M\sqrt{y\bar{\theta}}\right) ,%
\end{array}
\label{Eq6.27}
\end{equation}%
see, e.g., \cite{Lipton01} and references therein. The density $Q\left(
y\right) $ integrates to one:%
\begin{equation}
\begin{array}{c}
\int_{0}^{\infty }Q\left( \tau ,\theta ,t,y\right) dy=\int_{0}^{\infty
}e^{-u-v}\left( \frac{v}{u}\right) ^{\vartheta /2}I_{\vartheta }\left( 2%
\sqrt{uv}\right) dv=1,%
\end{array}
\label{Eq6.28}
\end{equation}%
where $u=M\bar{\theta},\ \ \ v=My$. We shall use this fact below.

This derivation can be extended to cover augmented Feller processes, as we
shall do next. It is worth noting that studying the Feller process with
time-dependent coefficients is rather tricky and remains to be wholly done;
see, however, \cite{Masoliver16, Giorno21} and references therein.

\subsubsection{Degenerate Augmented Feller Process}

Equations%
\begin{equation}
\begin{array}{c}
dx=ydt,\ \ \ x_{\tau }=\xi , \\ 
\\ 
dy=\left( \chi -\kappa y\right) dt+\varepsilon \sqrt{y}dZ,\ \ \ y_{\tau
}=\theta .%
\end{array}
\label{Eq6.29}
\end{equation}%
The Fokker-Planck equation:%
\begin{equation}
\begin{array}{c}
0=P_{t}-\frac{1}{2}\varepsilon ^{2}\left( yP\right) _{yy}+yP_{x}+\left(
\left( \chi -\kappa y\right) P\right) _{y} \\ 
\\ 
=P_{t}-\frac{1}{2}\varepsilon ^{2}yP_{yy}-\varepsilon
^{2}P_{y}+yP_{x}+\left( \chi -\kappa y\right) P_{y}-\kappa P \\ 
\\ 
=P_{t}-\frac{1}{2}\varepsilon ^{2}yP_{yy}+yP_{x}+\left( \chi -\varepsilon
^{2}-\kappa y\right) P_{y}-\kappa P.%
\end{array}
\label{Eq6.30}
\end{equation}%
The corresponding function $H$ has the form:%
\begin{equation}
\begin{array}{c}
H=\exp \left( \alpha \left( t\right) +ik\left( x-\xi \right) +i\Delta \left(
t\right) y-il\theta \right) ,%
\end{array}
\label{Eq6.31}
\end{equation}%
where%
\begin{equation}
\begin{array}{c}
\alpha ^{\prime }+i\Delta ^{\prime }y+\frac{1}{2}\varepsilon ^{2}\Delta
^{2}y+iky+i\Delta \left( \chi -\varepsilon ^{2}-\kappa y\right) -\kappa =0,
\\ 
\\ 
\alpha \left( \tau \right) =0,\ \ \ \Delta \left( \tau \right) =l.%
\end{array}
\label{Eq6.32}
\end{equation}%
Thus%
\begin{equation}
\begin{array}{c}
\alpha ^{\prime }+i\left( \chi -\varepsilon ^{2}\right) \Delta -\kappa =0,\
\ \ \alpha \left( \tau \right) =0, \\ 
\\ 
i\Delta ^{\prime }+\frac{1}{2}\varepsilon ^{2}\Delta ^{2}-i\kappa \Delta
+ik=0,\ \ \ \Delta \left( \tau \right) =l,%
\end{array}
\label{Eq6.33}
\end{equation}%
As before, we can linearize the Riccati equation for $\Delta $ via a
two-step linearization 
\begin{equation}
\begin{array}{c}
\Delta =\frac{2i\Lambda }{\varepsilon ^{2}},\ \ \ \Lambda =\frac{\Omega
^{\prime }}{\Omega },%
\end{array}
\label{Eq6.34}
\end{equation}%
which yields%
\begin{equation}
\begin{array}{c}
\alpha ^{\prime }-2\left( \frac{\chi }{\varepsilon ^{2}}-1\right) \left( \ln
\left( \Omega \right) \right) ^{\prime }-\kappa =0,\ \ \ \alpha \left( \tau
\right) =0, \\ 
\\ 
\Omega ^{\prime \prime }-\kappa \Omega ^{\prime }-\frac{i\varepsilon ^{2}k}{2%
}\Omega =0,\ \ \ \Omega \left( \tau \right) =1,\ \ \ \Omega ^{\prime }\left(
\tau \right) =-\frac{i\varepsilon ^{2}l}{2}.%
\end{array}
\label{Eq6.35}
\end{equation}%
We can represent $\Omega $ in the form:%
\begin{equation}
\begin{array}{c}
\Omega =c_{+}e^{\lambda _{+}\left( t-\tau \right) }+c_{-}e^{\lambda
_{+}\left( t-\tau \right) },%
\end{array}
\label{Eq6.36}
\end{equation}%
where $\lambda _{\pm }$ are solutions of the characteristic equation:%
\begin{equation}
\begin{array}{c}
\lambda ^{2}-\kappa \lambda -\frac{i\varepsilon ^{2}k}{2}=0, \\ 
\\ 
\lambda _{\pm }=\mu \pm \zeta , \\ 
\\ 
\mu =\frac{\kappa }{2},\ \ \ \zeta =\frac{\sqrt{\kappa ^{2}+2i\varepsilon
^{2}k}}{2},%
\end{array}
\label{Eq6.37}
\end{equation}%
and $c_{\pm }$ satisfy the following system of linear equations:%
\begin{equation}
\begin{array}{c}
c_{+}+c_{-}=1, \\ 
\\ 
\lambda _{+}c_{+}+\lambda _{-}c_{-}=-\frac{i\varepsilon ^{2}l}{2}.%
\end{array}
\label{Eq6.38}
\end{equation}%
so that%
\begin{equation}
\begin{array}{c}
c_{+}=-\frac{\left( 2\lambda _{-}+i\varepsilon ^{2}l\right) }{4\zeta },\ \ \
c_{-}=\frac{\left( 2\lambda _{+}+i\varepsilon ^{2}l\right) }{4\zeta }.%
\end{array}
\label{Eq6.39}
\end{equation}%
It is useful to note that%
\begin{equation}
\begin{array}{c}
\lambda _{+}\lambda _{-}=\mu ^{2}-\zeta ^{2}=-\frac{i\varepsilon ^{2}k}{2}.%
\end{array}
\label{Eq6.40}
\end{equation}%
For the sake of clarity, we introduce the following notation:%
\begin{equation}
\begin{array}{c}
\mathsf{E}_{0}=e^{\mu \left( t-\tau \right) },\ \ \ \mathsf{E}_{\pm }=e^{\pm
\zeta \left( t-\tau \right) }.%
\end{array}
\label{Eq6.41}
\end{equation}%
Thus,%
\begin{equation}
\begin{array}{c}
\Omega =\mathsf{E}_{0}\left( \frac{-\left( 2\lambda _{-}+i\varepsilon
^{2}l\right) \mathsf{E}_{+}+\left( 2\lambda _{+}+i\varepsilon ^{2}l\right) 
\mathsf{E}_{-}}{4\zeta }\right) ,%
\end{array}
\label{Eq6.42}
\end{equation}%
\begin{equation}
\begin{array}{c}
\Omega ^{\prime }=\mathsf{E}_{0}\left( \frac{-\lambda _{+}\left( 2\lambda
_{-}+i\varepsilon ^{2}l\right) \mathsf{E}_{+}+\lambda _{-}\left( 2\lambda
_{+}+i\varepsilon ^{2}l\right) \mathsf{E}_{-}}{4\zeta }\right) ,%
\end{array}
\label{Eq6.43}
\end{equation}%
\begin{equation}
\begin{array}{c}
\Delta =\frac{2i\left( -\lambda _{+}\left( 2\lambda _{-}+i\varepsilon
^{2}l\right) \mathsf{E}_{+}+\lambda _{-}\left( 2\lambda _{+}+i\varepsilon
^{2}l\right) \mathsf{E}_{-}\right) }{\varepsilon ^{2}\left( -\left( 2\lambda
_{-}+i\varepsilon ^{2}l\right) \mathsf{E}_{+}+\left( 2\lambda
_{+}+i\varepsilon ^{2}l\right) \mathsf{E}_{-}\right) },%
\end{array}
\label{Eq6.44}
\end{equation}%
\begin{equation}
\begin{array}{c}
\alpha =2\left( \frac{\chi }{\varepsilon ^{2}}-1\right) \ln \left( \frac{%
-\left( 2\lambda _{-}+i\varepsilon ^{2}l\right) \mathsf{E}_{+}+\left(
2\lambda _{+}+i\varepsilon ^{2}l\right) \mathsf{E}_{-}}{4\zeta }\right) +%
\frac{2\chi \mu \left( t-\tau \right) }{\varepsilon ^{2}}.%
\end{array}
\label{Eq6.45}
\end{equation}%
Accordingly, $H$ can be written in the form:%
\begin{equation}
\begin{array}{c}
H=\exp \left( \frac{2\chi \mu \left( t-\tau \right) }{\varepsilon ^{2}}%
+ik\left( x-\xi \right) \right. \\ 
\\ 
+2\left( \frac{\chi }{\varepsilon ^{2}}-1\right) \ln \left( \frac{2\left(
-\lambda _{-}\mathsf{E}_{+}+\lambda _{+}\mathsf{E}_{-}\right) -i\varepsilon
^{2}l\left( \mathsf{E}_{+}-\mathsf{E}_{-}\right) }{4\zeta }\right) \\ 
\\ 
\left. +\frac{2\left( 2\lambda _{+}\lambda _{-}\left( \mathsf{E}_{+}-\mathsf{%
E}_{-}\right) +i\varepsilon ^{2}l\left( \lambda _{+}\mathsf{E}_{+}-\lambda
_{-}\mathsf{E}_{-}\right) \right) y}{\varepsilon ^{2}\left( 2\left( -\lambda
_{-}\mathsf{E}_{+}+\lambda _{+}\mathsf{E}_{-}\right) -i\varepsilon
^{2}l\left( \mathsf{E}_{+}-\mathsf{E}_{-}\right) \right) }-il\theta \right) .%
\end{array}
\label{Eq6.46}
\end{equation}

Change of variables:%
\begin{equation}
\begin{array}{c}
\eta =\frac{\varepsilon ^{2}\left( \mathsf{E}_{+}-\mathsf{E}_{-}\right) l}{%
4\left( -\lambda _{-}\mathsf{E}_{+}+\lambda _{+}\mathsf{E}_{-}\right) }, \\ 
\\ 
l=\frac{4\left( -\lambda _{-}\mathsf{E}_{+}+\lambda _{+}\mathsf{E}%
_{-}\right) \eta }{\varepsilon ^{2}\left( \mathsf{E}_{+}-\mathsf{E}%
_{-}\right) }\equiv N\eta .%
\end{array}
\label{Eq6.47}
\end{equation}%
allows one to factorize $H$ as follows:%
\begin{equation}
\begin{array}{c}
\tilde{H}=\tilde{H}_{1}\tilde{H}_{2},%
\end{array}
\label{Eq6.48}
\end{equation}%
where%
\begin{equation}
\begin{array}{c}
\tilde{H}_{1}=\exp \left( \frac{2\chi \mu \left( t-\tau \right) }{%
\varepsilon ^{2}}+ik\left( x-\xi \right) +\frac{2\chi }{\varepsilon ^{2}}\ln
\left( \frac{\left( -\lambda _{-}\mathsf{E}_{+}+\lambda _{+}\mathsf{E}%
_{-}\right) }{2\zeta }\right) \right. \\ 
\\ 
\left. +\frac{2\lambda _{+}\lambda _{-}\left( \mathsf{E}_{+}-\mathsf{E}%
_{-}\right) y}{\varepsilon ^{2}\left( -\lambda _{-}\mathsf{E}_{+}+\lambda
_{+}\mathsf{E}_{-}\right) }\right) , \\ 
\\ 
\tilde{H}_{2}=2M\exp \left( 2\left( \frac{\chi }{\varepsilon ^{2}}-1\right)
\ln \left( 1-2i\eta \right) +2Mi\eta \left( \frac{y}{1-2i\eta }-\bar{\theta}%
\right) \right) .%
\end{array}
\label{Eq6.49}
\end{equation}%
and%
\begin{equation}
\begin{array}{c}
M=\frac{8\zeta ^{2}}{\varepsilon ^{2}\left( \mathsf{E}_{+}-\mathsf{E}%
_{-}\right) \left( -\lambda _{-}\mathsf{E}_{+}+\lambda _{+}\mathsf{E}%
_{-}\right) }, \\ 
\\ 
Z=\frac{\left( -\lambda _{-}\mathsf{E}_{+}+\lambda _{+}\mathsf{E}_{-}\right)
^{2}}{4\zeta ^{2}}=\frac{N}{2M}, \\ 
\\ 
\bar{\theta}=Z\theta .%
\end{array}
\label{Eq6.50}
\end{equation}%
Integration with respect to $\eta $ can be done analytically%
\begin{equation}
\begin{array}{c}
Q_{2}\left( \tau ,\theta ,t,y\right) =\frac{2M}{2\pi }\int_{-\infty
}^{\infty }\tilde{H}_{2}d\eta =Me^{-M\left( y+\bar{\theta}\right) }\left( 
\frac{y}{\bar{\theta}}\right) ^{\frac{\vartheta }{2}}I_{\vartheta }\left( 2M%
\sqrt{y\bar{\theta}}\right) .%
\end{array}
\label{Eq6.51}
\end{equation}%
which allows us to calculate $P$ via a single inverse Fourier transform: 
\begin{equation}
\begin{array}{c}
P=\frac{1}{2\pi }\int_{-\infty }^{\infty }\exp \left( \frac{2\chi \mu \left(
t-\tau \right) }{\varepsilon ^{2}}+ik\left( x-\xi \right) \right. \\ 
\\ 
\left. +\frac{2\chi }{\varepsilon ^{2}}\ln \left( \frac{\left( -\lambda _{-}%
\mathsf{E}_{+}+\lambda _{+}\mathsf{E}_{-}\right) }{2\zeta }\right) -Ry\right)
\\ 
\\ 
\times Me^{-M\left( y+\bar{\theta}\right) }\left( \frac{y}{\bar{\theta}}%
\right) ^{\frac{\vartheta }{2}}I_{\vartheta }\left( 2M\sqrt{y\bar{\theta}}%
\right) dk \\ 
\\ 
=\frac{1}{2\pi }\int_{-\infty }^{\infty }\exp \left( \frac{2\chi \mu \left(
t-\tau \right) }{\varepsilon ^{2}}+ik\left( x-\xi \right) \right. \\ 
\\ 
\left. +\frac{2\chi }{\varepsilon ^{2}}\ln \left( \frac{\left( -\lambda _{-}%
\mathsf{E}_{+}+\lambda _{+}\mathsf{E}_{-}\right) }{2\zeta }\right) \right)
\\ 
\\ 
\times Me^{-\left( M+R\right) y-M\bar{\theta}}\left( \frac{y}{\bar{\theta}}%
\right) ^{\frac{\vartheta }{2}}I_{\vartheta }\left( 2M\sqrt{y\bar{\theta}}%
\right) dk%
\end{array}
\label{Eq6.52}
\end{equation}%
where%
\begin{equation}
\begin{array}{c}
R=-\frac{2\lambda _{+}\lambda _{-}\left( \mathsf{E}_{+}-\mathsf{E}%
_{-}\right) }{\varepsilon ^{2}\left( -\lambda _{-}\mathsf{E}_{+}+\lambda _{+}%
\mathsf{E}_{-}\right) }.%
\end{array}
\label{Eq6.53}
\end{equation}

This expression is not as convenient to deal with due to the presence of the
term $\exp \left( -Ry\right) $. We wish to rewrite it in a form better
suited for further calculations. Further transformations to use the density
and transform extra dependence on $y$ to dependence on $\theta$. We proceed
as follows.\ Define%
\begin{equation}
\begin{array}{c}
Z_{1}=\left( \frac{M}{M+R}\right) ^{2}Z.%
\end{array}
\label{Eq6.54}
\end{equation}%
Then%
\begin{equation}
\begin{array}{c}
Me^{-\left( M+R\right) y-MZ\theta }\left( \frac{y}{Z\theta }\right) ^{\frac{%
\vartheta }{2}}I_{\vartheta }\left( 2M\sqrt{yZ\theta }\right) \\ 
\\ 
=\frac{M}{\left( M+R\right) }\left( M+R\right) e^{-\left( M+R\right) y-\frac{%
\left( M+R\right) ^{2}}{M}Z_{1}\theta }\left( \frac{M}{M+R}\right)
^{\vartheta } \\ 
\\ 
\times \left( \frac{y}{Z_{1}\theta }\right) ^{\frac{\vartheta }{2}%
}I_{\vartheta }\left( 2\left( M+R\right) \sqrt{yZ_{1}\theta }\right) \\ 
\\ 
=\left( \frac{M}{M+R}\right) ^{\vartheta +1}e^{-\left( \frac{\left(
M+R\right) ^{2}}{M}-\left( M+R\right) \right) Z_{1}\theta }\left( M+R\right)
e^{-\left( M+R\right) \left( y+Z_{1}\theta \right) } \\ 
\\ 
\times \left( \frac{y}{Z_{1}\theta }\right) ^{\frac{\vartheta }{2}%
}I_{\vartheta }\left( 2\left( M+R\right) \sqrt{yZ_{1}\theta }\right) \\ 
\\ 
=\left( \frac{M}{M+R}\right) ^{\vartheta +1}e^{-\frac{MRZ}{\left( M+R\right) 
}\theta }\left( M+R\right) e^{-\left( M+R\right) \left( y+Z_{1}\theta
\right) } \\ 
\\ 
\times \left( \frac{y}{Z_{1}\theta }\right) ^{\frac{\vartheta }{2}%
}I_{\vartheta }\left( 2\left( M+R\right) \sqrt{yZ_{1}\theta }\right) ,%
\end{array}
\label{Eq6.55}
\end{equation}%
so that%
\begin{equation}
\begin{array}{c}
P\left( t,x,y\right) =\frac{1}{2\pi }\int_{-\infty }^{\infty }\exp \left( 
\frac{2\chi \mu \left( t-\tau \right) }{\varepsilon ^{2}}+ik\left( x-\xi
\right) \right. \\ 
\\ 
\left. +\frac{2\chi }{\varepsilon ^{2}}\ln \left( \frac{M\left( -\lambda _{-}%
\mathsf{E}_{+}+\lambda _{+}\mathsf{E}_{-}\right) }{2\zeta \left( M+R\right) }%
\right) -\frac{MRZ\theta }{M+R}\right) \\ 
\\ 
\times \left( M+R\right) e^{-\left( M+R\right) \left( y+Z_{1}\theta \right)
}\left( \frac{y}{Z_{1}\theta }\right) ^{\frac{\vartheta }{2}}I_{\vartheta
}\left( 2\left( M+R\right) \sqrt{yZ_{1}\theta }\right) dk.%
\end{array}
\label{Eq6.56}
\end{equation}%
Explicitly,%
\begin{equation}
\begin{array}{c}
M+R=\frac{2\left( \left( \lambda _{+}^{2}+\lambda _{-}^{2}\right) -\lambda
_{+}\lambda _{-}\left( \mathsf{E}_{+}^{2}+\mathsf{E}_{-}^{2}\right) \right) 
}{\varepsilon ^{2}\left( \mathsf{E}_{+}-\mathsf{E}_{-}\right) \left(
-\lambda _{-}\mathsf{E}_{+}+\lambda _{+}\mathsf{E}_{-}\right) }, \\ 
\\ 
\frac{M}{M+R}=\frac{\left( \lambda _{+}-\lambda _{-}\right) ^{2}}{\left(
\lambda _{+}^{2}+\lambda _{-}^{2}\right) -\lambda _{+}\lambda _{-}\left( 
\mathsf{E}_{+}^{2}+\mathsf{E}_{-}^{2}\right) }, \\ 
\\ 
\frac{MRZ}{M+R}=-\frac{2\lambda _{+}\lambda _{-}\left( \mathsf{E}_{+}-%
\mathsf{E}_{-}\right) }{\varepsilon ^{2}\left( \lambda _{+}\mathsf{E}%
_{+}-\lambda _{-}\mathsf{E}_{-}\right) },%
\end{array}
\label{Eq6.57}
\end{equation}%
so that%
\begin{equation}
\begin{array}{c}
P\left( t,x,y\right) =\frac{1}{2\pi }\int_{-\infty }^{\infty }\exp \left( 
\frac{2\chi \mu \left( t-\tau \right) }{\varepsilon ^{2}}+ik\left( x-\xi
\right) \right. \\ 
\\ 
\left. +\frac{2\chi }{\varepsilon ^{2}}\ln \left( \frac{2\zeta }{\lambda _{+}%
\mathsf{E}_{+}-\lambda _{-}\mathsf{E}_{-}}\right) +\frac{2\lambda
_{+}\lambda _{-}\left( \mathsf{E}_{+}-\mathsf{E}_{-}\right) \theta }{%
\varepsilon ^{2}\left( \lambda _{+}\mathsf{E}_{+}-\lambda _{-}\mathsf{E}%
_{-}\right) }\right) \\ 
\\ 
\times \left( M+R\right) e^{-\left( M+R\right) \left( y+Z_{1}\theta \right)
}\left( \frac{y}{Z_{1}\theta }\right) ^{\frac{\vartheta }{2}}I_{\vartheta
}\left( 2\left( M+R\right) \sqrt{yZ_{1}\theta }\right) dk.%
\end{array}
\label{Eq6.58}
\end{equation}%
Since the integral over $y$ is equal to one, we can represent the marginal
distribution of $x$ in the form:%
\begin{equation}
\begin{array}{c}
Q^{\left( x\right) }\left( t,x\right) =\frac{1}{2\pi }\int_{-\infty
}^{\infty }\digamma \left( t,k\right) e^{ik\left( x-\xi \right) }dk,%
\end{array}
\label{Eq6.59}
\end{equation}%
where 
\begin{equation}
\begin{array}{c}
\digamma \left( t,k\right) =\exp \left( \frac{2\chi \mu \left( t-\tau
\right) }{\varepsilon ^{2}}+\frac{2\chi }{\varepsilon ^{2}}\ln \left( \frac{%
2\zeta }{\lambda _{+}\mathsf{E}_{+}-\lambda _{-}\mathsf{E}_{-}}\right) +%
\frac{2\lambda _{+}\lambda _{-}\left( \mathsf{E}_{+}-\mathsf{E}_{-}\right)
\theta }{\varepsilon ^{2}\left( \lambda _{+}\mathsf{E}_{+}-\lambda _{-}%
\mathsf{E}_{-}\right) }\right) .%
\end{array}
\label{Eq6.60}
\end{equation}%
Here $\mu ,\zeta $ are given by Eqs. (\ref{Eq6.37}). It is easy to check
that $Q^{\left( x\right) }\left( t,x\right) $ integrates to one:%
\begin{equation}
\begin{array}{c}
\int_{-\infty }^{\infty }Q\left( t,x\right) dx=\frac{1}{2\pi }\int_{-\infty
}^{\infty }\int_{-\infty }^{\infty }\digamma \left( t,k\right) \exp \left(
ik\left( x-\xi \right) \right) dkdx \\ 
\\ 
=\int_{-\infty }^{\infty }\digamma \left( t,k\right) \delta \left( k\right)
dk=\digamma \left( t,0\right) \\ 
\\ 
=\exp \left( \frac{2\chi \lambda _{-}\left( t-\tau \right) }{\varepsilon ^{2}%
}\right) =1.%
\end{array}
\label{Eq6.61}
\end{equation}%
The expected value of $x$ has the form:%
\begin{equation}
\begin{array}{c}
\bar{x}=\int_{-\infty }^{\infty }Q\left( t,x\right) xdx=\xi +\int_{-\infty
}^{\infty }Q\left( t,x\right) \left( x-\xi \right) dx \\ 
\\ 
=\xi +\frac{1}{2\pi }\int_{-\infty }^{\infty }\int_{-\infty }^{\infty
}\digamma \left( t,k\right) \exp \left( ik\left( x-\xi \right) \right)
\left( x-\xi \right) dkdx \\ 
\\ 
=\xi +\underset{\epsilon \rightarrow 0}{\lim }\left( \frac{d}{d\epsilon }%
\left( \frac{1}{2\pi }\int_{-\infty }^{\infty }\int_{-\infty }^{\infty
}\digamma \left( t,k\right) \exp \left( \left( ik+\epsilon \right) \left(
x-\xi \right) \right) dkdx\right) \right) \\ 
\\ 
=\xi +\underset{\epsilon \rightarrow 0}{\lim }\left( \frac{d}{d\epsilon }%
\left( \int_{-\infty }^{\infty }\digamma \left( t,k\right) \delta \left(
k-i\epsilon \right) dk\right) \right) \\ 
\\ 
=\xi +\left. \frac{d}{d\epsilon }\digamma \left( t,i\epsilon \right)
\right\vert _{\epsilon =0}.%
\end{array}
\label{Eq6.62}
\end{equation}%
A simple calculation left to the reader yields%
\begin{equation}
\begin{array}{c}
\bar{x}=\xi +\frac{\left( \kappa \left( t-\tau \right) -\left( 1-e^{-\kappa
\left( t-\tau \right) }\right) \right) \chi }{\kappa ^{2}}+\frac{\left(
1-e^{-\kappa \left( t-\tau \right) }\right) \theta }{\kappa },%
\end{array}
\label{Eq6.63}
\end{equation}%
which agrees with Eq. (\ref{Eq5.52}).

It is worth noting that $Q^{\left( x\right) }\left( t,x\right) $ has fat
tails, since some of the exponential moments of $x$ have finite-time
explosions; see \cite{Andersen07, Friz10} and references therein.\footnote{%
Of course, it is not surprising that such explosions exist since Riccati
equations are well-known to have solutions exploding in finite time.
Consider the following Riccati initial-value problem%
\begin{equation*}
\begin{array}{c}
f^{\prime }+af^{2}+bf+c=0,\ \ \ f\left( \tau \right) =d,%
\end{array}%
\end{equation*}%
and assume that $b^{2}-4ac<0$. The solution of this initial-value problem
has the form%
\begin{equation*}
\begin{array}{c}
f\left( \tau ,t\right) =-\frac{\left\vert \zeta \right\vert }{a}\tan \left(
\left\vert \zeta \right\vert \left( t-\tau \right) -\arctan \left( \frac{%
2ad+b}{2\left\vert \zeta \right\vert }\right) \right) -\frac{b}{2a},%
\end{array}%
\end{equation*}%
where $\left\vert \zeta \right\vert =\left. \sqrt{4ac-b^{2}}\right/ 2$. The
corresponding blow-up time $t^{\ast }$ has the form%
\begin{equation*}
\begin{array}{c}
t^{\ast }=\tau +\frac{\pi }{2\left\vert \zeta \right\vert }+\frac{1}{%
\left\vert \zeta \right\vert }\arctan \left( \frac{2ad+b}{2\left\vert \zeta
\right\vert }\right) .%
\end{array}%
\end{equation*}%
} Specifically, we wish to study if $\mathcal{I}_{p}\left( \tau ,t\right) $
of the form: \ 
\begin{equation}
\begin{array}{c}
\mathcal{I}_{p}\left( \tau ,t\right) =\int_{-\infty }^{\infty }Q\left( \tau
,\xi ,t,x\right) e^{px}dx \\ 
\\ 
=\frac{e^{p\xi }}{2\pi }\int_{-\infty }^{\infty }\int_{-\infty }^{\infty
}\digamma \left( t,k\right) \exp \left( ik\left( x-\xi \right) +p\left(
x-\xi \right) \right) dkdx \\ 
\\ 
=e^{p\xi }\int_{-\infty }^{\infty }\digamma \left( t,k\right) \delta \left(
ik+p\right) dk=e^{p\xi }\digamma \left( t,ip\right) ,%
\end{array}
\label{Eq6.64}
\end{equation}%
blows up for some finite $t>\tau $. We have%
\begin{equation}
\begin{array}{c}
\digamma \left( t,ip\right) =\exp \left( \frac{2\chi \mu \left( t-\tau
\right) }{\varepsilon ^{2}}+\frac{2\chi }{\varepsilon ^{2}}\ln \left( \frac{%
2\zeta }{\lambda _{+}\mathsf{E}_{+}-\lambda _{-}\mathsf{E}_{-}}\right)
\right. \\ 
\\ 
\left. +\frac{2\lambda _{+}\lambda _{-}\left( \mathsf{E}_{+}-\mathsf{E}%
_{-}\right) \theta }{\varepsilon ^{2}\left( \lambda _{+}\mathsf{E}%
_{+}-\lambda _{-}\mathsf{E}_{-}\right) }\right) ,%
\end{array}
\label{Eq6.65}
\end{equation}%
where%
\begin{equation}
\begin{array}{c}
\mu =\frac{\kappa }{2},\ \ \ \zeta =\frac{\sqrt{\kappa ^{2}-2\varepsilon
^{2}p}}{2},\ \ \ \lambda _{+}\lambda _{-}=\mu ^{2}-\zeta ^{2}=\frac{%
\varepsilon ^{2}p}{2},%
\end{array}
\label{Eq6.66}
\end{equation}%
so that%
\begin{equation}
\begin{array}{c}
\mathcal{I}_{p}=\left( \frac{2\zeta }{\lambda _{+}\mathsf{E}_{+}-\lambda _{-}%
\mathsf{E}_{-}}\right) ^{\frac{2\chi }{\varepsilon ^{2}}} \\ 
\\ 
\times \exp \left( \frac{2\chi }{\varepsilon ^{2}}\mu \left( t-\tau \right)
+p\xi +\frac{p\left( \mathsf{E}_{+}-\mathsf{E}_{-}\right) \theta }{\left(
\lambda _{+}\mathsf{E}_{+}-\lambda _{-}\mathsf{E}_{-}\right) }\right) ,%
\end{array}
\label{Eq6.67}
\end{equation}%
when $\zeta >0$ is real, and%
\begin{equation}
\begin{array}{c}
\mathcal{I}_{p}=\left( \frac{\left\vert \zeta \right\vert }{\mu \sin \left(
\left\vert \zeta \right\vert \left( t-\tau \right) \right) +\left\vert \zeta
\right\vert \cos \left( \left\vert \zeta \right\vert \left( t-\tau \right)
\right) }\right) ^{\frac{2\chi }{\varepsilon ^{2}}} \\ 
\\ 
\times \exp \left( \frac{2\chi }{\varepsilon ^{2}}\mu \left( t-\tau \right)
+p\xi +\frac{p\sin \left( \left\vert \zeta \right\vert \left( t-\tau \right)
\right) \theta }{\left( \mu \sin \left( \left\vert \zeta \right\vert \left(
t-\tau \right) \right) +\left\vert \zeta \right\vert \cos \left( \left\vert
\zeta \right\vert \left( t-\tau \right) \right) \right) }\right) ,%
\end{array}
\label{Eq6.68}
\end{equation}%
when $\zeta =i\left\vert \zeta \right\vert $ is imaginary.

For $p\in \left[ -\infty ,\hat{p}\right] $, $\zeta $ is real, for $p\in %
\left[ \hat{p},\infty \right] $, it is imaginary. Here%
\begin{equation}
\begin{array}{c}
\hat{p}=\frac{\kappa ^{2}}{2\varepsilon ^{2}}>0.%
\end{array}
\label{Eq6.69}
\end{equation}%
There is no blowup when $\zeta $ is real. When $\zeta $ is imaginary, the
blowup time $t^{\ast }$ is the smallest positive root of the equation%
\begin{equation}
\begin{array}{c}
\kappa \sin \left( \sqrt{2\varepsilon ^{2}p-\kappa ^{2}}\left( t^{\ast
}-\tau \right) \right) +\sqrt{2\varepsilon ^{2}p-\kappa ^{2}}\cos \left( 
\sqrt{2\varepsilon ^{2}p-\kappa ^{2}}\left( t^{\ast }-\tau \right) \right)
=0,%
\end{array}
\label{Eq6.70}
\end{equation}%
\begin{equation}
\begin{array}{c}
t^{\ast }=\tau +\frac{\pi -\arctan \left( \frac{\sqrt{2\varepsilon
^{2}p-\kappa ^{2}}}{\kappa }\right) }{\sqrt{2\varepsilon ^{2}p-\kappa ^{2}}}.%
\end{array}
\label{Eq6.71}
\end{equation}%
It is clear that $\mathcal{I}_{-1}$ does not blowup. This fact in used in
the next section.

The marginal distribution of $y$, $Q^{\left( y\right) }\left( t,y\right) $
is the standard Feller distribution given by Eq. (\ref{Eq6.27}).

\subsubsection{Non-Degenerate Augmented Feller Process}

We start with the following affine equations%
\begin{equation}
\begin{array}{c}
dx_{t}=-\frac{1}{2}y_{t}dt+\sqrt{y_{t}}dW_{t},\ \ \ x_{\tau }=\xi , \\ 
\\ 
dy_{t}=\left( \chi -\kappa y_{t}\right) dt+\varepsilon \sqrt{y_{t}}dZ_{t},\
\ \ y_{\tau }=\theta ,%
\end{array}
\label{Eq6.74}
\end{equation}%
extremely important in the context of the Heston model; see \cite{Heston93}.
The specific choice of the drift for $x_{t}$ is explained in Section \ref%
{Sec7}. The corresponding Fokker-Planck equation can be written as follows:%
\begin{equation}
\begin{array}{c}
0=P_{t}-\frac{1}{2}yP_{xx}-\rho \varepsilon \left( yP\right) _{xy}-\frac{1}{2%
}\varepsilon ^{2}\left( yP\right) _{yy}-\frac{1}{2}yP_{x}+\left( \left( \chi
-\kappa y\right) P\right) _{y} \\ 
\\ 
=P_{t}-\frac{1}{2}yP_{xx}-\rho \varepsilon yP_{xy}-\rho \varepsilon P_{x}-%
\frac{1}{2}\varepsilon ^{2}yP_{yy}-\varepsilon ^{2}P_{y}-\frac{1}{2}%
yP_{x}+\left( \chi -\kappa y\right) P_{y}-\kappa P \\ 
\\ 
=P_{t}-\frac{1}{2}yP_{xx}-\rho \varepsilon yP_{xy}-\frac{1}{2}\varepsilon
^{2}yP_{yy}-\left( \frac{1}{2}y+\rho \varepsilon \right) P_{x}+\left( \chi
-\varepsilon ^{2}-\kappa y\right) P_{y}-\kappa P.%
\end{array}
\label{Eq6.75}
\end{equation}%
Function $H$:%
\begin{equation}
\begin{array}{c}
H=\exp \left( \alpha \left( t\right) +ik\left( x-\xi \right) +i\Delta \left(
t\right) y-il\theta \right) .%
\end{array}
\label{Eq6.76}
\end{equation}%
Equations for $\alpha ,\Delta $:%
\begin{equation}
\begin{array}{c}
i\Delta ^{\prime }+\frac{1}{2}\varepsilon ^{2}\Delta ^{2}+\left( \rho
\varepsilon k-i\kappa \right) \Delta +\frac{1}{2}k\left( k-i\right) =0,\ \ \
\Delta \left( \tau \right) =l, \\ 
\\ 
\alpha ^{\prime }+i\left( \chi -\varepsilon ^{2}\right) \Delta -\left( i\rho
\varepsilon k+\kappa \right) =0,\ \ \ \alpha \left( \tau \right) =0.%
\end{array}
\label{Eq6.77}
\end{equation}%
Linearization 
\begin{equation}
\begin{array}{c}
\Delta \rightarrow \Delta =\frac{2i\Lambda }{\varepsilon ^{2}}=\frac{2i}{%
\varepsilon ^{2}}\frac{\Omega ^{\prime }}{\Omega },%
\end{array}
\label{Eq6.78}
\end{equation}%
yields%
\begin{equation}
\begin{array}{c}
\alpha ^{\prime }-2\left( \frac{\chi }{\varepsilon ^{2}}-1\right) \left( \ln
\left( \Omega \right) \right) ^{\prime }-\left( i\rho \varepsilon k+\kappa
\right) =0,\ \ \ \alpha \left( \tau \right) =0, \\ 
\\ 
\Omega ^{\prime \prime }-i\left( \rho \varepsilon k-i\kappa \right) \Omega
^{\prime }-\frac{1}{4}\varepsilon ^{2}k\left( k-i\right) \Omega =0,\ \ \
\Omega \left( \tau \right) =1,\ \ \ \Omega ^{\prime }\left( \tau \right) =-%
\frac{i\varepsilon ^{2}l}{2}.%
\end{array}
\label{Eq6.79}
\end{equation}%
The characteristic equation and its solution%
\begin{equation}
\begin{array}{c}
\lambda ^{2}-\left( \kappa +i\rho \varepsilon k\right) \lambda -\frac{1}{4}%
\varepsilon ^{2}k\left( k-i\right) =0, \\ 
\\ 
\lambda _{\pm }=\mu \pm \zeta , \\ 
\\ 
\mu =\frac{1}{2}\left( \kappa +i\rho \varepsilon k\right) ,\ \ \ \zeta =%
\frac{1}{2}\sqrt{\varepsilon ^{2}\bar{\rho}^{2}k^{2}+i\varepsilon \left(
2\rho \kappa -\varepsilon \right) k+\kappa ^{2}}, \\ 
\\ 
\lambda _{+}\lambda _{-}=\mu ^{2}-\zeta ^{2}=-\frac{1}{4}\varepsilon
^{2}k\left( k-i\right) .%
\end{array}
\label{Eq6.80}
\end{equation}%
where $\bar{\rho}^{2}=1-\rho ^{2}$. Subsequent calculations are very similar
to the ones performed in the previous subsection, so we omit them for
brevity. The final expressions for $P\left( t,x,y\right) $ and $Q\left(
t,x\right) $ are given by Eqs (\ref{Eq6.58}), (\ref{Eq6.59}), with $\mu
,\zeta $ are given by Eqs (\ref{Eq6.80}). These expressions are more
efficient than the original formulae for the t.p.d.f. of the stochastic
volatility model derived by \cite{Lipton01}.

As before, $Q\left( t,x\right) $ has fat tails. We wish to study the
behavior of $\mathcal{I}_{p}\left( \tau ,t\right) $ given by Eq. (\ref%
{Eq6.64}). The corresponding $\mu ,\zeta $ have the form:

\begin{equation}
\begin{array}{c}
\mu =\frac{1}{2}\left( -\rho \varepsilon p+\kappa \right) ,\ \ \ \zeta =%
\frac{1}{2}\sqrt{-\varepsilon ^{2}\bar{\rho}^{2}p^{2}-\varepsilon \left(
2\rho \kappa -\varepsilon \right) p+\kappa ^{2}}, \\ 
\\ 
\lambda _{+}\lambda _{-}=\mu ^{2}-\zeta ^{2}=\frac{\varepsilon ^{2}p\left(
p-1\right) }{4}.%
\end{array}
\label{Eq6.81}
\end{equation}%
Thus,%
\begin{equation}
\begin{array}{c}
\mathcal{I}_{p}=\left( \frac{2\zeta }{\lambda _{+}\mathsf{E}_{+}-\lambda _{-}%
\mathsf{E}_{-}}\right) ^{\frac{2\chi }{\varepsilon ^{2}}} \\ 
\\ 
\times \exp \left( \frac{2\chi }{\varepsilon ^{2}}\mu \left( t-\tau \right)
+p\xi +\frac{p\left( p-1\right) \left( \mathsf{E}_{+}-\mathsf{E}_{-}\right)
\theta }{2\left( \lambda _{+}\mathsf{E}_{+}-\lambda _{-}\mathsf{E}%
_{-}\right) }\right) ,%
\end{array}
\label{Eq6.82}
\end{equation}%
when $\zeta >0$ is real, and%
\begin{equation}
\begin{array}{c}
\mathcal{I}_{p}=\left( \frac{\left\vert \zeta \right\vert }{\mu \sin \left(
\left\vert \zeta \right\vert \left( t-\tau \right) \right) +\left\vert \zeta
\right\vert \cos \left( \left\vert \zeta \right\vert \left( t-\tau \right)
\right) }\right) ^{\frac{2\chi }{\varepsilon ^{2}}} \\ 
\\ 
\times \exp \left( \frac{2\chi }{\varepsilon ^{2}}\mu \left( t-\tau \right)
+p\xi +\frac{p\left( p-1\right) \sin \left( \left\vert \zeta \right\vert
\left( t-\tau \right) \right) \theta }{2\left( \mu \sin \left( \left\vert
\zeta \right\vert \left( t-\tau \right) \right) +\left\vert \zeta
\right\vert \cos \left( \left\vert \zeta \right\vert \left( t-\tau \right)
\right) \right) }\right) ,%
\end{array}
\label{Eq6.83}
\end{equation}%
when $\zeta =i\left\vert \zeta \right\vert $ is imaginary.

We need to determine when $\zeta $ becomes imaginary. The corresponding
quadratic equation and its roots have the form:%
\begin{equation}
\begin{array}{c}
\bar{\rho}^{2}\varepsilon ^{2}p^{2}+2\varepsilon \left( \rho \kappa -\frac{%
\varepsilon }{2}\right) p-\kappa ^{2}=0,%
\end{array}
\label{Eq6.84}
\end{equation}%
its roots are as follows:%
\begin{equation}
\begin{array}{c}
p_{\pm }=\frac{-\left( \rho \kappa -\frac{\varepsilon }{2}\right) \pm \sqrt{%
\kappa ^{2}-\rho \kappa \varepsilon +\frac{\varepsilon ^{2}}{4}}}{\bar{\rho}%
^{2}\varepsilon }.%
\end{array}
\label{Eq6.85}
\end{equation}%
Since%
\begin{equation}
\begin{array}{c}
\bar{\rho}^{2}\varepsilon ^{2}+2\varepsilon \left( \rho \kappa -\frac{%
\varepsilon }{2}\right) -\kappa ^{2}=-\rho ^{2}\varepsilon ^{2}+2\rho
\varepsilon \kappa -\kappa ^{2}=-\left( \rho \varepsilon -\kappa \right)
^{2}\leq 0,%
\end{array}
\label{Eq6.86}
\end{equation}%
we have%
\begin{equation}
\begin{array}{c}
p_{+}>1,\ \ \ p_{-}<1.%
\end{array}
\label{Eq6.87}
\end{equation}%
For $p\in \left[ p_{-},p_{+}\right] $, $\zeta $ is real, for $p\notin \left[
p_{-},p_{+}\right] $, it is imaginary. There is no blowup when $\zeta $ is
real. When $\zeta $ is imaginary, the blowup time $t^{\ast }$ is the
smallest positive root of the equation%
\begin{equation}
\begin{array}{c}
\mu \sin \left( \left\vert \zeta \right\vert \left( t^{\ast }-\tau \right)
\right) +\left\vert \zeta \right\vert \cos \left( \left\vert \zeta
\right\vert \left( t^{\ast }-\tau \right) \right) =0,%
\end{array}%
\label{Eq6.89}
\end{equation}

\subsection{Example: Augmented OU Process\label{Sec63}}

In this Section, we consider several instances when an augmented OU process
becomes non-Gaussian.

\subsubsection{Degenerate Augmented OU Process, I}

Let $y_{t}$ be a stochastic process and $z_{t}$ be its moving average. Then%
\begin{equation}
\begin{array}{c}
z_{t}=\varkappa \int_{-\infty }^{t}e^{-\varkappa \left( t-s\right) }y_{s}ds.%
\end{array}
\label{Eq6.90}
\end{equation}%
A simple calculation yields:%
\begin{equation}
\begin{array}{c}
dz_{t}=\varkappa \left( y_{t}-z_{t}\right) dt.%
\end{array}
\label{Eq6.91}
\end{equation}

We assume that the process $y_{t}$ is path-dependent, because its volatility 
$\sigma $ depends on its moving average $z_{t}$:%
\begin{equation}
\begin{array}{c}
\sigma _{t}=\sqrt{a_{0}+a_{1}\left( y_{t}-z_{t}\right) },%
\end{array}
\label{Eq6.92}
\end{equation}%
where $a_{0}>0$, $a_{1}<0$, in order to capture the skew. Thus, we can write
the governing degenerate system of SDEs as follows: 
\begin{equation}
\begin{array}{c}
dy_{t}=-\frac{1}{2}\left( a_{0}+a_{1}\left( y_{t}-z_{t}\right) \right) dt+%
\sqrt{a_{0}+a_{1}\left( y_{t}-z_{t}\right) }dW_{t},\ \ \ y_{\tau }=\theta ,
\\ 
\\ 
dz_{t}=\varkappa \left( y_{t}-z_{t}\right) dt,\ \ \ z_{\tau }=\varrho .%
\end{array}
\label{Eq6.93}
\end{equation}

The Fokker-Planck equation%
\begin{equation}
\begin{array}{c}
P_{t}-\frac{1}{2}\left( a_{0}+a_{1}\left( y-z\right) \right) \left(
P_{yy}+P_{y}\right) +\varkappa \left( y-z\right) P_{z}-\left(
a_{1}+\varkappa \right) P=0, \\ 
\\ 
P\left( \tau ,\theta ,\varrho ,\tau ,y,z\right) =\delta \left( y-\theta
\right) \delta \left( z-\varrho \right) .%
\end{array}
\label{Eq6.94}
\end{equation}%
A representative Kelvin mode%
\begin{equation}
\begin{array}{c}
H=\exp \left( \alpha +i\Gamma y-ik\theta +i\Delta z-il\varrho \right) .%
\end{array}
\label{Eq6.95}
\end{equation}%
The system of ODEs for $\alpha ,\Gamma ,\Delta $ has the form:%
\begin{equation}
\begin{array}{c}
\alpha ^{\prime }+\frac{a_{0}}{2}\left( \Gamma ^{2}-i\Gamma \right) -\left( 
\frac{a_{1}}{2}+\varkappa \right) =0,\ \ \ \alpha \left( \tau \right) =0, \\ 
\\ 
i\Gamma ^{\prime }+\frac{a_{1}}{2}\left( \Gamma ^{2}-i\Gamma \right)
+i\varkappa \Delta =0,\ \ \ \Gamma \left( \tau \right) =k, \\ 
\\ 
i\Delta ^{\prime }-\frac{a_{1}}{2}\left( \Gamma ^{2}-i\Gamma \right)
-i\varkappa \Delta =0,\ \ \ \Delta \left( \tau \right) =l.%
\end{array}
\label{Eq6.96}
\end{equation}

Eqs (\ref{Eq6.96}) are matrix Riccati equations. In general, such equations
are very difficult to solve. However, the case under consideration is one of
relatively rare instances when a matrix Riccati equation can be solved
explicitly. We start with an observation:%
\begin{equation}
\begin{array}{c}
\Gamma ^{\prime }+\Delta ^{\prime }=0,\ \ \ \Gamma \left( \tau \right)
+\Delta \left( \tau \right) =k+l,%
\end{array}
\label{Eq6.97}
\end{equation}%
\begin{equation}
\begin{array}{c}
\Delta =k+l-\Gamma .%
\end{array}
\label{Eq6.98}
\end{equation}%
Accordingly,%
\begin{equation}
\begin{array}{c}
i\Gamma ^{\prime }+\frac{a_{1}}{2}\left( \Gamma ^{2}-i\Gamma \right)
-i\varkappa \Gamma +i\varkappa \left( k+l\right) =0,\ \ \ \Gamma \left( \tau
\right) =k.%
\end{array}
\label{Eq6.99}
\end{equation}%
As usual,%
\begin{equation}
\begin{array}{c}
\Gamma =\frac{2i\Omega ^{\prime }}{a_{1}\Omega },%
\end{array}
\label{Eq6.100}
\end{equation}%
\begin{equation}
\begin{array}{c}
\Omega ^{\prime \prime }-\frac{\left( a_{1}+2\varkappa \right) }{2}\Omega
^{\prime }-\frac{ia_{1}\varkappa \left( k+l\right) }{2}=0, \\ 
\\ 
\Omega \left( \tau \right) =1,\ \ \ \Omega ^{\prime }\left( \tau \right) =-%
\frac{ia_{1}k}{2},%
\end{array}
\label{Eq6.101}
\end{equation}%
\begin{equation}
\begin{array}{c}
\Omega =\frac{\mathsf{E}_{0}\left( -\left( 2\lambda _{-}+ia_{1}k\right) 
\mathsf{E}_{+}+\left( 2\lambda _{+}+ia_{1}k\right) \mathsf{E}_{-}\right) }{%
4\zeta },%
\end{array}
\label{Eq6.102}
\end{equation}%
\begin{equation}
\begin{array}{c}
\Omega ^{\prime }=\frac{\mathsf{E}_{0}\left( -\lambda _{+}\left( 2\lambda
_{-}+ia_{1}k\right) \mathsf{E}_{+}+\lambda _{-}\left( 2\lambda
_{+}+ia_{1}k\right) \mathsf{E}_{-}\right) }{4\zeta },%
\end{array}
\label{Eq6.102a}
\end{equation}%
\begin{equation}
\begin{array}{c}
\lambda _{\pm }=\mu \pm \zeta ,\ \ \ \mu =\frac{\left( a_{1}+2\varkappa
\right) }{4},\ \ \ \zeta =\frac{\sqrt{\left( a_{1}+2\varkappa \right)
^{2}+8ia_{1}\varkappa \left( k+l\right) }}{4}.%
\end{array}
\label{Eq6.103}
\end{equation}%
Eq. (\ref{Eq6.98}) yields:%
\begin{equation}
\begin{array}{c}
\Delta =-\frac{2i\Omega ^{\prime }}{a_{1}\Omega }+k+l.%
\end{array}
\label{Eq6.104}
\end{equation}

Now,%
\begin{equation}
\begin{array}{c}
\Gamma ^{2}-i\Gamma =-\frac{2i\left( \Gamma ^{\prime }+\varkappa \Delta
\right) }{a_{1}},%
\end{array}
\label{Eq6.105}
\end{equation}%
so that%
\begin{equation}
\begin{array}{c}
\alpha ^{\prime }=\frac{ia_{0}}{a_{1}}\left( \Gamma ^{\prime }+\varkappa
\Delta \right) +\left( \frac{a_{1}}{2}+\varkappa \right) ,\ \ \ \alpha
\left( \tau \right) =0.%
\end{array}
\label{Eq6.106}
\end{equation}%
Accordingly,%
\begin{equation}
\begin{array}{c}
\alpha =\frac{ia_{0}}{a_{1}}\left( \Gamma -k\right) +\frac{2a_{0}\varkappa }{%
a_{1}^{2}}\ln \left( \Omega \right) +\frac{ia_{0}\varkappa }{a_{1}}\left(
k+l\right) \left( t-\tau \right) +\left( \frac{a_{1}}{2}+\varkappa \right)
\left( t-\tau \right) \\ 
\\ 
=-\frac{2a_{0}\Omega ^{\prime }}{a_{1}^{2}\Omega }-\frac{a_{0}}{a_{1}}ik+%
\frac{2a_{0}\varkappa }{a_{1}^{2}}\ln \left( \Omega \right) +\frac{%
a_{0}\varkappa \left( t-\tau \right) }{a_{1}}i\left( k+l\right) +\left( 
\frac{a_{1}}{2}+\varkappa \right) \left( t-\tau \right) .%
\end{array}
\label{Eq6.107}
\end{equation}

Finally,%
\begin{equation}
\begin{array}{c}
H=\exp \left( \alpha +i\Gamma y-ik\theta +i\Delta z-il\varrho \right) \\ 
\\ 
=\exp \left( \left( \frac{a_{1}}{2}+\varkappa \right) \left( t-\tau \right)
\right) \exp \left( -\frac{2\Omega ^{\prime }}{a_{1}\Omega }\left( y-z+\frac{%
a_{0}}{a_{1}}\right) +\frac{2a_{0}\varkappa }{a_{1}^{2}}\ln \left( \Omega
\right) \right. \\ 
\\ 
\left. +i\left( k+l\right) \left( z+\frac{a_{0}\varkappa \left( t-\tau
\right) }{a_{1}}\right) -ik\left( \theta +\frac{a_{0}}{a_{1}}\right)
-il\varrho \right) .%
\end{array}
\label{Eq6.108}
\end{equation}%
We consider $y,z$, such that%
\begin{equation}
\begin{array}{c}
y-z+\frac{a_{0}}{a_{1}}>0,%
\end{array}
\label{Eq6.109}
\end{equation}%
in order to make sure that $\sigma $ given by Eq. (\ref{Eq6.92}) and the
integrand (\ref{Eq6.108}) are well defined.

\subsubsection{Degenerate Augmented OU Process, II}

On occasion, problems seemingly not of the type given by Eq. (\ref{Eq6.1})
can be cast in the proper form via a suitable trick. Consider, for example,
the following system of SDEs:%
\begin{equation}
\begin{array}{c}
dx_{t}=y_{t}^{2}dt,\ \ \ x_{\tau }=\xi , \\ 
\\ 
dy_{t}=\left( \chi -\kappa y_{t}\right) dt+\varepsilon dZ_{t},\ \ \ y_{\tau
}=\theta .%
\end{array}
\label{Eq6.110}
\end{equation}%
Superficially, it does not belong to the class of processes we have studied
above. However, by introducing a new variable $z_{t}=y_{t}^{2}$, we can
expand Eqs (\ref{Eq6.110}) as follows:%
\begin{equation}
\begin{array}{c}
dx_{t}=z_{t}dt,\ \ \ x_{\tau }=\xi , \\ 
\\ 
dy_{t}=\left( \chi -\kappa y_{t}\right) dt+\varepsilon dZ_{t},\ \ \ y_{\tau
}=\theta , \\ 
\\ 
dz_{t}=\left( \varepsilon ^{2}+2\chi y_{t}-2\kappa z_{t}\right)
+2\varepsilon y_{t}dZ_{t},\ \ \ z_{\tau }=\theta ^{2}.%
\end{array}
\label{Eq6.111}
\end{equation}%
These equations are \textquotedblleft almost\textquotedblright\ in the
suitable form. The only snag is that we cannot claim that $y_{t}=\sqrt{z_{t}}
$ since $y_{t}$ is not always positive.

The corresponding Fokker-Planck problem can be written as follows:%
\begin{equation}
\begin{array}{c}
P_{t}-\frac{1}{2}\varepsilon ^{2}P_{yy}-2\varepsilon
^{2}yP_{yz}-2\varepsilon ^{2}zP_{zz} \\ 
\\ 
+zP_{x}+\left( \chi -\kappa y_{t}\right) P_{y}+\left( \varepsilon ^{2}+2\chi
y_{t}-2\kappa z_{t}\right) P_{z}-3\kappa P=0, \\ 
\\ 
P\left( \tau ,x,y,z\right) =\delta \left( x-\xi \right) \delta \left(
y-\theta \right) \delta \left( z-\theta ^{2}\right) .%
\end{array}
\label{Eq6.112}
\end{equation}%
As usual, $H$ has the form:%
\begin{equation}
\begin{array}{c}
H\left( t,x,y,z,k,l,m\right) =\exp \left( \alpha \left( t\right) +ik\left(
x-\xi \right) +i\Delta \left( t\right) y-il\theta +i\Xi \left( t\right)
z-im\theta ^{2}\right) .%
\end{array}
\label{Eq6.113}
\end{equation}%
The corresponding set of ODEs for $\alpha ,\Delta ,\Xi $ is as follows:%
\begin{equation}
\begin{array}{c}
\alpha ^{\prime }+\frac{\varepsilon ^{2}}{2}\Delta ^{2}+i\chi \Delta
+i\varepsilon ^{2}\Xi -3\kappa =0,\ \ \ \alpha \left( \tau \right) =0, \\ 
\\ 
i\Delta ^{\prime }+2\varepsilon ^{2}\Delta \Xi -i\kappa \Delta +2i\chi \Xi
=0,\ \ \ \Delta \left( \tau \right) =l, \\ 
\\ 
i\Xi ^{\prime }+2\varepsilon ^{2}\Xi ^{2}-2i\kappa \Xi +ik=0,\ \ \ \Xi
\left( \tau \right) =m.%
\end{array}
\label{Eq6.114}
\end{equation}%
These equations are matrix Riccati equations.

Once again, the corresponding matrix Riccati equation can be solved
explicitly. Since the third equation is separable, and hence can be viewed
as a scalar Riccati equation, we can start with a familiar ansatz: 
\begin{equation}
\begin{array}{c}
\Xi =\frac{i\Omega ^{\prime }}{2\varepsilon ^{2}\Omega }.%
\end{array}
\label{Eq6.115}
\end{equation}%
The corresponding characteristic equation and its solutions are as follows:%
\begin{equation}
\begin{array}{c}
\lambda ^{2}-2\kappa \lambda -2\varepsilon ^{2}ik=0, \\ 
\\ 
\mu =\kappa ,\ \ \ \zeta =\sqrt{\kappa ^{2}+2i\varepsilon ^{2}k}.%
\end{array}
\label{Eq6.116}
\end{equation}%
As usual, we can write:%
\begin{equation}
\begin{array}{c}
\Omega =e^{\mu \left( t-\tau \right) }\left( \frac{-\left( \lambda
_{-}+2i\varepsilon ^{2}m\right) e^{\zeta \left( t-\tau \right) }+\left(
\lambda _{+}+2i\varepsilon ^{2}m\right) e^{-\zeta \left( t-\tau \right) }}{%
2\zeta }\right) ,%
\end{array}
\label{Eq6.117}
\end{equation}%
\begin{equation}
\begin{array}{c}
\Omega ^{\prime }=e^{\mu \left( t-\tau \right) }\left( \frac{-\lambda
_{+}\left( \lambda _{-}+2i\varepsilon ^{2}m\right) e^{\zeta \left( t-\tau
\right) }+\lambda _{-}\left( \lambda _{+}+2i\varepsilon ^{2}m\right)
e^{-\zeta \left( t-\tau \right) }}{2\zeta }\right) ,%
\end{array}
\label{Eq6.118}
\end{equation}%
\begin{equation}
\begin{array}{c}
\Xi =\frac{i\left( -\lambda _{+}\left( \lambda _{-}+2i\varepsilon
^{2}m\right) e^{\zeta \left( t-\tau \right) }+\lambda _{-}\left( \lambda
_{+}+2i\varepsilon ^{2}m\right) e^{-\zeta \left( t-\tau \right) }\right) }{%
2\varepsilon ^{2}\left( -\left( \lambda _{-}+2i\varepsilon ^{2}m\right)
e^{\zeta \left( t-\tau \right) }+\left( \lambda _{+}+2i\varepsilon
^{2}m\right) e^{-\zeta \left( t-\tau \right) }\right) }.%
\end{array}
\label{Eq6.119}
\end{equation}

To linearize Eqs. (\ref{Eq6.114}) as a whole, we use the following ansatz:%
\begin{equation}
\begin{array}{c}
\Xi =\frac{i\Omega ^{\prime }}{2\varepsilon ^{2}\Omega },\ \ \ \Delta =\frac{%
i\delta }{\Omega },\ \ \ \alpha =\frac{\lambda }{\Omega }+g\left( t-\tau
\right) +\frac{1}{2}\ln \left( \Omega \right) +3\kappa \left( t-\tau \right)
,%
\end{array}
\label{Eq6.120}
\end{equation}%
where%
\begin{equation}
\begin{array}{c}
\Omega =\mathsf{E}_{0}\left( c_{+}\mathsf{E}_{+}+c_{-}\mathsf{E}_{-}\right) ,
\\ 
\\ 
\delta =\mathsf{E}_{0}\left( a_{0}+a_{+}\mathsf{E}_{+}+a_{-}\mathsf{E}%
_{-}\right) , \\ 
\\ 
\lambda =\mathsf{E}_{0}\left( b_{0}+b_{+}\mathsf{E}_{+}+b_{-}\mathsf{E}%
_{-}\right) .%
\end{array}
\label{Eq6.121}
\end{equation}%
and $g$ is a constant to be determined. Our ansatz is useful since terms
proportional to $\sim 1,\mathsf{E}_{+},\mathsf{E}_{-}$ balance each other,
which allows us to find the coefficients explicitly. Initial conditions
complete the picture. The actual calculation is omitted for brevity. The
result is as follows:%
\begin{equation}
\begin{array}{c}
c_{+}=-\frac{\left( \lambda _{-}+2i\varepsilon ^{2}m\right) }{2\zeta },\ \ \
c_{-}=\frac{\left( \lambda _{+}+2i\varepsilon ^{2}m\right) }{2\zeta }, \\ 
\\ 
a_{0}=-il+\frac{\chi \lambda _{+}}{\varepsilon ^{2}\zeta }c_{+}-\frac{\chi
\lambda _{-}}{\varepsilon ^{2}\zeta }c_{-},\ \ \ a_{+}=-\frac{\chi \lambda
_{+}}{\varepsilon ^{2}\zeta }c_{+},\ \ \ a_{-}=\frac{\chi \lambda _{-}}{%
\varepsilon ^{2}\zeta }c_{-}, \\ 
\\ 
g=\frac{\chi ^{2}\lambda _{+}\lambda _{-}}{2\varepsilon ^{2}\zeta ^{2}},\ \
\ b_{0}=\frac{\chi \mu }{\zeta ^{2}}a_{0}, \\ 
\\ 
b_{+}=-c_{+}b_{0}+\frac{\varepsilon ^{2}}{4\zeta }a_{0}^{2}-\frac{\chi
^{2}\mu ^{2}}{\varepsilon ^{2}\zeta ^{3}}c_{+}c_{-},\ \ \ b_{-}=-c_{-}b_{0}-%
\frac{\varepsilon ^{2}}{4\zeta }a_{0}^{2}+\frac{\chi ^{2}\mu ^{2}}{%
\varepsilon ^{2}\zeta ^{3}}c_{+}c_{-}.%
\end{array}
\label{Eq6.122}
\end{equation}%
Although these expressions can be simplified further and then substituted in
the function $H$ to obtain the corresponding t.p.d.f, we shall not pursue
this line of research here. We shall report the outcome of our investigation
elsewhere.

\subsubsection{Non-Degenerate Augmented OU Process}

In this Section, we study an affine process of the form:%
\begin{equation}
\begin{array}{c}
dx_{t}=-\frac{1}{2}y_{t}^{2}dt+y_{t}dW_{t},\ \ \ x_{\tau }=\xi , \\ 
\\ 
dy_{t}=\left( \chi -\kappa y_{t}\right) dt+\varepsilon dZ_{t},\ \ \ y_{\tau
}=\theta ,%
\end{array}
\label{Eq6.124}
\end{equation}%
which is at the heart of the Stein \& Stein model; see \cite{Stein91} who
considered the special case of zero correlation, $dW_{t}dZ_{t}=0$ and \cite%
{Schobel99} who studied the general case of arbitrary correlation, $%
dW_{t}dZ_{t}=\rho dt$. Further details are given in Section \ref{Sec7}.

Precisely as before, we can introduce a new variable $z_{t}=y_{t}^{2}$, and
expand Eqs (\ref{Eq6.124}) as follows:%
\begin{equation}
\begin{array}{c}
dx_{t}=-\frac{1}{2}z_{t}dt+y_{t}dW_{t},\ \ \ x_{\tau }=\xi , \\ 
\\ 
dy_{t}=\left( \chi -\kappa y_{t}\right) dt+\varepsilon dZ_{t},\ \ \ y_{\tau
}=\theta , \\ 
\\ 
dz_{t}=\left( \varepsilon ^{2}+2\chi y_{t}-2\kappa z_{t}\right)
+2\varepsilon y_{t}dZ_{t},\ \ \ z_{\tau }=\theta ^{2}.%
\end{array}
\label{Eq6.125}
\end{equation}%
It is clear that Eqs (\ref{Eq6.125}) are affine.

The corresponding Fokker-Planck problem can be written as follows:%
\begin{equation}
\begin{array}{c}
P_{t}-\frac{1}{2}zP_{xx}-\rho \varepsilon yP_{xy}-2\rho \varepsilon zP_{xz}-%
\frac{1}{2}\varepsilon ^{2}P_{yy}-2\varepsilon ^{2}yP_{yz}-2\varepsilon
^{2}zP_{zz} \\ 
\\ 
-\frac{1}{2}zP_{x}+\left( \chi -\kappa y\right) P_{y}+\left( \varepsilon
^{2}+2\chi y-2\kappa z\right) P_{z}-3\kappa P=0, \\ 
\\ 
P\left( \tau ,x,y,z\right) =\delta \left( x-\xi \right) \delta \left(
y-\theta \right) \delta \left( z-\theta ^{2}\right) .%
\end{array}
\label{Eq6.126}
\end{equation}%
We can use $H$ given by Eq. (\ref{Eq6.113}) and write the set of ODEs for $%
\alpha ,\Delta ,\Xi $ is as follows:%
\begin{equation}
\begin{array}{c}
\alpha ^{\prime }+\frac{\varepsilon ^{2}}{2}\Delta ^{2}+i\chi \Delta
+i\varepsilon ^{2}\Xi -3\kappa =0,\ \ \ \alpha \left( \tau \right) =0, \\ 
\\ 
i\Delta ^{\prime }+2\varepsilon ^{2}\Delta \Xi +\left( \rho \varepsilon
k-i\kappa \right) \Delta +2i\chi \Xi =0,\ \ \ \Delta \left( \tau \right) =l,
\\ 
\\ 
i\Xi ^{\prime }+2\varepsilon ^{2}\Xi ^{2}+2\left( \rho \varepsilon k-i\kappa
\right) \Xi +\frac{k\left( k-i\right) }{2}=0,\ \ \ \Xi \left( \tau \right)
=m.%
\end{array}
\label{Eq6.127}
\end{equation}

As mentioned above, this system can be linearized and solved analytically,
which was discovered earlier by\ \cite{Stein91, Schobel99}. We can repeat
the result obtained in the previous Section verbatim, except for Eq. (\ref%
{Eq6.116}). The corresponding characteristic equation and its solutions can
be written as follows:%
\begin{equation}
\begin{array}{c}
\lambda ^{2}-2\left( \kappa +i\rho \varepsilon k\right) \lambda -\varepsilon
^{2}k\left( k-i\right) =0, \\ 
\\ 
\mu =\kappa +i\rho \varepsilon k,\ \ \ \zeta =\sqrt{\bar{\rho}%
^{2}\varepsilon ^{2}k^{2}-i\varepsilon ^{2}k+\kappa ^{2}+2i\rho \kappa
\varepsilon }.%
\end{array}
\label{Eq6.128}
\end{equation}%
The rest of the formal analysis is the same.\ But the asymptotic behavior of
the t.p.d.f. is, of course, different.

\section{Financial Instruments\label{Sec7}}

\subsection{Arithmetic and Geometric Brownian Motions and Their
Generalizations\label{Sec71}}

The original approach to modeling financial assets was developed by
Bachelier, who assumed that prices $y_{t}$ of such instruments are governed
by an arithmetic Brownian motion; see \cite{Bachelier00}:%
\begin{equation}
\begin{array}{c}
dS_{t}=rS_{t}dt+\hat{\sigma}dW_{t},\ \ \ S_{\tau }=s.%
\end{array}
\label{Eq7.56}
\end{equation}%
Here, $r$ is the risk-neutralized drift, $\hat{\sigma}$ is the volatility,
and $W_{t}$ is a Wiener process; $r$, $\hat{\sigma}$ are dimensional
quantities, $\left[ r\right] =T^{-1}$, $\left[ \sigma \right] =\$T^{-1/2}$.
The process for $S_{t}$ given by Eq. (\ref{Eq7.56}) is affine; in fact, it
is an OU process with mean-repulsion instead of mean-reversion. Accordingly,
the corresponding t.p.d.f. is given by a modified Eq. (\ref{Eq5.107a}):%
\begin{equation}
\begin{array}{c}
P\left( \tau ,s,t,S\right) =\frac{1}{\sqrt{\frac{\pi \hat{\sigma}^{2}\left(
e^{2r\left( t-\tau \right) }-1\right) }{\kappa }}}\exp \left( -\frac{\left(
S-e^{r\left( t-\tau \right) }s\right) ^{2}}{\frac{\hat{\sigma}^{2}\left(
e^{2r\left( t-\tau \right) }-1\right) }{r}}\right) .%
\end{array}
\label{Eq7.56a}
\end{equation}

By virtue of this formula, one can price the so-called European options,
which depend solely on the terminal value of $S$ at option's maturity. For
instance, the price of call and put options with maturity $t$ and strike $K$
with payoffs of the form:%
\begin{equation}
\begin{array}{c}
V\left( t,S_{t}\right) =\max \left\{ \nu \left( S_{t}-K\right) ,0\right\} ,%
\end{array}
\label{Eq7.56b}
\end{equation}%
where $\nu =1$ for a call, and $\nu =-1$ for a put. The corresponding
formulas have the form:%
\begin{equation}
\begin{array}{c}
V\left( \tau ,s\right) =\nu \left( s-e^{-r\left( t-\tau \right) }K\right) 
\mathrm{N}\left( \nu \frac{s-e^{-r\left( t-\tau \right) }K}{\hat{\sigma}%
\left( \tau ,t\right) }\right) +\hat{\sigma}\left( \tau ,t\right) \mathrm{n}%
\left( \frac{s-e^{-r\left( t-\tau \right) }K}{\hat{\sigma}\left( \tau
,t\right) }\right) ,%
\end{array}
\label{Eq7.56c}
\end{equation}%
where%
\begin{equation}
\begin{array}{c}
\hat{\sigma}\left( \tau ,t\right) =\hat{\sigma}\sqrt{\frac{e^{2r\left(
t-\tau \right) }-1}{2re^{2r\left( t-\tau \right) }}.}%
\end{array}
\label{Eq7.56d}
\end{equation}

Subsequently, the academic community concluded that using a geometric
Brownian motion as a driver is more appropriate; see \cite{Boness64,
Samuelson65, Black73, Merton73}. At present, the basic assumption is that
the price $S_{t}$ of an underlying financial instrument follows a geometric
Brownian motion process with constant coefficients:%
\begin{equation}
\begin{array}{c}
\frac{dS_{t}}{S_{t}}=rdt+\sigma dW_{t},\ \ \ S_{\tau }=s.%
\end{array}
\label{Eq7.61}
\end{equation}%
Here, $r$ is the risk-neutralized drift, and $\sigma $ is the volatility.
These are dimensional quantities, $\left[ r\right] =T^{-1}$, $\left[ \sigma %
\right] =T^{-1/2}$.

The well-known logarithmic change of variables,%
\begin{equation}
\begin{array}{c}
y_{t}=\ln \left( S_{t}\right) ,%
\end{array}
\label{Eq7.62}
\end{equation}%
results in the following process for $y_{t}$:%
\begin{equation}
\begin{array}{c}
dy_{t}=\left( r-\frac{1}{2}\sigma ^{2}\right) dt+\sigma dW_{t},\ \ \ y_{\tau
}=\ln \left( s\right) =\theta .%
\end{array}
\label{Eq7.63}
\end{equation}%
The t.p.d.f. for the process $y_{t}$ is Gaussian:%
\begin{equation}
\begin{array}{c}
P\left( \tau ,\theta ,t,y\right) =\frac{1}{\sqrt{2\pi \sigma ^{2}\left(
t-\tau \right) }}\exp \left( -\frac{\left( y-\theta -\left( r-\sigma
^{2}/2\right) \left( t-\tau \right) \right) ^{2}}{2\sigma ^{2}\left( t-\tau
\right) }\right) .%
\end{array}
\label{Eq7.64}
\end{equation}

Call and put prices with maturity $t$ and strike $K=e^{k}$ have terminal
payoffs of the form:%
\begin{equation}
\begin{array}{c}
V\left( t,y_{t}\right) =\max \left\{ \nu \left( e^{y_{t}}-e^{k}\right)
,0\right\} .%
\end{array}
\label{Eq7.65}
\end{equation}%
Eq. (\ref{Eq7.64}) can be used to derive the celebrated \cite{Black73}
closed-form formula for their prices at time $\tau $:%
\begin{equation}
\begin{array}{c}
V\left( \tau ,\theta \right) =\nu \left( e^{\theta }\mathrm{N}\left( \nu 
\frac{\theta -k+\left( r+\sigma ^{2}/2\right) \left( t-\tau \right) }{\sigma 
\sqrt{t-\tau }}\right) -e^{-r\left( t-\tau \right) +k}\mathrm{N}\left( \nu 
\frac{\theta -k+\left( r-\sigma ^{2}/2\right) \left( t-\tau \right) }{\sigma 
\sqrt{t-\tau }}\right) \right) ,%
\end{array}
\label{Eq7.66}
\end{equation}%
where $\mathrm{N}\left( .\right) $ is the cumulative normal function.

Since the hedging and speculation needs of market participants cannot be
satisfied by European options alone, the whole industry emerged to design,
price, and hedge the so-called exotic options, with payoffs depending on the
entire price trajectory between inception and maturity. Prices of the
fundamental financial instruments, such as forwards and European calls and
puts, depend on the underlying prices only at maturity. However, the prices
of many other instruments depend on the entire underlying price history
between the instrument's inception and maturity. Typical examples are
barrier, American, Asian, lookback, and passport options; see, e.g., \cite%
{Lipton99a, Lipton01} and references therein. Moreover, the prices of bonds
also depend on the history of the interest rates and credit spreads
throughout their life. This section shows how to price some path-dependent
financial instruments using the methodology developed in the previous
sections.

In addition, it was realized very soon after the seminal paper by \cite%
{Black73} was published, that in practice, it provides a rather poor
description of reality. Hence, considerable efforts were dedicated to
developing more adequate models. Such models include the jump-diffusion,
local volatility, path-dependent volatility, stochastic volatility,
local-stochastic volatility, rough volatility, and culminate in the
universal volatility model; see \cite{Merton76, Stein91, Bick93, Heston93,
Derman94, Dupire94, Rubinstein94, Hobson98, Jex99, Lewis00, Lipton01,
Boyarchenko02, Hagan02, Lipton02, Bergomi15, Reghai15, Gatheral18, Gershon22}%
, and references therein.

Replacing a constant volatility for a geometric Brownian motion with
stochastic volatility driven by a Feller process results in the popular
Heston model; see \cite{Heston93}. the governing SDEs:%
\begin{equation}
\begin{array}{c}
\frac{dS_{t}}{S_{t}}=\sqrt{v_{t}}dW_{t},\ \ \ S_{\tau }=s, \\ 
\\ 
dv_{t}=\left( \chi -\kappa v_{t}\right) dt+\varepsilon \sqrt{v_{t}}dZ_{t},\
\ \ v_{\tau }=\nu ,%
\end{array}
\label{Eq7.67}
\end{equation}%
where $dW_{t}dZ_{t}=\rho dt$. The logarithmic change of variables, given by
Eq. (\ref{Eq7.67}), yields Eqs (\ref{Eq6.74}).

Replacing a constant volatility with stochastic volatility driven by an OU
process results in the (less popular) Stein \& Stein model; see \cite%
{Stein91, Schobel99}. The corresponding SDEs have the form:%
\begin{equation}
\begin{array}{c}
\frac{dS_{t}}{S_{t}}=\sqrt{v_{t}}dW_{t},\ \ \ S_{\tau }=s, \\ 
\\ 
dv_{t}=\left( \chi -\kappa v_{t}\right) dt+\varepsilon dZ_{t},\ \ \ v_{\tau
}=\nu ,%
\end{array}
\label{Eq7.68}
\end{equation}%
The logarithmic change of variables reduces Eqs (\ref{Eq7.68}) to Eqs (\ref%
{Eq6.124}).

We shall show how to use formulas derived in Sections \ref{Sec5}, \ref{Sec6}
in the context of financial engineering.

\subsection{Asian Options with Geometric Averaging\label{Sec72}}

The most basic path-dependent options are fixed strike Asian calls and puts,
whose payoff depends on the underlying value averaged between the inception
and maturity. Such options are popular for commodity and energy trading and
in many other circumstances. The average $A_{\tau ,t}$ on the interval $%
\left[ \tau ,t\right] $ can be defined in several ways. The simplest and, as
a result, the most popular is an arithmetic average:%
\begin{equation}
\begin{array}{c}
A_{\tau ,t}=\frac{1}{t-\tau }\dint\limits_{\tau }^{t}S_{s}ds.%
\end{array}
\label{Eq7.2}
\end{equation}%
A less frequent, but technically easier to deal with, alternative is a
geometric average:%
\begin{equation}
A_{\tau ,t}=\exp \left( \frac{1}{t-\tau }\dint\limits_{\tau }^{t}\ln \left(
S_{s}\right) ds\right) .  \label{Eq7.3}
\end{equation}%
The payoff of an Asian option with maturity $t$ and fixed strike $K$ is%
\begin{equation}
V\left( t,A_{\tau ,t}\right) =\max \left\{ \nu \left( A_{\tau ,t}-K\right)
,0\right\} ,  \label{Eq7.4}
\end{equation}%
as before, $\nu =1$ for a call, and $\nu =-1$ for a put. For the floating
strike, the payoff is%
\begin{equation}
V\left( t,S_{t},A_{\tau ,t}\right) =\max \left\{ \nu \left( S_{t}-kA_{\tau
,t}\right) ,0\right\} ,  \label{Eq7.5}
\end{equation}%
where the non-dimensional parameter $k$ is called weighting; typically, $k=1$%
.

Analytical pricing of Asian options with arithmetic averaging is notoriously
tricky; see \cite{Geman95, Rogers95, Lipton99b, Lipton01}. At the same time,
pricing Asian options with geometric averaging can be done quickly; see \cite%
{Barrucci01, Lipton01, DiFrancesco05}, and references therein. Here we show
how to price such options using formula (\ref{Eq5.23}) derived in Section %
\ref{Sec5}. An alternative approach based on the path integral method is
discussed in \cite{Devreese10}. We define%
\begin{equation}
\begin{array}{c}
y_{t}=\ln \left( S_{t}\right) ,\ \ \ x_{t}=\dint\limits_{\tau }^{t}y_{s}ds.%
\end{array}
\label{Eq7.6}
\end{equation}%
Then%
\begin{equation}
\begin{array}{c}
dx_{t}=y_{t}dt,\ \ \ x_{\tau }=0, \\ 
\\ 
dy_{t}=\left( r-\frac{\sigma ^{2}}{2}\right) dt+\sigma dW_{t},\ \ \ y_{\tau
}=\ln \left( S_{\tau }\right) \equiv \theta .%
\end{array}
\label{Eq7.7}
\end{equation}%
The value of the option can be written as follows:%
\begin{equation}
\begin{array}{c}
V\left( \tau ,S_{\tau }\right) =e^{-r\left( t-\tau \right)
}\dint\limits_{x^{\ast }}^{\nu \infty }P\left( x_{t}\right) \left( \exp
\left( \frac{x_{t}}{t-\tau }\right) -\exp \left( \ln K\right) \right) dx_{t},%
\end{array}
\label{Eq7.8}
\end{equation}%
where%
\begin{equation}
\begin{array}{c}
x^{\ast }=\left( t-\tau \right) \ln K.%
\end{array}
\label{Eq7.9}
\end{equation}%
Since Eq. (\ref{Eq7.7}) is a special case of Eq. (\ref{Eq5.26}), we can use
Eqs (\ref{Eq5.33}) to obtain the marginal distribution for $x_{t}$, which is
a Gaussian distribution of the form:%
\begin{equation}
\begin{array}{c}
P\left( x,y\right) =\frac{\exp \left( -\frac{\left( x-p\right) ^{2}}{2\sigma
_{x}^{2}}\right) }{\sqrt{2\pi \sigma _{x}^{2}}}, \\ 
\\ 
\sigma _{x}^{2}=\frac{\sigma ^{2}\left( t-\tau \right) ^{3}}{3},\ \ \ p=\ln
\left( S_{\tau }\right) \left( t-\tau \right) +\frac{1}{2}\left( r-\frac{%
\sigma ^{2}}{2}\right) \left( t-\tau \right) ^{2}.%
\end{array}
\label{Eq7.10}
\end{equation}%
Thus,%
\begin{equation}
\begin{array}{c}
V\left( \tau ,S_{\tau }\right) =\mathcal{J}_{1}\left( \tau ,S_{\tau }\right)
-\mathcal{J}_{2}\left( \tau ,S_{\tau }\right) ,%
\end{array}
\label{Eq7.11}
\end{equation}%
where%
\begin{equation}
\begin{array}{c}
\mathcal{J}_{1}\left( \tau ,S_{\tau }\right) =e^{-r\left( t-\tau \right)
}\dint\limits_{x^{\ast }}^{\nu \infty }\frac{\exp \left( -\frac{\left(
x-p\right) ^{2}}{2\sigma _{x}^{2}}+\frac{x}{\left( t-\tau \right) }\right) }{%
\sqrt{2\pi \sigma _{x}^{2}}}dx \\ 
\\ 
=\nu e^{-\frac{1}{2}\left( r+\frac{\sigma ^{2}}{6}\right) \left( t-\tau
\right) }S_{\tau }\mathrm{N}\left( \nu \frac{\ln \left( S_{\tau }/K\right) +%
\frac{1}{2}\left( r+\frac{\sigma ^{2}}{6}\right) \left( t-\tau \right) }{%
\sqrt{\sigma ^{2}\left( t-\tau \right) /3}}\right) ,%
\end{array}
\label{Eq7.12}
\end{equation}%
\begin{equation}
\begin{array}{c}
\mathcal{J}_{2}\left( \tau ,S_{\tau }\right) =e^{-r\left( t-\tau \right)
}\dint\limits_{x^{\ast }}^{\nu \infty }\frac{\exp \left( -\frac{\left(
x-p\right) ^{2}}{2\sigma _{x}^{2}}+\ln \left( K\right) \right) }{\sqrt{2\pi
\sigma _{x}^{2}}}dx \\ 
\\ 
=\nu e^{-r\left( t-\tau \right) }K\mathrm{N}\left( \nu \frac{\ln \left(
S_{\tau }/K\right) +\frac{1}{2}\left( r-\frac{\sigma ^{2}}{2}\right) \left(
t-\tau \right) }{\sqrt{\sigma ^{2}\left( t-\tau \right) /3}}\right) .%
\end{array}
\label{Eq7.13}
\end{equation}%
Finally, we obtain a well-known formula for the price of a fixed strike
Asian option with geometric averaging:%
\begin{equation}
\begin{array}{c}
V\left( \tau ,S_{\tau }\right) =\nu e^{-\frac{1}{2}\left( r+\frac{\sigma ^{2}%
}{6}\right) \left( t-\tau \right) }S_{\tau }\mathrm{N}\left( \nu \frac{\ln
\left( S_{\tau }/K\right) +\frac{1}{2}\left( r+\frac{\sigma ^{2}}{6}\right)
\left( t-\tau \right) }{\sqrt{\sigma ^{2}\left( t-\tau \right) /3}}\right)
\\ 
\\ 
-\nu e^{-r\left( t-\tau \right) }K\mathrm{N}\left( \nu \frac{\ln \left(
S_{\tau }/K\right) +\frac{1}{2}\left( r-\frac{\sigma ^{2}}{2}\right) \left(
t-\tau \right) }{\sqrt{\sigma ^{2}\left( t-\tau \right) /3}}\right) .%
\end{array}
\label{Eq7.14}
\end{equation}%
Of course, a similar formula holds when $r,\sigma $ are time-dependent. Our
derivation, although very simple, seems to be new.

\subsection{Volatility and Variance Swaps and Swaptions\label{Sec73}}

\subsubsection{Volatility Swaps and Swaptions}

Recall that the Stein-Stein stochastic volatility model assumes that the
volatility is driven by an OU process; see \cite{Stein91}. We wish to find
Green's function associated with the following augmented SDE:%
\begin{equation}
\begin{array}{c}
dx_{t}=y_{t}dt,\ \ \ x_{\tau }=0, \\ 
\\ 
dy_{t}=\left( \hat{\chi}-\kappa y_{t}\right) dt+\hat{\varepsilon}dW_{t},\ \
\ y_{\tau }=\hat{\theta},%
\end{array}
\label{Eq7.15}
\end{equation}%
which describes the evolution of the volatility $\sigma _{t}\equiv y_{t}$
and its integral $x_{t}$. It is clear that Eqs (\ref{Eq7.15}) are identical
to Eqs (\ref{Eq5.34}).

We get the bivariate Gaussian distribution for $\left( x,y\right) $ with
covariance matrix $\mathfrak{H}$ given by Eq. (\ref{Eq5.51}), and means $%
\left( p,q\right) $ given by Eq. (\ref{Eq5.52}): 
\begin{equation}
\begin{array}{c}
p=\frac{\left( \kappa \left( t-\tau \right) -\left( 1-e^{-\kappa \left(
t-\tau \right) }\right) \right) \hat{\chi}}{\kappa ^{2}}+\frac{\left(
1-e^{-\kappa \left( t-\tau \right) }\right) \hat{\theta}}{\kappa }, \\ 
\\ 
q=\frac{\left( 1-e^{-\kappa \left( t-\tau \right) }\right) \hat{\chi}}{%
\kappa }+e^{-\kappa \left( t-\tau \right) }\hat{\theta}.%
\end{array}
\label{Eq7.16}
\end{equation}%
Since the marginal distribution of $x_{t}$ given by Eq. (\ref{Eq5.53}) is
Gaussian, the fair strike of a volatility swap with maturity $t$ is simply
the expected value of $x_{t}/\left( t-\tau \right) $:%
\begin{equation}
\begin{array}{c}
VolSwap=\frac{\hat{\chi}}{\kappa }+\left( \hat{\theta}-\frac{\hat{\chi}}{%
\kappa }\right) \Xi _{1}\left( \tau ,t\right) ,%
\end{array}
\label{Eq7.18}
\end{equation}%
where 
\begin{equation}
\begin{array}{c}
\Xi _{1}\left( \tau ,t\right) =\frac{\left( 1-e^{-\kappa \left( t-\tau
\right) }\right) }{\kappa \left( t-\tau \right) }.%
\end{array}
\label{Eq7.18a}
\end{equation}%
Here $\left[ VolSwap\right] =\left[ \hat{\chi}/\kappa \right] =\left[ \theta %
\right] =T^{-1/2}$.

Of course, one can calculate the expected value of $x_{t}/\left( t-\tau
\right) $ via more straightforward means. To this end, Eq. (\ref{Eq7.18})
can be derived directly by taking expectations of SDE (\ref{Eq7.15}).
However, as we shall see in the following subsection, Eq. (\ref{Eq5.53})
allows one to solve more interesting problems, such as calculating prices of
bonds and bond options, see below.

Moreover, by virtue of the marginal distribution $Q^{\left( x\right) }\left(
\tau ,\theta ,t,x\right) $ given by Eq. (\ref{Eq5.53}), we can price
volatility swaptions with payoffs of the form:%
\begin{equation}
\begin{array}{c}
V\left( t,x_{t}\right) =\max \left\{ \nu \left( x-\hat{\varpi}\right)
,0\right\} .%
\end{array}
\label{Eq7.18b}
\end{equation}%
The prices $V\left( \tau ,\theta \right) $ are as follows: 
\begin{equation}
\begin{array}{c}
V\left( \tau ,\theta \right) =\nu \dint\limits_{\hat{\varpi}}^{\nu \infty
}\left( x-\hat{\varpi}\right) Q^{\left( x\right) }\left( \tau ,\theta
,t,x\right) dx \\ 
\\ 
=\frac{\nu }{\sqrt{2\pi h_{0}\left( \tau ,t\right) }}\dint\limits_{\hat{%
\varpi}}^{\nu \infty }\left( x-\hat{\varpi}\right) \exp \left( \frac{\left(
x-p\left( \tau ,t\right) \right) ^{2}}{2h_{0}\left( \tau ,t\right) }\right)
dx \\ 
\\ 
=\left( \nu \left( p\left( \tau ,t\right) -\hat{\varpi}\right) \mathrm{N}%
\left( \nu \frac{\left( p\left( \tau ,t\right) -\hat{\varpi}\right) }{\sqrt{%
h_{0}\left( \tau ,t\right) }}\right) +\sqrt{h_{0}\left( \tau ,t\right) }%
\mathrm{n}\left( \frac{\left( p\left( \tau ,t\right) -\hat{\varpi}\right) }{%
\sqrt{h_{0}\left( \tau ,t\right) }}\right) \right) .%
\end{array}
\label{Eq7.18c}
\end{equation}%
It is clear that formula (\ref{Eq7.18c}) is a variant of the Bachelier
formula (\ref{Eq7.56c}); see \cite{Bachelier00}.

\subsubsection{Variance Swaps and Swaptions}

In contrast to volatility, which, despite common misconceptions, can be
negative, variance must be nonnegative. Accordingly, the easiest way to
model is it to use the augmented Feller process with $\vartheta >0$; see Eq.
(\ref{Eq6.29}).

By using Eq. (\ref{Eq6.63}), we can immediately obtain the following
expression for the fair value of a variance swap for the Feller process:%
\begin{equation}
\begin{array}{c}
VarSwap=\frac{\chi }{\kappa }+\left( \theta -\frac{\chi }{\kappa }\right)
\Xi _{1}\left( \tau ,t\right) ,%
\end{array}
\label{Eq7.19}
\end{equation}%
where $\left[ VarSwap\right] =\left[ \chi /\kappa \right] =\left[ \theta %
\right] =T^{-1}$. We use hats in this Section to avoid confusing parameters
for the OU and Feller processes. Formulas (\ref{Eq7.18}), (\ref{Eq7.19})
look the same but deal with the volatility and variance, respectively, so
the corresponding parameters have different meanings.

Alternatively, we can use the augmented OU process, see Eqs (\ref{Eq6.110}).
Averaging away stochastic terms, we get the following formula for the fair
price of the variance swap:%
\begin{equation}
\begin{array}{c}
VarSwap=\frac{\hat{\chi}^{2}}{\kappa ^{2}}+\frac{2\hat{\chi}\left( \kappa 
\hat{\theta}-\hat{\chi}\right) }{\kappa ^{2}}\Xi _{1}\left( \tau ,t\right) +%
\frac{\left( \kappa \hat{\theta}-\hat{\chi}\right) ^{2}}{\kappa ^{2}}\Xi
_{2}\left( \tau ,t\right) ,%
\end{array}
\label{Eq7.19a}
\end{equation}%
where 
\begin{equation}
\begin{array}{c}
\Xi _{2}\left( \tau ,t\right) =\frac{\left( 1-e^{-2\kappa \left( t-\tau
\right) }\right) }{2\kappa \left( t-\tau \right) }.%
\end{array}
\label{Eq7.19b}
\end{equation}%
We see that Eqs (\ref{Eq7.19}), (\ref{Eq7.19a}) provide different
expressions for the fair value of a variance swap, although these
expressions asymptotically agree. This fact reflects the so-called model
risk - by using different models, one gets different answers to the same
question.

By virtue of Eq. (\ref{Eq6.59}), we can calculate the price of a variance
swaption:%
\begin{equation}
\begin{array}{c}
V\left( \tau ,\theta \right) =\frac{1}{2\pi }\dint\limits_{\varpi }^{\nu
\infty }\dint\limits_{-\infty }^{\infty }\nu \left( x-\varpi \right)
\digamma \left( t,k\right) e^{ikx}dkdx \\ 
\\ 
=\frac{1}{2\pi }\dint\limits_{-\infty }^{\infty }\digamma \left( t,k\right)
\left( \nu \dint\limits_{\varpi }^{\nu \infty }\left( x-\varpi \right)
e^{ikx}dx\right) dk \\ 
\\ 
=\frac{1}{2\pi }\underset{\epsilon \rightarrow 0}{\lim }\dint\limits_{-%
\infty }^{\infty }\digamma \left( t,k\right) e^{ik\varpi }\left( -\frac{%
\partial }{\partial \epsilon }\dint\limits_{\varpi }^{\nu \infty }e^{\left(
ik-\nu \epsilon \right) x}dx\right) \\ 
\\ 
=\frac{1}{2\pi }\underset{\epsilon \rightarrow 0}{\lim }\dint\limits_{-%
\infty }^{\infty }\frac{\digamma \left( t,k\right) e^{ik\varpi }}{\left(
ik-\nu \epsilon \right) ^{2}}dk,%
\end{array}
\label{Eq7.19c}
\end{equation}%
where $\digamma \left( t,k\right) $ is given by Eq. (\ref{Eq6.60}).

For a long time, variance and volatility swaps occupied a niche in the
financial products universe. Recently, they received a somewhat unexpected
boost from cryptocurrency trading. It turned out that these swaps are
beneficial for hedging the so-called impermanent loss generated by automated
market makers; see \cite{Lipton21, Lipton22, Cartea23, Fukusawa23}, among
others.

\subsection{Path-Dependent Volatility Model\label{Sec74}}

\cite{Hobson98} initially proposed path-dependent volatility models;
subsequently, they were studied by many authors; see \cite{Davis04,
DiFrancesco04, DiFrancesco05, Lipton23} among others. They present a viable
alternative to the more popular local volatility modes developed by \cite%
{Bick93, Derman94, Dupire94, Rubinstein94}.

The main advantage of path-dependent volatility models compared to their
local volatility brethren is that the former deal with volatility functions
depending on a non-dimensional argument, such as $S_{t}/A_{t}$, where $S_{t}$
is the stock price, and $A_{t}$ is its average, say, $\sigma =\sigma \left(
S_{t}/A_{t}\right) $, while the latter use volatilities depending on a
dimensional argument $S_{t}$, $\sigma =\sigma \left( S_{t}\right) $, which
is conceptually unsound and results in model dynamics deviating from the one
observed in the market. The problem with path-dependent models is that it is
very hard to build an analytically tractable version of the model, which
complicates gaining the necessary intuition or benchmarking numerical
solution. In this Section, we develop such a model using results derived in\
Section \ref{Sec63}.

The model we propose is the first one to have a semi-analytical solution.
The dynamics is adapted from Section \ref{Sec63}, Eq. (\ref{Eq6.93}) as
follows:%
\begin{equation}
\begin{array}{c}
\frac{dS_{t}}{S_{t}}=\sqrt{c_{0}+c_{1}\ln \left( \frac{S_{t}}{A_{t}}\right) }%
dW_{t},\ \ \ S_{\tau }=s, \\ 
\\ 
A_{t}=\exp \left( \varkappa \int_{-\infty }^{t}e^{-\varkappa \left(
t-t^{\prime }\right) }\ln S_{t^{\prime }}dt^{\prime }\right) ,\ \ \ A_{\tau
}=a.%
\end{array}
\label{Eq7.69}
\end{equation}%
We do not describe in detail how $S_{t}$, and, hence, $A_{t}$, behave when $%
t<\tau $, since it becomes unimportant provided that $\varkappa \left(
t-\tau \right) $ is sufficiently large. For instance, we can assume that $%
S_{t}\equiv s$, when $t<\tau $, then $a=s$. In Eq. (\ref{Eq7.69}), we use
the same type of arithmetic averaging as in Eq. (\ref{Eq7.3}).

In logarithmic variables $x_{t}=\ln \left( S_{t}\right) $, $y_{t}=\ln \left(
A_{t}\right) $, Eqs (\ref{Eq7.69}) assume the form given by Eqs (\ref{Eq6.93}%
). Accordingly, the t.p.d.f. has the form (\ref{Eq5.7}) with \thinspace $H$
given by Eq (\ref{Eq6.108}). Thus, we can price European options by
calculating explicit integrals. Several useful shortcuts for calculating the
corresponding integrals will be presented elsewhere due to the lack of space.

\subsection{Bonds and Bond Options\label{Sec75}}

\subsubsection{Vasicek Model}

We can use formulas derived in the previous subsection to price bonds and
bond options in the popular Vasicek and Hull-White models; see \cite%
{Vasicek77, Hull90}. Recall that Vasicek postulated the following dynamics
for the short interest rate $y_{t}$:%
\begin{equation}
\begin{array}{c}
dy_{t}=\left( \chi -\kappa y_{t}\right) dt+\varepsilon dW_{t},\ \ \ y_{\tau
}=\theta ,%
\end{array}
\label{Eq7.21}
\end{equation}%
alternatively,%
\begin{equation}
\begin{array}{c}
dy_{t}=\kappa \left( \zeta -y_{t}\right) dt+\varepsilon dW_{t},\ \ \ y_{\tau
}=\theta ,%
\end{array}
\label{Eq7.21a}
\end{equation}%
where $\kappa \zeta =\chi $.

At time $\tau $, the price of a bond maturing at time $t$, which we denote
by $\mathtt{Z}\left( \tau ,\theta ,t\right) $, boils down to solving the
following classical backward problem:%
\begin{equation}
\begin{array}{c}
\mathtt{Z}_{\tau }\left( \tau ,\theta ,t\right) +\left( \chi -\kappa \theta
\right) \mathtt{Z}_{\theta }\left( \tau ,\theta ,t\right) +\frac{1}{2}%
\varepsilon ^{2}\mathtt{Z}_{\theta \theta }\left( \tau ,\theta ,t\right)
-\theta \mathtt{Z}\left( \tau ,\theta ,t\right) =0, \\ 
\\ 
\mathtt{Z}\left( t,\theta ,t\right) =1.%
\end{array}
\label{Eq7.22}
\end{equation}%
The standard affine ansatz yields%
\begin{equation}
\begin{array}{c}
\mathtt{Z}\left( \tau ,\theta ,t\right) =\exp \left( \mathtt{A}\left( \tau
,t\right) -\mathtt{B}\left( \tau ,t\right) \theta \right) \\ 
\\ 
\mathtt{B}\left( \tau ,t\right) =\frac{1-e^{-\kappa \left( t-\tau \right) }}{%
\kappa }, \\ 
\\ 
\mathtt{A}\left( \tau ,t\right) =\left( \frac{\chi }{\kappa }-\frac{%
\varepsilon ^{2}}{2\kappa ^{2}}\right) \left( \mathtt{B}\left( \tau
,t\right) -\left( t-\tau \right) \right) -\frac{\varepsilon ^{2}}{4\kappa }%
\mathtt{B}^{2}\left( \tau ,t\right) \\ 
\\ 
=\left( \frac{1-e^{-\kappa \left( t-\tau \right) }}{\kappa }-\left( t-\tau
\right) \right) \frac{\chi }{\kappa }-\frac{3\varepsilon ^{2}}{4\kappa ^{3}}+%
\frac{\varepsilon ^{2}}{\kappa ^{3}}e^{-\kappa \left( t-\tau \right) }-\frac{%
\varepsilon ^{2}}{4\kappa ^{3}}e^{-2\kappa \left( t-\tau \right) }+\frac{%
\varepsilon ^{2}}{2\kappa ^{2}}\left( t-\tau \right) \\ 
\\ 
=\left( \frac{1-e^{-\kappa \left( t-\tau \right) }}{\kappa }-\left( t-\tau
\right) \right) \frac{\chi }{\kappa }+\frac{h_{0}\left( \tau ,t\right) }{2}.%
\end{array}
\label{Eq7.23}
\end{equation}

We can use formulae derived in the previous section to come up with an
alternative derivation. Introduce $x_{t}=\int_{\tau }^{t}y_{s}ds$. The
distribution of $\left( x_{t},y_{t}\right) $ is given by Eq. (\ref{Eq5.23})
with the covariance matrix $\mathfrak{H}$, given by Eq. (\ref{Eq5.51}) and
the expected value $\left( p,q\right) ^{T}$ given by Eq. (\ref{Eq5.52}).
Accordingly, the price of a bond can be written as follows:%
\begin{equation}
\begin{array}{c}
\mathtt{Z}\left( \tau ,\theta ,t\right) =\mathbb{E}\left\{
e^{-x_{t}}\right\} =\frac{1}{\sqrt{2\pi h_{0}}}\dint\limits_{-\infty
}^{\infty }e^{-x-\frac{\left( x-p\left( \tau ,t\right) \right) ^{2}}{%
2h_{0}\left( \tau ,t\right) }}dx \\ 
\\ 
=e^{-p\left( \tau ,t\right) +\frac{h_{0}\left( \tau ,t\right) }{2}}=\exp
\left( \mathtt{A}\left( \tau ,t\right) -\mathtt{B}\left( \tau ,t\right)
\theta \right) ,%
\end{array}
\label{Eq7.24}
\end{equation}%
so that Eqs. (\ref{Eq7.23}) and (\ref{Eq7.24}) are in agreement.

We now show how to use the knowledge of the joint Gaussian distribution for $%
\left( x_{t},y_{t}\right) $ to price an option on zero coupon bond maturing
at time $\bar{t}>t$. The payoff of a European option with strike $K$ has the
form:%
\begin{equation}
V\left( t,y_{t}\right) =\max \left\{ \nu \left( \exp \left( \mathtt{A}\left(
t,\bar{t}\right) -\mathtt{B}\left( t,\bar{t}\right) y_{t}\right) -\exp
\left( \ln K\right) \right) ,0\right\}  \label{Eq7.26}
\end{equation}%
where, as before, $\nu =1$ for a call, and $\nu =-1$ for a put. At maturity $%
t$, the payoff is independent on $x$, however, at inception it does depend
on the realized value of $x_{\tau ,t}$. By using Eqs (\ref{Eq5.23}), (\ref%
{Eq5.51}), (\ref{Eq5.52}), we can write $V\left( \tau ,\theta \right) $
(recall that here $\xi \equiv 0$) as follows:%
\begin{equation}
\begin{array}{c}
V\left( \tau ,\theta \right) =\mathcal{J}_{1}\left( \tau ,\theta \right) -%
\mathcal{J}_{2}\left( \tau ,\theta \right) ,%
\end{array}
\label{Eq7.27}
\end{equation}%
where%
\begin{equation}
\begin{array}{c}
\mathcal{J}_{1}\left( \tau ,\theta \right) =\frac{1}{2\pi \det \left( 
\mathfrak{H}\left( \tau ,t\right) \right) ^{1/2}}\dint\limits_{-\infty
}^{\infty }\dint\limits_{-\nu \infty }^{\mathsf{y}^{\ast }}\exp \left( -%
\frac{1}{2\det \left( \mathfrak{H}\left( \tau ,t\right) \right) }\left(
h_{2}\left( \tau ,t\right) \left( x-p\left( \tau ,t\right) \right)
^{2}\right. \right. \\ 
\\ 
\left. -2h_{1}\left( \tau ,t\right) \left( x-p\left( \tau ,t\right) \right)
\left( y-q\left( \tau ,t\right) \right) +h_{0}\left( \tau ,t\right) \left(
y-q\left( \tau ,t\right) \right) ^{2}\right) \\ 
\\ 
\left. -x+\mathtt{A}\left( t,\bar{t}\right) -\mathtt{B}\left( t,\bar{t}%
\right) y\right) dxdy,%
\end{array}
\label{Eq7.28}
\end{equation}%
\begin{equation}
\begin{array}{c}
\mathcal{J}_{2}\left( \tau ,\theta \right) =\frac{1}{2\pi \det \left( 
\mathfrak{H}\left( \tau ,t\right) \right) ^{1/2}}\dint\limits_{-\infty
}^{\infty }\dint\limits_{-\nu \infty }^{\mathsf{y}^{\ast }}\exp \left( -%
\frac{1}{2\det \left( \mathfrak{H}\left( \tau ,t\right) \right) }\left(
h_{2}\left( \tau ,t\right) \left( x-p\left( \tau ,t\right) \right)
^{2}\right. \right. \\ 
\\ 
\left. -2h_{1}\left( \tau ,t\right) \left( x-p\left( \tau ,t\right) \right)
\left( y-q\left( \tau ,t\right) \right) +h_{0}\left( \tau ,t\right) \left(
y-q\left( \tau ,t\right) \right) ^{2}\right) \\ 
\\ 
\left. -x+\ln K\right) dxdy,%
\end{array}
\label{Eq7.28a}
\end{equation}%
with $h_{i}$ given by Eq. (\ref{Eq5.51}), $\det \left( \mathfrak{H}\right)
=h_{0}h_{2}-h_{1}^{2}$. Here $\mathsf{y}^{\ast }$ is defined as follows:%
\begin{equation}
\begin{array}{c}
\mathsf{y}^{\ast }=\frac{\mathtt{A}\left( t,\bar{t}\right) -\ln K}{\mathtt{B}%
\left( t,\bar{t}\right) }.%
\end{array}
\label{Eq7.29}
\end{equation}

First, we consider $\mathcal{J}_{1}$. Completing the square, we get:%
\begin{equation}
\begin{array}{c}
-\frac{1}{2\det \left( \mathfrak{H}\left( \tau ,t\right) \right) }\left(
h_{2}\left( \tau ,t\right) \left( x-p\left( \tau ,t\right) \right)
^{2}-2h_{1}\left( \tau ,t\right) \left( x-p\left( \tau ,t\right) \right)
\left( y-q\left( \tau ,t\right) \right) \right. \\ 
\\ 
\left. +h_{0}\left( \tau ,t\right) \left( y-q\left( \tau ,t\right) \right)
^{2}\right) -x+\mathtt{A}\left( t,\bar{t}\right) -\mathtt{B}\left( t,\bar{t}%
\right) y \\ 
\\ 
=-\frac{1}{2\det \left( \mathfrak{H}\left( \tau ,t\right) \right) }\left(
h_{2}\left( \tau ,t\right) \left( \left( x-p\left( \tau ,t\right) \right) -%
\frac{\left( h_{1}\left( \tau ,t\right) \left( y-q\left( \tau ,t\right)
\right) -\det \left( \mathfrak{H}\left( \tau ,t\right) \right) \right) }{%
h_{2}\left( \tau ,t\right) }\right) ^{2}\right. \\ 
\\ 
\left. -\frac{\left( h_{1}\left( \tau ,t\right) \left( y-q\left( \tau
,t\right) \right) -\det \left( \mathfrak{H}\left( \tau ,t\right) \right)
\right) ^{2}}{h_{2}\left( \tau ,t\right) }+h_{0}\left( \tau ,t\right) \left(
y-q\left( \tau ,t\right) \right) ^{2}\right) \\ 
\\ 
-\mathtt{B}\left( t,\bar{t}\right) \left( y-q\left( \tau ,t\right) \right)
-p\left( \tau ,t\right) +\mathtt{A}\left( t,\bar{t}\right) -\mathtt{B}\left(
t,\bar{t}\right) q\left( \tau ,t\right) .%
\end{array}
\label{Eq7.52}
\end{equation}%
Integrating over $x$, we obtain the following expression for $\mathcal{J}%
_{1} $:%
\begin{equation}
\begin{array}{c}
\mathcal{J}_{1}\left( \tau ,\theta \right) =\frac{e^{-p+\mathtt{A}\left( t,%
\bar{t}\right) -\mathtt{B}\left( t,\bar{t}\right) q}}{\sqrt{2\pi h_{2}}}%
\dint\limits_{-\nu \infty }^{\mathsf{y}^{\ast }}\exp \left( -\frac{1}{2\det
\left( \mathfrak{H}\left( \tau ,t\right) \right) }\left( -\frac{\left(
h_{1}\left( \tau ,t\right) \left( y-q\left( \tau ,t\right) \right) -\det
\left( \mathfrak{H}\left( \tau ,t\right) \right) \right) ^{2}}{h_{2}\left(
\tau ,t\right) }\right. \right. \\ 
\\ 
\left. \left. +h_{0}\left( \tau ,t\right) \left( y-q\left( \tau ,t\right)
\right) ^{2}+2\det \left( \mathfrak{H}\left( \tau ,t\right) \right) \mathtt{B%
}\left( t,\bar{t}\right) \left( y-q\left( \tau ,t\right) \right) \right)
\right) dy.%
\end{array}
\label{Eq7.45}
\end{equation}%
Completing the square one more time, we get:%
\begin{equation}
\begin{array}{c}
-\frac{-\frac{\left( h_{1}\left( \tau ,t\right) \left( y-q\left( \tau
,t\right) \right) -\det \left( \mathfrak{H}\left( \tau ,t\right) \right)
\right) ^{2}}{h_{2}}+h_{0}\left( \tau ,t\right) \left( y-q\left( \tau
,t\right) \right) ^{2}+2\det \left( \mathfrak{H}\left( \tau ,t\right)
\right) \mathtt{B}\left( t,\bar{t}\right) \left( y-q\left( \tau ,t\right)
\right) }{2\det \left( \mathfrak{H}\left( \tau ,t\right) \right) } \\ 
\\ 
=-\frac{\left( y-q\left( \tau ,t\right) +h_{1}\left( \tau ,t\right) +\mathtt{%
B}\left( t,\bar{t}\right) h_{2}\left( \tau ,t\right) \right) ^{2}}{%
2h_{2}\left( \tau ,t\right) }+\frac{h_{0}\left( \tau ,t\right) }{2}+\mathtt{B%
}\left( t,\bar{t}\right) h_{1}\left( \tau ,t\right) +\frac{\mathtt{B}%
^{2}\left( t,\bar{t}\right) h_{2}\left( \tau ,t\right) }{2},%
\end{array}
\label{Eq7.54}
\end{equation}%
so that%
\begin{equation}
\begin{array}{c}
\mathcal{J}_{1}\left( \tau ,\theta \right) =\frac{e^{-p\left( \tau ,t\right)
+\mathtt{A}\left( t,\bar{t}\right) \mathsf{-}\mathtt{B}\left( t,\bar{t}%
\right) q\left( \tau ,t\right) +\frac{h_{0}\left( \tau ,t\right) }{2}+%
\mathtt{B}\left( t,\bar{t}\right) h_{1}\left( \tau ,t\right) +\frac{\mathtt{B%
}^{2}\left( t,\bar{t}\right) h_{2}\left( \tau ,t\right) }{2}}}{\sqrt{2\pi
h_{2}\left( \tau ,t\right) }} \\ 
\\ 
\times \dint\limits_{-\nu \infty }^{\mathsf{y}^{\ast }}\exp \left( -\frac{%
\left( y-q\left( \tau ,t\right) +h_{1}\left( \tau ,t\right) +\mathtt{B}%
\left( t,\bar{t}\right) h_{2}\left( \tau ,t\right) \right) ^{2}}{%
2h_{2}\left( \tau ,t\right) }\right) dy \\ 
\\ 
=\nu e^{-p\left( \tau ,t\right) +\mathtt{A}\left( t,\bar{t}\right) \mathsf{-}%
\mathtt{B}\left( t,\bar{t}\right) q\left( \tau ,t\right) +\frac{h_{0}\left(
\tau ,t\right) }{2}+\mathtt{B}\left( t,\bar{t}\right) h_{1}\left( \tau
,t\right) +\frac{\mathtt{B}^{2}\left( t,\bar{t}\right) h_{2}\left( \tau
,t\right) }{2}} \\ 
\\ 
\times \mathrm{N}\left( \frac{\nu \left( \mathsf{y}^{\ast }-q+h_{1}+\mathtt{B%
}\left( t,\bar{t}\right) h_{2}\right) }{\sqrt{h_{2}}}\right) .%
\end{array}
\label{Eq7.47}
\end{equation}%
It is easy to see that $\mathtt{Z}\left( \tau ,\theta ,\bar{t}\right) $ is
given by Eq. (\ref{Eq7.47}) with $\nu =1$ and $\mathsf{y}^{\ast }=\infty $,
so that 
\begin{equation}
\begin{array}{c}
\mathtt{Z}\left( \tau ,\theta ,\bar{t}\right) =e^{-p\left( \tau ,t\right) +%
\mathtt{A}\left( t,\bar{t}\right) \mathsf{-}\mathtt{B}\left( t,\bar{t}%
\right) q\left( \tau ,t\right) +\frac{h_{0}\left( \tau ,t\right) }{2}+%
\mathtt{B}\left( t,\bar{t}\right) h_{1}\left( \tau ,t\right) +\frac{\mathtt{B%
}^{2}\left( t,\bar{t}\right) h_{2}\left( \tau ,t\right) }{2}}.%
\end{array}
\label{Eq7.55}
\end{equation}%
Thus,%
\begin{equation}
\begin{array}{c}
\mathcal{J}_{1}\left( \tau ,\theta \right) =\nu \mathtt{Z}\left( \tau
,\theta ,\bar{t}\right) \mathrm{N}\left( \frac{\nu \left( \mathtt{A}\left( t,%
\bar{t}\right) -\ln K-\mathtt{B}\left( t,\bar{t}\right) q\left( \tau
,t\right) +\mathtt{B}\left( t,\bar{t}\right) h_{1}\left( \tau ,t\right) +%
\mathtt{B}^{2}\left( t,\bar{t}\right) h_{2}\left( \tau ,t\right) \right) }{%
\sqrt{h_{2}\left( \tau ,t\right) }\mathtt{B}\left( t,\bar{t}\right) }\right)
.%
\end{array}
\label{Eq7.48}
\end{equation}%
Eq. (\ref{Eq7.55}) is the Chapman-Kolmogorov equation for bond prices in the
Vasicek model. Its direct verification is left to the reader as a useful
exercise. By using (\ref{Eq7.55}), it is easy but tedious to show that%
\begin{equation}
\begin{array}{c}
\mathcal{J}_{1}\left( \tau ,\theta \right) =\nu \mathtt{Z}\left( \tau
,\theta ,\bar{t}\right) \mathrm{N}\left( \nu \left( \frac{\ln \left( \frac{%
\mathtt{Z}\left( \tau ,\theta ,\bar{t}\right) }{\mathtt{Z}\left( \tau
,\theta ,t\right) K}\right) }{\Sigma \left( \tau ,t,\bar{t}\right) }+\frac{%
\Sigma \left( \tau ,t,\bar{t}\right) }{2}\right) \right) .%
\end{array}
\label{Eq7.49}
\end{equation}%
where%
\begin{equation}
\begin{array}{c}
\Sigma \left( \tau ,t,\bar{t}\right) =\sqrt{h_{2}\left( \tau ,t\right) }%
\mathtt{B}\left( t,\bar{t}\right) .%
\end{array}
\label{Eq7.50}
\end{equation}

Second, we consider $\mathcal{J}_{2}$. We proceed in the same way as before.
Completing the square, we get:%
\begin{equation}
\begin{array}{c}
-\frac{1}{2\det \left( \mathfrak{H}\left( \tau ,t\right) \right) }\left(
h_{2}\left( \tau ,t\right) \left( x-p\left( \tau ,t\right) \right)
^{2}\right. \\ 
\\ 
\left. -2h_{1}\left( \tau ,t\right) \left( x-p\left( \tau ,t\right) \right)
\left( y-q\left( \tau ,t\right) \right) +h_{0}\left( \tau ,t\right) \left(
y-q\left( \tau ,t\right) \right) ^{2}\right) -x+\ln K \\ 
\\ 
=-\frac{1}{2\det \left( \mathfrak{H}\left( \tau ,t\right) \right) }\left(
h_{2}\left( \tau ,t\right) \left( \left( x-p\left( \tau ,t\right) \right) -%
\frac{\left( h_{1}\left( \tau ,t\right) \left( y-q\left( \tau ,t\right)
\right) -\det \left( \mathfrak{H}\left( \tau ,t\right) \right) \right) }{%
h_{2}\left( \tau ,t\right) }\right) ^{2}\right. \\ 
\\ 
\left. -\frac{\left( h_{1}\left( \tau ,t\right) \left( y-q\left( \tau
,t\right) \right) -\det \left( \mathfrak{H}\left( \tau ,t\right) \right)
\right) ^{2}}{h_{2}\left( \tau ,t\right) }+h_{0}\left( \tau ,t\right) \left(
y-q\left( \tau ,t\right) \right) ^{2}\right) -p\left( \tau ,t\right) +\ln K.%
\end{array}
\label{Eq7.30}
\end{equation}%
Integration over $x$ yields%
\begin{equation}
\begin{array}{c}
\mathcal{J}_{2}\left( \tau ,\theta \right) =\frac{e^{-p\left( \tau ,t\right)
+\ln K}}{\sqrt{2\pi h_{2}\left( \tau ,t\right) }}\dint\limits_{-\nu \infty
}^{\mathsf{y}^{\ast }}\exp \left( -\frac{-\frac{\left( h_{1}\left( \tau
,t\right) \left( y-q\left( \tau ,t\right) \right) -\det \left( \mathfrak{H}%
\left( \tau ,t\right) \right) \right) ^{2}}{h_{2}\left( \tau ,t\right) }%
+h_{0}\left( \tau ,t\right) \left( y-q\left( \tau ,t\right) \right) ^{2}}{%
2\det \left( \mathfrak{H}\left( \tau ,t\right) \right) }\right) dy.%
\end{array}
\label{Eq7.31}
\end{equation}%
Completing the square one more time, we get:%
\begin{equation}
\begin{array}{c}
-\frac{-\frac{\left( h_{1}\left( \tau ,t\right) \left( y-q\left( \tau
,t\right) \right) -\det \left( \mathfrak{H}\left( \tau ,t\right) \right)
\right) ^{2}}{h_{2}\left( \tau ,t\right) }+h_{0}\left( \tau ,t\right) \left(
y-q\left( \tau ,t\right) \right) ^{2}}{2\det \left( \mathfrak{H}\left( \tau
,t\right) \right) } \\ 
\\ 
=-\frac{\left( y-q\left( \tau ,t\right) +h_{1}\left( \tau ,t\right) \right)
^{2}}{2h_{2}\left( \tau ,t\right) }+\frac{h_{0}\left( \tau ,t\right) }{2}.%
\end{array}
\label{Eq7.32}
\end{equation}%
Thus,%
\begin{equation}
\begin{array}{c}
\mathcal{J}_{2}\left( \tau ,\theta \right) =\nu e^{-p+\frac{h_{0}}{2}+\ln K}%
\mathrm{N}\left( \frac{\nu \left( \mathsf{y}^{\ast }-q\left( \tau ,t\right)
+h_{1}\left( \tau ,t\right) \right) }{\sqrt{h_{2}\left( \tau ,t\right) }}%
\right) \\ 
\\ 
=\nu \mathtt{Z}\left( \tau ,\theta ,t\right) K\mathrm{N}\left( \frac{\nu
\left( \mathtt{A}\left( t,\bar{t}\right) -\ln K-\mathtt{B}\left( t,\bar{t}%
\right) q\left( \tau ,t\right) +\mathtt{B}\left( t,\bar{t}\right)
h_{1}\left( \tau ,t\right) \right) }{\mathtt{B}\left( t,\bar{t}\right) \sqrt{%
h_{2}\left( \tau ,t\right) }}\right) .%
\end{array}
\label{Eq7.33}
\end{equation}%
Once again, we can represent $\mathcal{J}_{2}\left( \tau ,\theta \right) $
in a more intuitive form: 
\begin{equation}
\begin{array}{c}
\mathcal{J}_{2}\left( \tau ,\theta \right) =\nu \mathtt{Z}\left( \tau
,\theta ,t\right) \mathrm{N}\left( \nu \left( \frac{\ln \left( \frac{\mathtt{%
Z}\left( \tau ,\theta ,\bar{t}\right) }{\mathtt{Z}\left( \tau ,\theta
,t\right) K}\right) }{\Sigma \left( \tau ,t,\bar{t}\right) }-\frac{\Sigma
\left( \tau ,t,\bar{t}\right) }{2}\right) \right) .%
\end{array}
\label{Eq7.35}
\end{equation}

Finally, we arrive at the following familiar expression for the bond option
price: 
\begin{equation}
\begin{array}{c}
V\left( \tau ,\theta \right) =\nu \left( \mathtt{Z}\left( \tau ,\theta ,\bar{%
t}\right) \mathrm{N}\left( \nu \left( \frac{\ln \left( \frac{\mathtt{Z}%
\left( \tau ,\theta ,\bar{t}\right) }{\mathtt{Z}\left( \tau ,\theta
,t\right) K}\right) }{\Sigma \left( \tau ,t,\bar{t}\right) }+\frac{\Sigma
\left( \tau ,t,\bar{t}\right) }{2}\right) \right) \right. \\ 
\\ 
\left. -\mathtt{Z}\left( \tau ,\theta ,t\right) K\mathrm{N}\left( \nu \left( 
\frac{\ln \left( \frac{\mathtt{Z}\left( \tau ,\theta ,\bar{t}\right) }{%
\mathtt{Z}\left( \tau ,\theta ,t\right) K}\right) }{\Sigma \left( \tau ,t,%
\bar{t}\right) }-\frac{\Sigma \left( \tau ,t,\bar{t}\right) }{2}\right)
\right) \right) .%
\end{array}
\label{Eq7.38}
\end{equation}

\subsubsection{CIR\ Model}

The CIR model postulates that the short rate follows the Feller process; see 
\cite{Cox85}. Accordingly, the bond price can be calculated by using Eq. (%
\ref{Eq6.59}) with $\xi =0$, and $k=-i$:%
\begin{equation}
\begin{array}{c}
B=\int_{-\infty }^{\infty }Q\left( t,x\right) e^{-x}dx=\frac{1}{2\pi }%
\int_{-\infty }^{\infty }\int_{-\infty }^{\infty }\digamma \left( t,k\right)
e^{\left( ik-1\right) x}dkdx=\digamma \left( t,-i\right) ,%
\end{array}
\label{Eq7.39}
\end{equation}%
where%
\begin{equation}
\begin{array}{c}
\digamma \left( t,-i\right) =\exp \left( \frac{2\chi \mu \left( t-\tau
\right) }{\varepsilon ^{2}}+\frac{2\chi }{\varepsilon ^{2}}\ln \left( \frac{%
2\zeta }{\lambda _{+}\mathsf{E}_{+}-\lambda _{-}\mathsf{E}_{-}}\right)
\right. \\ 
\\ 
\left. +\frac{2\lambda _{+}\lambda _{-}\left( \mathsf{E}_{+}-\mathsf{E}%
_{-}\right) \theta }{\varepsilon ^{2}\left( \lambda _{+}\mathsf{E}%
_{+}-\lambda _{-}\mathsf{E}_{-}\right) }\right) ,%
\end{array}
\label{Eq7.40}
\end{equation}%
with%
\begin{equation}
\begin{array}{c}
\mu =\frac{\kappa }{2},\ \ \ \zeta =\frac{\sqrt{\kappa ^{2}+2\varepsilon ^{2}%
}}{2}.%
\end{array}
\label{Eq7.41}
\end{equation}%
Thus,%
\begin{equation}
\begin{array}{c}
\mathtt{Z}\left( \tau ,\theta ,t\right) =e^{\mathtt{A}\left( \tau ,t\right) -%
\mathtt{B}\left( \tau ,t\right) \theta },%
\end{array}
\label{Eq7.42}
\end{equation}%
where%
\begin{equation}
\begin{array}{c}
\mathtt{A}\left( \tau ,t\right) =\frac{\chi \kappa \left( t-\tau \right) }{%
\varepsilon ^{2}}+\frac{2\chi }{\varepsilon ^{2}}\ln \left( \frac{2\zeta 
\mathsf{E}_{+}}{\left( \lambda _{+}\mathsf{E}_{+}-\lambda _{-}\mathsf{E}%
_{-}\right) }\right) , \\ 
\\ 
\mathtt{B}\left( \tau ,t\right) =\frac{\left( \mathsf{E}_{+}-\mathsf{E}%
_{-}\right) }{\left( \lambda _{+}\mathsf{E}_{+}-\lambda _{-}\mathsf{E}%
_{-}\right) },%
\end{array}
\label{Eq7.43}
\end{equation}%
which coincides with the standard expressions given by \cite{Cox85}.

\subsection{Further Extensions\label{Sec76}}

Formulas derived in Sections \ref{Sec5}, \ref{Sec6} can be used to solve
numerous problems of financial engineering within a consistent framework
based on Kevin waves. For instance, these formulas allow one to calculate
t.p.d.fs and price options in Heston, Stein \& Stein models, expand these
models by incorporating jumps, stochastic interest rates, and default
intensities driven by OU or Feller processes, describe risky bonds, and
solve several other complex but practically significant problems. Some
features of proposed mathematical formalism can be extended to other
processes, which become affine after suitable changes of variables.

Our novel mathematical tools and techniques developed using Kelvin waves can
be adapted and applied to studying mean-reverting trading strategies in
financial markets; see \cite{Lipton20}. We shall report the corresponding
results elsewhere.

\section{Conclusions\label{Sec8}}

In this paper, we have developed a unified approach to finding t.p.d.fs for
affine processes via the integral representation based on Kelvin waves. Our
methodology transcends disciplinary boundaries and reveals deep connections
between problems in hydrodynamics, molecular physics, stochastic processes,
and financial engineering.

Some of these problems are degenerate because they have more independent
variables than sources of uncertainty, and others are not. Our method can
handle both situations equally well.

In particular, we discovered an unexpected connection between the Langevin
equation for underdamped Brownian motion and the vorticity equation for
two-dimensional flows of a viscous incompressible fluid.

We used the Kelvin wave expansions to solve many essential financial
problems, such as pricing of Asian options with geometric averaging, finding
an analytically tractable model for processes with path-dependent
volatility, deriving convenient expressions for t.p.d.fs for processes with
stochastic volatility, and finding prices of bonds and bond options by
augmenting the process for the short rate with a process for its integral.

The method developed in the paper can address problems with varying degrees
of complexity and contributes to the understanding and modeling of
stochastic systems in diverse fields. It can be expanded in several
directions, mainly by studying jump-diffusion models or, if desired, by
considering general affine pseudo-differential equations.

\begin{acknowledgement}
We are grateful to our ADIA\ colleagues Majed Alromaithi, Marcos Lopez de
Prado, Koushik Balasubramanian, Andrey Itkin, Oleksiy Kondratiev, Dmitry
Muravey, Adil Reghai, other Q-team colleagues, and last but not least,
Marsha Lipton for her encouragement and council.
\end{acknowledgement}

\appendix

\section{Killed Gaussian Process\label{AppA}}

We no longer assume that the governing system of SDEs is degenerate,
although we do not exclude such possibility and write the governing system
of SDEs as follows:%
\begin{equation}
\begin{array}{c}
d\mathbf{z}_{t}=\left( \mathbf{b}+\mathfrak{B}\mathbf{z}_{t}\right) dt+%
\mathbf{\Sigma }d\mathbf{W}_{t},\ \ \ \mathbf{z}_{\tau }=\mathbf{\zeta }.%
\end{array}
\label{EqA.1}
\end{equation}%
Below, we assume that the corresponding coefficients are time-dependent. We
also assume that the process is killed with intensity $\iota $ linearly
depending of $\mathbf{z}$, namely, 
\begin{equation}
\begin{array}{c}
\iota =c+\mathbf{c}\cdot \mathbf{z},%
\end{array}
\label{EqA.2}
\end{equation}%
where $c$ is a scalar, and $\mathbf{c}^{\left( z\right) }$ is a $\left(
I\times 1\right) $ column vector.

The Fokker-Plank equation for a killed process assuming an undefined or
"killed" state at some random time has the form:%
\begin{equation}
\begin{array}{c}
P_{t}\left( \tau ,\mathbf{\zeta ,}t,\mathbf{z}\right) -\frac{1}{2}\dsum
\dsum \mathfrak{A}P_{\mathbf{zz}}\left( \tau ,\mathbf{\zeta ,}t,\mathbf{z}%
\right) \\ 
\\ 
+\left( \mathbf{b}+\mathfrak{B}\mathbf{z}\right) \cdot P_{\mathbf{z}}\left(
\tau ,\mathbf{\zeta ,}t,\mathbf{z}\right) +\left( b+c+\mathbf{c\cdot z}%
\right) P\left( \tau ,\mathbf{\zeta ,}t,\mathbf{z}\right) =0, \\ 
\\ 
P\left( \tau ,\mathbf{z},\tau ,\mathbf{\zeta }\right) =\delta \left( \mathbf{%
z}-\mathbf{\zeta }\right) .%
\end{array}
\label{EqA.3}
\end{equation}%
Here $\mathfrak{A}$ is the covariance matrix, 
\begin{equation}
\begin{array}{c}
\mathfrak{A}=\left( a_{nn^{\prime }}\right) =\sum\limits_{k=1}^{N}\sigma
_{nk}\sigma _{n^{\prime }k}=\mathbf{\Sigma \Sigma }^{T},%
\end{array}
\label{EqA.4}
\end{equation}%
and $b=\mathrm{Tr}\left( \mathfrak{B}\right) $.

Explicitly,%
\begin{equation}
\begin{array}{c}
P_{t}-\frac{1}{2}\sum\limits_{n=1}^{N}\sum\limits_{n^{\prime
}=1}^{N}a_{nn^{\prime }}P_{x_{n}x_{n^{\prime }}}+\sum\limits_{n=1}^{N}\left(
b_{n}+\sum\limits_{n=1}^{N}b_{nn^{\prime }}x_{n^{\prime }}\right) P_{x_{n}}
\\ 
\\ 
+\left( b+c+\sum\limits_{n=1}^{N}c_{n}^{\left( x\right) }x_{n}\right) P=0,%
\end{array}
\label{EqA.5}
\end{equation}

We use the familiar Kelvin ansatz and represent $P$ in the form:%
\begin{equation}
\begin{array}{c}
P\left( t,\mathbf{z}\right) =\frac{1}{\left( 2\pi \right) ^{I}}\int_{-\infty
}^{\infty }...\int_{-\infty }^{\infty }H\left( t,\mathbf{z},\mathbf{m}%
\right) d\mathbf{m}, \\ 
\\ 
H\left( t,\mathbf{z},\mathbf{m}\right) =\exp \left( \Psi \left( t,\mathbf{z},%
\mathbf{m}\right) \right) , \\ 
\\ 
\Psi \left( t,\mathbf{z},\mathbf{m}\right) =\alpha \left( t\right) +i\mathbf{%
\Upsilon }\left( t\right) \cdot \mathbf{z}-i\mathbf{m}\cdot \mathbf{\zeta },%
\end{array}
\label{EqA.6}
\end{equation}%
where $\mathbf{m}\mathcal{=}\left( m_{i}\right) $ is $\left( I\times
1\right) $ column vector, $\mathbf{\Upsilon }=\left( \upsilon _{i}\right) ~$%
is $\left( I\times 1\right) $ column vector, and%
\begin{equation}
\begin{array}{c}
\alpha \left( \tau \right) =0,\ \ \ \mathbf{\Upsilon }\left( \tau \right) =%
\mathbf{m}.%
\end{array}
\label{EqA.7}
\end{equation}%
As before:%
\begin{equation}
\begin{array}{c}
\frac{H_{t}}{H}=\Psi _{t}=\left( \alpha ^{\prime }\left( t\right) +i\mathbf{%
\Upsilon }^{\prime }\left( t\right) \cdot \mathbf{z}\right) , \\ 
\\ 
\frac{H_{\mathbf{z}}}{H}=\Psi _{\mathbf{z}}=i\mathbf{\Upsilon }\left(
t\right) , \\ 
\\ 
\ \frac{H_{\mathbf{zz}}}{H}=\Psi _{\mathbf{z}}^{2}=-\mathbf{\Upsilon }\left(
t\right) \mathbf{\Upsilon }^{T}\left( t\right) .%
\end{array}
\label{EqA.8}
\end{equation}%
The coupled equations for $\alpha ,\mathbf{\Upsilon }$ have the form:%
\begin{equation}
\begin{array}{c}
\alpha ^{\prime }\left( t\right) +i\mathbf{\Upsilon }^{\prime }\left(
t\right) \cdot \mathbf{z}+\frac{1}{2}\mathbf{\Upsilon }\left( t\right) \cdot 
\mathfrak{A}\mathbf{\Upsilon }\left( t\right) +i\mathbf{\Upsilon }\left(
t\right) \cdot \left( \mathbf{b}+\mathfrak{B}\mathbf{z}\right) +b+c+\mathbf{c%
}\cdot \mathbf{z}=0.%
\end{array}
\label{EqA.9}
\end{equation}%
Accordingly,%
\begin{equation}
\begin{array}{c}
\alpha ^{\prime }\left( t\right) +\frac{1}{2}\mathbf{\Upsilon }\left(
t\right) \cdot \mathfrak{A}\mathbf{\Upsilon }\left( t\right) +i\mathbf{%
\Upsilon }\left( t\right) \cdot \mathbf{b}+b+c=0,\ \ \ \alpha \left( \tau
\right) =0,%
\end{array}
\label{EqA.10}
\end{equation}%
and%
\begin{equation}
\begin{array}{c}
\mathbf{\Upsilon }^{\prime }\left( t\right) +\mathfrak{B}^{T}\mathbf{%
\Upsilon }\left( t\right) -i\mathbf{c}=0,\ \ \ \mathbf{\Upsilon }\left( \tau
\right) =\mathbf{m}.%
\end{array}
\label{EqA.11}
\end{equation}%
Let $\mathfrak{L}\left( t\right) $ is the fundamental solution of the
homogeneous ODE, i.e., the matrix such that%
\begin{equation}
\begin{array}{c}
\mathbf{\mathfrak{L}}^{\prime }\left( \tau ,t\right) +\mathfrak{B}^{T}%
\mathbf{\mathfrak{L}}\left( \tau ,t\right) =0,\ \ \ \mathbf{\mathfrak{L}}%
\left( \tau ,\tau \right) =\mathfrak{I},%
\end{array}
\label{EqA.12}
\end{equation}%
The solution of Eq. (\ref{EqA.11}) has the form:%
\begin{equation}
\begin{array}{c}
\mathbf{\Upsilon }\left( t\right) =\mathbf{\mathfrak{L}}\left( \tau
,t\right) \mathbf{m}+i\mathfrak{L}\left( \tau ,t\right) \int_{\tau }^{t}%
\mathbf{\mathfrak{L}}^{-1}\left( \tau ,s\right) \mathbf{c}\left( s\right)
ds\equiv \mathbf{\mathfrak{L}}\left( \tau ,t\right) \left( \mathbf{m}+i%
\mathbf{e}\left( \tau ,t\right) \right) , \\ 
\\ 
\mathbf{e}\left( \tau ,t\right) =\int_{\tau }^{t}\mathbf{\mathfrak{L}}%
^{-1}\left( \tau ,s\right) \mathbf{c}\left( s\right) ds.%
\end{array}
\label{EqA.13}
\end{equation}%
Thus,%
\begin{equation}
\begin{array}{c}
\alpha =-\frac{1}{2}\mathbf{m}\cdot \mathfrak{C}^{-1}\mathbf{m}-i\mathbf{%
m\cdot d}-\varpi ,%
\end{array}
\label{EqA.14}
\end{equation}%
where $\mathfrak{C}^{-1}$ is an $I\times I$ positive-definite matrix of the
form:%
\begin{equation}
\begin{array}{c}
\mathfrak{C}^{-1}\left( \tau ,t\right) =\int_{\tau }^{t}\mathfrak{L}%
^{T}\left( \tau ,s\right) \mathfrak{A}\left( s\right) \mathfrak{L}\left(
\tau ,s\right) ds,%
\end{array}
\label{EqA.15}
\end{equation}%
while $\mathbf{d}$ is an $\left( I\times 1\right) $ column vector,%
\begin{equation}
\begin{array}{c}
\mathbf{d}\left( \tau ,t\right) =\int_{\tau }^{t}\mathfrak{L}^{T}\left( \tau
,s\right) \left( \mathbf{b}\left( s\right) +\mathfrak{A}\left( s\right) 
\mathfrak{L}\left( \tau ,s\right) \mathbf{e}\left( s\right) \right) ds,%
\end{array}
\label{EqA.16}
\end{equation}%
and $\varpi =\varpi _{0}+\varpi _{1}$ is a scalar,%
\begin{equation}
\begin{array}{c}
\varpi _{0}\left( \tau ,t\right) =\int_{\tau }^{t}b\left( s\right) ds, \\ 
\\ 
\varpi _{1}\left( \tau ,t\right) =\int_{\tau }^{t}\left( c\left( s\right) -%
\frac{1}{2}\mathbf{e}\left( \tau ,s\right) \cdot \mathfrak{L}^{T}\left( \tau
,s\right) \mathfrak{A}\left( s\right) \mathfrak{L}\left( \tau ,s\right) 
\mathbf{e}\left( s\right) -\mathbf{e}\left( \tau ,s\right) \cdot \mathfrak{L}%
^{T}\left( \tau ,s\right) \mathbf{b}\left( s\right) \right) ds.%
\end{array}
\label{EqA.17}
\end{equation}%
Accordingly,%
\begin{equation}
\begin{array}{c}
\Psi \left( t,\mathbf{z},\mathbf{m}\right) =-\frac{1}{2}\mathbf{m}\cdot 
\mathfrak{C}^{-1}\left( \tau ,t\right) \mathbf{m}+i\mathbf{m}\cdot \left( 
\mathfrak{L}^{T}\left( \tau ,t\right) \mathbf{z}-\mathbf{d}\left( \tau
,t\right) -\mathbf{\zeta }\right) \\ 
\\ 
-\mathfrak{L}\left( \tau ,t\right) \mathbf{e}\left( \tau ,t\right) \cdot 
\mathbf{z}-\varpi \left( \tau ,t\right) .%
\end{array}
\label{EqA.18}
\end{equation}%
Thus, 
\begin{equation}
\begin{array}{c}
P\left( \tau ,\mathbf{\zeta },\mathbf{\theta ,}t,\mathbf{z},\mathbf{y}%
\right) =\frac{\det \left( \mathfrak{C}\left( \tau ,t\right) \right)
^{1/2}\exp \left( -\mathfrak{L}\left( \tau ,t\right) \mathbf{e}\left( \tau
,t\right) \cdot \mathbf{z}-\varpi _{0}\left( \tau ,t\right) -\varpi
_{1}\left( \tau ,t\right) \right) }{\left( 2\pi \right) ^{I/2}} \\ 
\\ 
\times \int_{-\infty }^{\infty }...\int_{-\infty }^{\infty }G\left( t,%
\mathbf{m}\right) \exp \left( i\mathbf{m}\cdot \left( \mathfrak{L}^{T}\left(
\tau ,t\right) \mathbf{z}-\mathbf{d}\left( \tau ,t\right) -\mathbf{\zeta }%
\right) \right) d\mathbf{m},%
\end{array}
\label{EqA.19}
\end{equation}%
where $G\left( t,\mathbf{m}\right) $ is the density of a multivariate
Gaussian distribution in the $\mathbf{m}$-space. It is clear that $P\left(
\tau ,\mathbf{\zeta ,}t,\mathbf{z}\right) $ is proportional to the
characteristic function of $G$ evaluated at the point $\left( \mathfrak{L}%
^{T}\left( \tau ,t\right) \mathbf{z}-\mathbf{d}\left( \tau ,t\right) -%
\mathbf{\zeta }\right) $,so that%
\begin{equation}
\begin{array}{c}
P\left( \tau ,\mathbf{\zeta ,}t,\mathbf{z}\right) =\frac{\det \left( 
\mathfrak{C}\left( \tau ,t\right) \right) ^{1/2}\exp \left( -\mathfrak{L}%
\left( \tau ,t\right) \mathbf{e}\left( \tau ,t\right) \cdot \mathbf{z}%
-\varpi _{0}\left( \tau ,t\right) -\varpi _{1}\left( \tau ,t\right) \right) 
}{\left( 2\pi \right) ^{I/2}} \\ 
\\ 
\times \exp \left( -\frac{1}{2}\left( \mathfrak{L}^{T}\left( \tau ,t\right) 
\mathbf{z}-\mathbf{d}\left( \tau ,t\right) -\mathbf{\zeta }\right) \mathcal{%
\cdot }\mathfrak{C}\left( \tau ,t\right) \left( \mathfrak{L}^{T}\left( \tau
,t\right) \mathbf{z}-\mathbf{d}\left( \tau ,t\right) -\mathbf{\zeta }\right)
\right) .%
\end{array}
\label{EqA.20}
\end{equation}%
We can present (\ref{EqA.20}) it the form:%
\begin{equation}
\begin{array}{c}
P\left( \tau ,\mathbf{\zeta ,}t,\mathbf{z}\right) =Q\left( t,\mathbf{z},\tau
\right) \mathrm{N}\left( \mathbf{q}\left( \tau ,t\right) \mathbf{,}\mathfrak{%
H}\left( \tau ,t\right) \right) ,%
\end{array}
\label{EqA.21}
\end{equation}%
where%
\begin{equation}
\begin{array}{c}
\mathfrak{H}\left( \tau ,t\right) =\left( \mathfrak{L}^{T}\left( \tau
,t\right) \right) ^{-1}\mathfrak{C}^{-1}\left( \tau ,t\right) \mathfrak{L}%
^{-1}\left( \tau ,t\right) , \\ 
\\ 
\mathbf{q}\left( \tau ,t\right) =\left( \mathfrak{L}^{T}\left( \tau
,t\right) \right) ^{-1}\left( \mathbf{d}\left( \tau ,t\right) +\mathbf{\zeta 
}\right) , \\ 
\\ 
Q\left( t,\mathbf{z},\tau \right) =\exp \left( -\mathfrak{L}\left( \tau
,t\right) \mathbf{e}\left( \tau ,t\right) \cdot \mathbf{z}-\varpi _{1}\left(
\tau ,t\right) \right) .%
\end{array}
\label{EqA.22}
\end{equation}%
As could be expected, the probability is no longer conserved due to a
prefactor $Q$, reflecting the fact that the process is killed with intensity 
$\iota $, is present. It is interesting to note that $Q$ depends on $\mathbf{%
z}$, but does not depend on $\mathbf{\zeta }$. Completing the square, we can
represent $P$ is the form:

\begin{equation}
\begin{array}{c}
P\left( \tau ,\mathbf{\zeta ,}t,\mathbf{z}\right) =R\left( \tau ,\mathbf{%
\zeta ,}t\right) \mathrm{N}\left( \mathbf{r}\left( \tau ,t\right) \mathbf{,}%
\mathfrak{H}\left( \tau ,t\right) \right) ,%
\end{array}
\label{EqA.23}
\end{equation}%
where 
\begin{equation}
\begin{array}{c}
\mathbf{r}\left( \tau ,t\right) =\left( \mathfrak{L}^{T}\left( \tau
,t\right) \right) ^{-1}\left( \mathbf{d}\left( \tau ,t\right) +\mathbf{\zeta
-}\mathfrak{C}^{-1}\left( \tau ,t\right) \mathbf{e}\left( \tau ,t\right)
\right) \\ 
\\ 
=\mathbf{q}\left( \tau ,t\right) -\mathfrak{H}\left( \tau ,t\right) 
\mathfrak{L}\left( \tau ,t\right) \mathbf{e}\left( \tau ,t\right) , \\ 
\\ 
R\left( \tau ,\mathbf{\zeta ,}t\right) =\exp \left( -\mathbf{e}\left( \tau
,t\right) \cdot \left( \mathbf{d}\left( \tau ,t\right) +\mathbf{\zeta }%
\right) +\frac{1}{2}\mathbf{e}\left( \tau ,t\right) \cdot \mathfrak{C}%
^{-1}\left( \tau ,t\right) \mathbf{e}\left( \tau ,t\right) -\varpi
_{1}\left( \tau ,t\right) \right) .%
\end{array}
\label{EqA.24}
\end{equation}%
It is clear that $R$ depends on $\mathbf{\zeta }$, but does not depend on $%
\mathbf{z}$.

As an example, consider the Vasicek model. The dynamics of $r$ is given by
Eq. (\ref{Eq7.21}), while Eq. (\ref{Eq7.22}), describes the bond pricing
problem. Straightforward calculation yields:%
\begin{equation}
\begin{array}{c}
\mathbf{e}\left( \tau ,t\right) =\int_{\tau }^{t}\mathbf{\mathfrak{L}}%
^{-1}\left( \tau ,s\right) \mathbf{c}\left( s\right) ds \\ 
\\ 
=\int_{\tau }^{t}\left( 
\begin{array}{cc}
1 & 0 \\ 
\frac{1-e^{-\kappa \left( s-\tau \right) }}{\kappa } & e^{-\kappa \left(
s-\tau \right) }%
\end{array}%
\right) \left( 
\begin{array}{c}
0 \\ 
1%
\end{array}%
\right) ds=\left( 
\begin{array}{c}
0 \\ 
\mathtt{B}\left( \tau ,t\right)%
\end{array}%
\right) ,%
\end{array}
\label{EqA.25}
\end{equation}%
\begin{equation}
\begin{array}{c}
\mathbf{r}\left( \tau ,t\right) =\left( 
\begin{array}{c}
\frac{\left( \left( t-\tau \right) -\mathtt{C}\left( \tau ,t\right) \right) 
}{\kappa }\chi +\xi +\mathtt{B}\left( \tau ,t\right) \left( \mathtt{C}\left(
\tau ,t\right) \chi +\theta \right) \\ 
\\ 
\mathtt{B}\left( \tau ,t\right) \chi +e^{-\kappa \left( t-\tau \right)
}\theta%
\end{array}%
\right) -\left( 
\begin{array}{c}
h_{1}\mathtt{C}\left( \tau ,t\right) \\ 
h_{2}\mathtt{C}\left( \tau ,t\right)%
\end{array}%
\right) ,%
\end{array}
\label{EqA.26}
\end{equation}%
where $\mathtt{B}\left( \tau ,t\right) $ is given by Eq. (\ref{Eq7.23})%
\begin{equation}
\begin{array}{c}
\mathtt{C}\left( \tau ,t\right) =\frac{e^{\kappa \left( t-\tau \right) }-1}{%
\kappa },%
\end{array}
\label{EqA.27}
\end{equation}%
while $h_{i}$ are given by Eq. (\ref{Eq5.51}).

Simple but very tedious calculation shows that%
\begin{equation}
\begin{array}{c}
R\left( \tau ,\mathbf{\zeta ,}t\right) =\mathtt{A}\left( 0,T\right) -\mathtt{%
B}\left( 0,T\right) \theta ,%
\end{array}
\label{EqA.30}
\end{equation}%
where $\mathtt{A,B}$ are given by Eq. (\ref{Eq7.23}). Thus, we recover the
familiar expression for the bond price.

\section{Killed Non-Gaussian Process\label{AppB}}

The non-Gaussian governing SDE has the form:%
\begin{equation}
\begin{array}{c}
d\mathbf{z}_{t}=\left( \mathbf{b}+\mathfrak{B}\mathbf{z}_{t}\right) dt+%
\mathbf{\Sigma }\left( \mathrm{diag}\left( \mathbf{d}^{\left( 0\right) }+%
\mathfrak{D}\mathbf{z}_{t}\right) \right) ^{1/2}d\mathbf{W}_{t},%
\end{array}
\label{EqB.1}
\end{equation}%
where $\mathbf{z}$, $\mathbf{b}$, $\mathbf{d}^{\left( 0\right) }$ are $%
\left( I\times 1\right) $ vectors, and $\mathbf{\Sigma }$, $\mathfrak{B}$, $%
\mathfrak{D}$ are $\left( I\times I\right) $ matrices. It is convenient to
introduce vectors $\mathbf{d}^{\left( i\right) }$ equal to the $i$-th column
of $\mathfrak{D}$. The corresponding $\left( I\times I\right) $ covariance
matrix $\mathfrak{A}$ linearly depends on $\mathbf{z}$:%
\begin{equation}
\begin{array}{c}
\mathfrak{A=}\mathbf{\Sigma }\left( \mathrm{diag}\left( \mathbf{d}+\mathfrak{%
D}\mathbf{z}_{t}\right) \right) \mathbf{\Sigma }^{T}=\mathfrak{A}%
^{0}+\dsum\limits_{i=1}^{I}z_{i}\mathfrak{A}^{i},%
\end{array}
\label{EqB.2}
\end{equation}%
where%
\begin{equation}
\begin{array}{c}
\mathfrak{A}^{0}=\mathbf{\Sigma }\mathrm{diag}\left( \mathbf{d}^{\left(
0\right) }\right) \Sigma ^{T},\ \ \ \mathfrak{A}^{i}=\mathbf{\Sigma }\mathrm{%
diag}\left( \mathbf{d}^{\left( i\right) }\right) \Sigma ^{T}.%
\end{array}
\label{EqB.3}
\end{equation}%
As in the previous appendix, we no longer restrict ourselves to the
degenerate (Klein-Kramers and Kolmogorov) case, and allow \textbf{$\Sigma $}
to be a full-rank matrix. Once again, we assume that the process is killed
with intensity $\iota $ linearly depending of $\mathbf{z}$, namely, 
\begin{equation}
\begin{array}{c}
\iota =c+\mathbf{c}\cdot \mathbf{z},%
\end{array}
\label{EqB.4}
\end{equation}%
where $c$ is a scalar, and $\mathbf{c}^{\left( z\right) }$ is a $\left(
I\times 1\right) $ column vector.

The corresponding Fokker-Plank problem has the form: 
\begin{equation}
\begin{array}{c}
P_{t}\left( \tau ,\mathbf{\zeta ,}t,\mathbf{z}\right) -\frac{1}{2}\dsum
\dsum \left( \mathfrak{A}^{0}+\dsum\limits_{i=1}^{I}z_{i}\mathfrak{A}%
^{i}\right) P_{\mathbf{zz}}\left( \tau ,\mathbf{\zeta ,}t,\mathbf{z}\right)
\\ 
\\ 
+\left( \mathbf{\hat{b}}+\mathfrak{B}\mathbf{z}_{t}\right) \cdot P_{\mathbf{z%
}}\left( \tau ,\mathbf{\zeta ,}t,\mathbf{z}\right) +\left( b+c+\mathbf{c}%
\cdot \mathbf{z}\right) P\left( \tau ,\mathbf{\zeta ,}t,\mathbf{z}\right) =0,
\\ 
\\ 
P\left( \tau ,\mathbf{z},\tau ,\mathbf{\zeta }\right) =\delta \left( \mathbf{%
z}-\mathbf{\zeta }\right) ,%
\end{array}
\label{EqB.5}
\end{equation}%
The equations for $\alpha ,\Upsilon $ generalize Eqs (\ref{Eq6.10}). They
can be written in the form:%
\begin{equation}
\begin{array}{c}
\alpha ^{\prime }\left( t\right) +\frac{1}{2}\mathbf{\Upsilon }\left(
t\right) \cdot \mathfrak{A}^{\left( 0\right) }\left( t\right) \mathbf{%
\Upsilon }\left( t\right) +i\mathbf{\Upsilon }\left( t\right) \cdot \mathbf{b%
}\left( t\right) +b\left( t\right) +c\left( t\right) =0,\ \ \ \alpha \left(
\tau \right) =0, \\ 
\\ 
i\mathbf{\Upsilon }_{i}^{\prime }\left( t\right) +\frac{1}{2}\mathbf{%
\Upsilon }\left( t\right) \cdot \mathfrak{A}^{\left( i\right) }\mathbf{%
\Upsilon }\left( t\right) +\sum\limits_{i=1}^{I}i\mathfrak{B}_{ij}\mathbf{%
\Upsilon }_{j}\left( t\right) +c_{i}=0,\ \ \ \mathbf{\Upsilon }_{i}\left(
\tau \right) =m_{i}.%
\end{array}
\label{EqB.6}
\end{equation}%
As in the case without killing, finding an analytical solution to a
multi-dimensional Riccati equation is generally impossible. However, in the
time-independent one-dimensional case, it can be done. Solution becomes
particularly simple in the special case when $\mathfrak{A}^{\left( 0\right)
}=0$. The most important case is the killed one-dimensional Feller
process.\qquad


\begin{thebibliography}{Friedlander \& Lipton-Lifschitz (2003)}
\bibitem[Aksenov (1995)]{Aksenov95} Aksenov, A.V. 1995 Symmetries of linear
partial differential equations and fundamental solutions. \textit{Dokl. Math.%
} \textbf{51 (3)}, 329--331.

\bibitem[Andersen \& Piterbarg (2007)]{Andersen07} Andersen, L.B.G. \&
Piterbarg, V. 2007 Moment explosions in stochastic volatility models. 
\textit{Finance Stoch} \textbf{11}, 29--50.

\bibitem[Bachelier (1900)]{Bachelier00} Bachelier, L. 1900 Th\'{e}orie de la
sp\'{e}culation. \textit{Annales de l'Ecole Normale Sup\'{e}rieure }\textbf{%
17}, 21-86.

\bibitem[Barrucci \textit{et al.} (2001)]{Barrucci01} Barucci, E., Polidoro,
S. \& Vespri, V. 2001 Some results on partial differential equations and
Asian options. \textit{Math. Models Methods Appl. Sci.} \textbf{11(3)},
475--497.

\bibitem[Bayly (1986)]{Bayly86} Bayly, B. J. 1986 Three-dimensional
instability of elliptical flow.\ \textit{Phys. Rev. Lett.} \textbf{57},
2160-2163.

\bibitem[Bayly \textit{et al.} (1996)]{Bayly96} Bayly, B.J., Holm, D.D. \&
Lifschitz, A., 1996. Three-dimensional stability of elliptical vortex
columns in external strain flows. \textit{Phil. Trans. R. Soc. Lond. Ser. A} 
\textbf{354(1709)}, 895-926.

\bibitem[Black \& Scholes (1973)]{Black73} Black, F. \& Scholes, M. 1973 The
pricing of options and corporate liabilities. \textit{Journal of Political
Economy }\textbf{81}, no.3, 637-659.

\bibitem[Berest (1993)]{Berest93} Berest, Yu.Yu. 1993 Group analysis of
linear differential equations in distributions and the construction of
fundamental solutions. \textit{Differ. Equations} \textbf{29 (11)},
1700--1711.

\bibitem[Bergomi (2015)]{Bergomi15} Bergomi, L. 2015 \textit{Stochastic
volatility modeling.} Boca Raton: CRC press.

\bibitem[Boyarchenko \& Levendorsky (2002)]{Boyarchenko02} Boyarchenko, S.
\& Levendorskii, S. 2002 \textit{Non-Gaussian Merton-Black-Scholes Theory}.
River Edge, NJ: World Scientific.

\bibitem[Bick \& Reisman (1993)]{Bick93} Bick, A. \& Reisman, H. 1993
Generalized implied volatility. \textit{Preprint.}

\bibitem[Bluman \& Kumei (1989).]{Bluman1989} Bluman, G. \& Kumei, S. 1989 
\textit{Symmetries and Differential Equations.} Berlin: Springer.

\bibitem[Boness (1964)]{Boness64} Boness, A. J. 1964 Elements of a theory of
a stock option value. \textit{Journal of Political Economy }\textbf{72(2)},
163-175.

\bibitem[Carr \textit{et al.} (2002)]{Carr02} Carr, P., Lipton, A. \& Madan,
D. 2002 The reduction method for valuing derivative securities. \textit{%
Working Paper, New York University.}

\bibitem[Cartea \textit{et al.} (2023)]{Cartea23} Cartea, \'{A}., Drissi, F.
\& Monga, M. 2023 Predictable Losses of Liquidity Provision in Constant
Function Markets and Concentrated Liquidity Markets. Available at SSRN
4541034.

\bibitem[Chandresekhar (1943)]{Chandrasekhar43} Chandresekhar, S. 1943
Stochastic problems in physics and astronomy. \textit{Rev. Modern Phys.} 
\textbf{15}, 1--89.

\bibitem[Chandrasekhar (1961)]{Chandrasekhar61} Chandrasekhar, S. 1961 
\textit{Hydrodynamic and hydromagnetic stability.} Oxford: Clarendon.

\bibitem[Chapman (1928)]{Chapman28} Chapman, S. 1928 On the Brownian
displacements and thermal diffusion of grains suspended in a nonuniform
fluid. \textit{Proc R Soc Lond Ser A} \textbf{119(781)}, 34--54

\bibitem[Cox \textit{et al.} (1985)]{Cox85} Cox, J. C. , Ingersoll Jr., J.
E. \& Ross, S. A. 1985 A theory of the term structure of interest rates. 
\textit{Econometrica} \textbf{53}, 385-408.

\bibitem[Craddock (2012)]{Craddock12} Craddock, M. 2012 Lie symmetry methods
for multi-dimensional parabolic PDEs and diffusions. \textit{J. Differential
Equations} \textbf{252(1)}, 56--90.

\bibitem[Craddock \& Platen (2004)]{Craddock04} Craddock, M. \& Platen, E.
2004 Symmetry group methods for fundamental solutions. \textit{J.
Differential Equations} \textbf{207(2)}, 285--302.

\bibitem[Craik \& Criminale (1986)]{Craik86} Craik, A. D. D. \& Criminale,
W. O. 1986 Evolution of wavelike disturbances in shear flows: A class of
exact solutions of the Navier--Stokes equations.\ \textit{Proc. R. Soc.
London, Ser. A} \textbf{406}, 13-26.

\bibitem[Dai \& Singleton (2000)]{Dai00} Dai, Q. \& Singleton, K. J. 2000
Specification analysis of affine term structure models. \textit{Journal of
Finance} \textbf{55}, 1943-1978.

\bibitem[Davis (2004)]{Davis04} Davis, M. H. A. 2004 Complete-market models
of stochastic volatility. \textit{Proc. R. Soc. Lond. Ser. A} \textbf{460},
11--26.

\bibitem[Derman \& Kani (1994)]{Derman94} Derman, E. \& Kani, I. 1994 Riding
on a smile. \textit{Risk Magazine }\textbf{7(2)}, 32-39.

\bibitem[Devreese \textit{et al.} (2010)]{Devreese10} Devreese, J.P.A.,
Lemmens, D. \& Tempere, J. 2010 Path integral approach to Asian options in
the Black-Scholes model. \textit{Physica A} \textbf{389 (4)}, 780--788.

\bibitem[Di Francesco \& Pascucci (2004)]{DiFrancesco04} Di Francesco, M. \&
Pascucci, A. 2004 On the complete model with stochastic volatility by Hobson
and Rogers. \textit{Proc. R. Soc. Lond. Ser. A} \textbf{460(2051)},
3327--3338.

\bibitem[Di Francesco \& Pascucci (2005)]{DiFrancesco05} Di Francesco, M. \&
Pascucci, A. 2005 On a class of degenerate parabolic equations of Kolmogorov
type. \textit{Applied Mathematics Research eXpress} \textbf{3}, 77-116.

\bibitem[Duffie \textit{et al.} (2003)]{Duffie03} Duffie, D., Filipovic, D.
\& Schachermayer, W. 2003 Affine processes and applications in finance. 
\textit{Ann. Appl. Probab.} \textbf{13(3)}, 984--1053.

\bibitem[Duffie \& Kan (1996)]{Duffie96} Duffie, J.D. \& Kan, R. 1996 A
yield-factor model of interest rates. \textit{Mathematical Finance} \textbf{6%
}, 379-406.

\bibitem[Duffie \textit{et al.} (2000)]{Duffie00} Duffie, D., Pan, J. \&
Singleton, K. 2000 Transform analysis and asset pricing for affine
jump-diffusions. \textit{Econometrica} \textbf{68(6)}, 1343-1376.

\bibitem[Duong \& Tran (2018)]{Duong18} Duong, M.H. \& Tran, H.M. 2018 On
the fundamental solution and a variational formulation for a degenerate
diffusion of Kolmogorov type. \textit{Discrete \& Continuous Dynamical
Systems: Series A} \textbf{38(7)}.

\bibitem[Dupire (1994)]{Dupire94} Dupire, B. 1994 Pricing with a smile. 
\textit{Risk Magazine }\textbf{7(1)}, 18-20.

\bibitem[Ebeling \textit{et al.} (2008)]{Ebeling08} Ebeling, W.,
Gudowska-Nowak, E. \& Sokolov, I.M. 2008 On stochastic dynamics in physics
-- remarks on history and terminology. \textit{Acta Physica Polonica B} 
\textbf{39(5)}, 1003-1017.

\bibitem[Feller (1951)]{Feller51} Feller, W. 1951 Two singular diffusion
problems. \textit{Ann. Math.} \textbf{54(1)}, 173--182.

\bibitem[Feller (1952)]{Feller52} Feller, W. 1952 The parabolic differential
equations and the associated semi-groups of transformations. \textit{Ann.
Math.} \textbf{55}, 468--518.

\bibitem[Filipovic (2009)]{Filipovic09} Filipovic, D. 2009 \textit{%
Term-structure models.} Berlin: Springer-Verlag.

\bibitem[Fokker (1914)]{Fokker14} Fokker, A.D. 1914 Die mittlere Energie
rotierender elektrischer Dipole im Strahlungsfeld. \textit{Annalen der Physik%
} \textbf{348(5)}, 810--820.

\bibitem[Friedlander \& Vishik (1991)]{Friedlander91} Friedlander, S. \&
Vishik, M. 1991 Instability criteria for the flow of an inviscid
incompressible fluid. \textit{Phys. Rev. Lett.} \textbf{66}, 2204-2206

\bibitem[Friedlander \& Lipton-Lifschitz (2003)]{Friedlander03} Friedlander,
S. \& Lipton-Lifschitz, A. 2003 Localized instabilities in fluids. In 
\textit{Handbook of Mathematical Fluid Dynamics} \textbf{2}, 289-354.
North-Holland: Amsterdam.

\bibitem[Friz \& Keller-Ressel (2010)]{Friz10} Friz, P. \& Keller-Ressel, M.
2010. Moment explosions in stochastic volatility models. \textit{%
Encyclopedia of quantitative finance}, 1247-1253.

\bibitem[Fukusawa \textit{et al.} (2023)]{Fukusawa23} Fukasawa, M., Maire,
B. \& Wunsch, M. 2023 Weighted variance swaps hedge against impermanent
loss. \textit{Quantitative Finance} \textbf{23(6)}, 1-11.

\bibitem[Gatheral \textit{et al.} (2018)]{Gatheral18} Gatheral, J., Jaisson,
T. \& Rosenbaum, M. 2018 Volatility is rough. \textit{Quant. Finance }%
\textbf{18(6)}, 933-949.

\bibitem[Geman \& Eydeland (1995)]{Geman95} Geman, H. \& Eydeland, A. 1995
Asian options revisited: inverting the Laplace transform. \textit{Risk
Magazine }\textbf{8(4)}, 65-67.

\bibitem[Gershon \textit{et al.} (2022)]{Gershon22} Gershon, D., Lipton, A.,
Rosenbaum, M. \& Wiener, Z. eds. 2022 \textit{Options-45 Years Since the
Publication of the Black-Scholes-Merton Model: The Gershon Fintech Center
Conference}. Singapore: World Scientific.

\bibitem[Giorno \& Nobile (2021)]{Giorno21} Giorno, V. \& Nobile, A.G. 2021
Time-inhomogeneous Feller-type diffusion process in population dynamics. 
\textit{Mathematics} \textbf{9(16)}, 1879.

\bibitem[Guyon (2014)]{Guyon14} Guyon, J. 2014 Path-Dependent volatility. 
\textit{Risk Magazine} \textbf{27(10)}.

\bibitem[Hagan \textit{et al.} (2002)]{Hagan02} Hagan, P., Kumar, D.,
Lesniewski, A. \& Woodward, D. 2002 Managing smile risk. \textit{Wilmott
Magazine} \textbf{(9)}, 84--108.

\bibitem[H\"{a}nggi \textit{et al.} (1990)]{Hanggi90} H\"{a}nggi, P.,
Talkner, P. \& Borkovec, M. 1990 Reaction-rate theory: fifty years after
Kramers. \textit{Reviews of modern physics} \textbf{62(2)}, 251-341.

\bibitem[Heston (1993)]{Heston93} Heston, S. L. 1993 A closed-form solution
for options with stochastic volatility with applications to bond and
currency options. \textit{Review of Financial Studies }\textbf{6}, 327-343.

\bibitem[Hobson \& Rogers (1998)]{Hobson98} Hobson, D. G. \& Rogers, L. C.
G. 1998 Complete models with stochastic volatility. \textit{Math. Finance} 
\textbf{8(1)}, 27--48.

\bibitem[H\"{o}rmander (1967)]{Hormander67} H\"{o}rmander, L. 1967
Hypoelliptic second order differential equations. \textbf{Acta Math.} 
\textit{119}, 147--171.

\bibitem[Hull \& White (1990)]{Hull90} Hull, J. \& A. White, A. 1990 Pricing
interest rate derivative securities. \textit{Review of Financial Studies }%
\textbf{3}, 573-592.

\bibitem[Ibragimov (1985)]{Ibragimov85} Ibragimov, N.H. 1985 \textit{%
Transformation Groups Applied to Mathematical Physics. }Dordrecht: D. Reidel.

\bibitem[Ivasishen \& Medynsky (2010)]{Ivasishen10} Ivasishen, S.D. \&
Medynsky, I.P. 2010 The Fokker Planck Kolmogorov equations for some
degenerate diffusion processes. \textit{Theory of Stochastic Processes} 
\textbf{16(1)}, 57-66.

\bibitem[Jex \textit{et al.} (1999)]{Jex99} Jex, M., Henderson, R. and Wang,
D. 1999 Pricing exotics under the smile. \textit{Risk Magazine} \textbf{%
12(11)}, 72--75.

\bibitem[Kelvin (1887)]{Kelvin87} Kelvin, Lord 1887 Stability of fluid
motion: rectilinear motion of viscous fluid parallel plates. \textit{Phil.
Mag.} \textbf{24}, 188-196.

\bibitem[Klein (1921)]{Klein21} Klein, O. 1921 Zur statistischen Theorie der
Suspension und L\"{o}sungen. Inaugural-Dissertation. Uppsala: Almqvist \&
Wiksells.

\bibitem[Kolmogoroff (1931)]{Kolmogorov31} Kolmogoroff, A. 1931 \"{U}ber die
analytischen Methoden in der Wahrscheinlichkeitsrechnung. \textit{Math. Ann.}
\textbf{104(1)}, 415--458.

\bibitem[Kolmogoroff (1933)]{Kolmogorov33} Kolmogoroff, A. 1933 Zur Theorie
der stetigen zuf\"{a}lligen Prozesse. \textit{Math. Ann.} \textbf{108},
149--160

\bibitem[Kolmogoroff (1934)]{Kolmogorov34} Kolmogoroff, A. 1934 Zufallige
Bewegungen (Zur Theorie der Brownschen Bewegung), \textit{Ann. Math.} 
\textbf{35(1)}, 116--117.

\bibitem[Kovalenko \textit{et al.} (2014)]{Kovalenko14} Kovalenko, S.,
Stogniy, V. \& Tertychnyi, M. 2014 Lie symmetries of fundamental solutions
of one (2+ 1)-dimensional ultra-parabolic Fokker--Planck--Kolmogorov
equation. \textit{arXiv preprint} arXiv:1408.0166.

\bibitem[Kramers (1940)]{Kramers40} Kramers, H.A. 1940 Brownian motion in a
field of force and the diffusion model of chemical reactions. \textit{%
Physica. Elsevier BV.} \textbf{7(4)}, 284--304.

\bibitem[Kuptsov (1972)]{Kuptsov72} Kuptsov, L. P. 1972 The fundamental
solutions of a certain class of elliptic-parabolic second order equations. 
\textit{Differential Equations} \textbf{8}, 1649--1660.

\bibitem[Lanconelli \textit{et al.} (2002)]{Lanconelli02} Lanconelli, E.,
Pascucci, A. \& Polidoro, S. 2002 Linear and nonlinear ultraparabolic
equations of Kolmogorov type arising in diffusion theory and in finance. In: 
\textit{Nonlinear Problems in Mathematical Physics and Related Topics}, vol.
II, 243-265. New York: Kluwer/Plenum.

\bibitem[Langevin (1908)]{Langevin08} Langevin, P. 1908 Sur la th\'{e}orie
du mouvement brownien. \textit{C. R. Acad. Sci. Paris.} \textbf{146},
530--533.

\bibitem[Lewis (2000)]{Lewis00} Lewis, A. L. 2000 \textit{Option Valuation
under Stochastic Volatility with Mathematica Code. }Newport Beach: Finance
Press.

\bibitem[Lifschitz (1991)]{Lifschitz91a} Lifschitz, A. 1991 Short wavelength
instabilities of incompressible three-dimensional flows and generation of
vorticity. \textit{Physics Letters A} \textbf{157(8-9)}, 481-487.

\bibitem[Lifschitz (1995)]{Lifschitz95} Lifschitz, A. 1995 Exact description
of the spectrum of elliptical vortices in hydrodynamics and
magnetohydrodynamics.\ \textit{Phys. Fluids A} \textbf{7}, 1626-1636.

\bibitem[Lifschitz \& Hameiri (1991)]{Lifschitz91b} Lifschitz, A. \&
Hameiri, E. 1991 Local stability conditions in fluid dynamics.\ \textit{%
Phys. Fluids A} \textbf{3}, 2644-2651.

\bibitem[Lipton-Lifschitz (1999)]{Lipton99a} Lipton-Lifschitz, A. 1999
Predictability and unpredictability in financial markets. \textit{Physica D:
Nonlinear Phenomena} \textbf{133(1-4)}, 321-347.

\bibitem[Lipton (1999)]{Lipton99b} Lipton, A. 1999 Similarities via
self-similarities. \textit{Risk Magazine} \textbf{12(9)}, 101--105.

\bibitem[Lipton (2001)]{Lipton01} Lipton, A. 2001 \textit{Mathematical
Methods For Foreign Exchange: A Financial Engineer's Approach.} Singapore:
World Scientific.

\bibitem[Lipton (2002)]{Lipton02} Lipton, A. 2002 The volatility smile
problem. \textit{Risk Magazine} \textbf{15(2)}, 61--65.

\bibitem[Lipton (2018)]{Lipton18} Lipton, A. 2018 \textit{Financial
Engineering: Selected Works of Alexander Lipton.} Singapore: World
Scientific.

\bibitem[Lipton \textit{et al.} (2014)]{Lipton14} Lipton, A., Gal, A. \&
Lasis, A. 2014 Pricing of vanilla and first-generation exotic options in the
local stochastic volatility framework: survey and new results. \textit{%
Quantitative Finance} \textbf{14(11)}, 1899-1922.

\bibitem[Lipton \& Lopez de Prado (2020)]{Lipton20} Lipton, A. \& Lopez de
Prado, M. 2020 A closed-form solution for optimal ornstein--uhlenbeck driven
trading strategies. \textit{International Journal of Theoretical and Applied
Finance} \textbf{23(08)}, 2050056.

\bibitem[Lipton \& Reghai (2023)]{Lipton23} Lipton, A. \& Reghai, A. 2023
SPX, VIX and scale-invariant LSV. \textit{Wilmott Magazine} \textbf{2023(126)%
}, 78-84.

\bibitem[Lipton \& Sepp (2008)]{Lipton08} Lipton, A. \& Sepp, A. 2008
Stochastic volatility models and Kelvin waves.\textit{\ J. Phys. A: Math.
Theor.} \textbf{41}, 344012 (23pp)

\bibitem[Lipton \& Sepp (2022)]{Lipton22} Lipton, A. \& Sepp, A. 2022
Automated market-making for fiat currencies. \textit{Risk Magazine} \textbf{%
35(5)}.

\bibitem[Lipton \& Treccani (2021)]{Lipton21} Lipton, A. \& Treccani, A.
2021 \textit{Blockchain and distributed ledgers: Mathematics, technology,
and economics.} Singapore: World Scientific.

\bibitem[Masoliver (2016)]{Masoliver16} Masoliver, J. 2016 Nonstationary
Feller process with time-varying coefficients. \textit{Phys. Rev. E} \textbf{%
93}, 012122.

\bibitem[Merton (1973)]{Merton73} Merton, R. C. 1973 Theory of rational
option pricing. \textit{Bell Journal of Economics and Management Science }%
\textbf{4}, 141-183.

\bibitem[Merton (1976)]{Merton76} Merton, R. C. 1976 Option pricing when
underlying stock returns are discontinuous. \textit{Journal of Financial
Economics }\textbf{3}, 125-144.

\bibitem[Olver (1986)]{Olver86} Olver, P. J. 1986 Applications of Lie Groups
to Differential Equations, 1st ed. New York: Springer--Verlag.

\bibitem[Orr (1907)]{Orr07} Orr, W. McF. 1907 The stability or instability
of the steady motions of a perfect fluid. \textit{Irish Acad. A} \textbf{27}%
, 9-69.

\bibitem[Ovsiannikov (1982)]{Ovsiannikov82} Ovsiannikov, L.V. 1982 Group
Analysis of Differential Equations. New York: Academic Press.

\bibitem[Pascucci (2005)]{Pascucci05} Pascucci, A. 2005 Kolmogorov Equations
in Physics and in Finance. In: \textit{Progress in Nonlinear Differential
Equations and their Applications}, \textbf{63}, 313-324. Basel: Birkh\"{a}%
user.

\bibitem[Planck (1917)]{Planck17} Planck, M. 1917 \"{U}ber einen Satz der
statistischen Dynamik und seine Erweiterung in der Quantentheorie. \textit{%
Sitzungsberichte der K\"{o}niglich Preussischen Akademie der Wissenschaften}%
, 324--341.

\bibitem[Reghai (2015)]{Reghai15} Reghai, A. 2015 \textit{Quantitative
Finance, Back to Basics}. New York: Palgrave MacMillan.

\bibitem[Risken (1989)]{Risken89} Risken, H. 1989 \textit{The Fokker--Planck
Equation: Method of Solution and Applications.} New York: Springer-Verlag.

\bibitem[Rogers \& Shi (1995)]{Rogers95} Rogers, L.C.G. \& Shi, Z. 1995 The
value of an Asian option. \textit{Journal of Applied Probability} \textbf{%
32(4)}, 1077--1088.

\bibitem[Rubinstein (1994)]{Rubinstein94} Rubinstein, M. 1994 Implied
binomial trees. \textit{Journal of Finance }\textbf{49}, 771-818.

\bibitem[Samuelson (1965)]{Samuelson65} Samuelson, P. A. 1965 Rational
theory of warrant pricing. \textit{Industrial Management Review }\textbf{6},
13-32.

\bibitem[Sch\"{o}bel \& Zhu (1999)]{Schobel99} Sch\"{o}bel, R. \& Zhu, J.
1999 Stochastic volatility with an Ornstein--Uhlenbeck process: an
extension. \textit{Review of Finance} \textbf{3(1)}, 23-46.

\bibitem[Stein \& Stein (1991)]{Stein91} Stein, E. M. \& Stein, J.\ C. 1991
Stock price distributions with stochastic volatility: an analytic approach. 
\textit{Review of Financial Studies }\textbf{4}, 727-752.

\bibitem[Uhlenbeck \& Ornstein (1930)]{Uhlenbeck30} Uhlenbeck, G. E. \&
Ornstein, L. S. 1930 On the theory of Brownian motion. \textit{Physical
Review }\textbf{36}, 823-841.

\bibitem[Vasicek (1977)]{Vasicek77} Vasicek, O.\ A. 1977 An equilibrium
characterization of the term structure. \textit{Journal of Financial
Economics }\textbf{5}, 177-188.

\bibitem[Weber (1951)]{Weber51} Weber, M. 1951 The fundamental solution of a
degenerate partial differential equation of parabolic type. \textit{Trans.
Amer. Math. Soc.} \textbf{71}, 24--37.
\end{thebibliography}
\end{document}